\documentclass[a4paper,12pt]{book}

\usepackage{amssymb}
\usepackage{latexsym}
\usepackage{amsfonts}
\usepackage{amsthm}
\usepackage{amsmath}
\usepackage{graphicx, caption}
\usepackage{booktabs}
\usepackage{rotating}
\usepackage{pdflscape}
\usepackage{longtable}
\usepackage{geometry}

\usepackage[T1]{fontenc}
\usepackage{ae,aecompl}
\usepackage{txfonts}
\usepackage{hyperref}

  \theoremstyle{definition}

\def\Msun{\ifmmode{\rm M_\odot}\else$\rm M_\odot$\fi}
\def\msun{\ifmmode{\rm M_\odot}\else$\rm M_\odot$\fi}
\def\Rsun{\ifmmode{\rm R_\odot}\else$\rm R_\odot$\fi}
\def\rsun{\ifmmode{\rm R_\odot}\else$\rm R_\odot$\fi}

\newcommand{\lsimeq}{\mbox{$\, \stackrel{\scriptstyle <}{\scriptstyle\sim}\,$}}
\newcommand{\gsimeq}{\mbox{$\, \stackrel{\scriptstyle >}{\scriptstyle\sim}\,$}}

\usepackage{graphics}
\usepackage{tabto}
\usepackage{natbib}
\usepackage{sectsty}
\usepackage{caption}
\usepackage{tabto}
\usepackage{setspace}
\usepackage{textcomp}
\usepackage{color}

\chapternumberfont{\LARGE} 
\chaptertitlefont{\Large}
\sectionfont{\large}
\subsectionfont{\it}

\newcommand{\blanknonumber}{\newpage\thispagestyle{empty}}

\begin{document}

\frontmatter

\begin{titlepage}
\begin{center}
\vspace*{\fill}
\huge
         {\bf The Genesis of Magnetic Fields\\in\\White Dwarfs }
\\
\vspace{3cm}
\Large
                          {\bf Gordon P. Briggs\\May 2018}
\vspace{5mm}
\\
                   {\large {\bf Mathematical Sciences Institute}}\\
\vspace{-2mm}
                   {\large {\bf Australian National University}}
\\
\vspace{4cm}
\includegraphics[width=0.45\columnwidth]{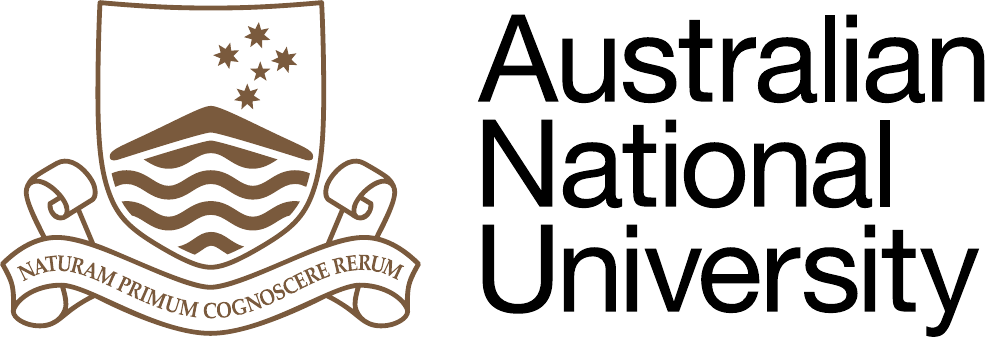}
\\
\vspace{3cm}
\normalsize
         {\bf A thesis submitted for the degree of\\{\large Doctor of Philosophy}\\
         of the Australian National University}
\vspace{1cm}
\newpage
\blanknonumber
\vspace*{5cm}
\large
\textcopyright{ Gordon P. Briggs}\\
May 2018\\
All rights reserved
\end{center}
\end{titlepage}
\blanknonumber
\pagestyle{empty}
\blanknonumber
\vspace*{40mm}
\Large
\begin{center}\emph{For my Mother}
\cleardoublepage
\vspace*{40mm}
\it{`` ... the day of the Lord will come as a thief in the night; in which the heavens shall pass away with a great noise, and the elements shall melt with fervent heat, the earth also and the works that are therein shall be burned up.''}\\
\vspace{12pt}\rm \hspace{115mm}2 Peter 3:10
\normalsize
\vfill\vfill\vfill
\end{center}
\blanknonumber
\chapter{Declaration}\label{declaration}
\normalsize
\pagestyle{plain}
I hereby declare that the work in this thesis is my own except where indicated below. The work was carried out between March 2010 and May 2018 at the Australian National University (ANU), Canberra.  It has not been submitted in whole or in part for any other degree at this or any other university.

\vspace{\baselineskip}\noindent
This thesis is submitted as a Thesis by Compilation in accordance with the relevant ANU policies. The chapters refer to material submitted to, accepted or published by a peer-reviewed astrophysics journal.  The extent of the contribution of this candidate to the research and authorship is detailed below.

\vspace{\baselineskip}\noindent
My collaborators have been Lilia Ferrario (LF, chair of supervisory pane), Christopher A. Tout (CAT, member of the supervisory panel), Dayal T. Wickramasinghe (DTW, member of the supervisory panel), Jarrod Hurley, Adela Kawka, Stephane Vennes and Ernst Paunzen.\\

\begin{itemize}

\item The BSE code was originally developed by Jarrod Hurley.  I made extensive modifications to the code over the years to allow the modelling of the properties of merging white dwarfs and magnetic cataclysmic variables.

\item \textbf{Chapter 2: Merging Binaries and Magnetic White Dwarfs.}  This chapter is a reproduction of the paper published in Monthly Notices of the Royal Astronomical Society, viz: {\color{blue}\textit{Briggs, Ferrario, Tout, Wickramasinghe \& Hurley, MNRAS (2015), 447(2): 1713--1723. Merging binary stars and the magnetic white dwarfs}}. GPB developed all the codes, wrote the paper and made the plots. LF made suggestions, corrected mistakes and added some relevant material.  CAT and DTW reviewed the paper and made further suggestions.
\newpage
\item \textbf{Chapter 3: Genesis of the Magnetic Field.}  This chapter is a reproduction of a paper in publication in Monthly Notices of the Royal Astronomical Society, viz: {\color{blue}\textit{Briggs, Ferrario, Tout \& Wickramasinghe, MNRAS (2018), (In press). Genesis of magnetic fields in isolated white dwarfs}}. I developed all the codes, wrote the paper and made the plots. LF made suggestions, corrected mistakes and added some relevant material.  CAT and DTW reviewed the paper and made further suggestions.

\vspace{\baselineskip}\noindent
\item \textbf{Chapter 4: Origin of Magnetic Fields in Cataclysmic Variables.} This chapter is a reproduction of the paper submitted for publication in Monthly Notices of the Royal Astronomical Society, viz: {\color{blue}\textit{Briggs, Ferrario, Tout \& Wickramasinghe, MNRAS (2018), Origin of Magnetic Fields in Cataclysmic Variables}}. I developed all the codes, wrote the paper and made the plots. LF made suggestions, corrected mistakes and added some relevant material.  CAT and DTW reviewed the paper and made further suggestions.

\vspace{\baselineskip}\noindent
\item \textbf{Chapter 5: A Double Degenerate White Dwarf System.} This chapter is a reproduction of the paper published in Monthly Notices of the Royal Astronomical Society, viz: {\color{blue}\textit{Kawka, Briggs, Vennes, Ferrario,  Paunzen \& Wickramasinghe, MNRAS (2017), 466(1): 1127--1139. A fast spinning magnetic white dwarf in the double-degenerate, super--Chandrasekhar system NLTT 12758}}. The main body of the work was carried out by the co-authors.  I calculated the evolution of the system and wrote the section {\color{blue}\textit{5.4.4 Evolution of NLTT 12758}}.
\vspace{\baselineskip}\noindent

\end{itemize}
\noindent
{I also carried out the work on the following conference paper while LF presented it.}

\begin{itemize}
\item \color{blue}{\textit{Briggs, Ferrario, Tout \& Wickramasinghe, Contributions of the Astronomical Observatory Skalnat\'e Pleso, vol. 48, no. 1, p. 271-272.}}
\end{itemize}
\noindent
This conference paper is not included here.

\vspace{36pt}
\begin{flushright}

Gordon P. Briggs\\
May 2018
\end{flushright}
\hspace*{\fill}
\blanknonumber

\chapter*{Acknowledgements}\label{acknowledgements}
\addcontentsline{toc}{chapter}{Acknowledgements}

I wish to acknowledge the help and encouragement of my supervisors Chris Tout, Dayal Wickramasinghe and especially Lilia Ferrario who coached me along the way about the science of white dwarfs and common envelope evolution.   I thank Jarrod Hurley for providing the SSE and BSE computer codes and Adela Kawka for inviting me to collaborate and contribute to the paper on NLTT-12758.  A special thank you goes to my little four legged friend Rufus, who kept me company and took me out for doggochinnos when I was stressed.
\newline\\
I also gratefully acknowledge the receipt of an Australian Postgraduate Award.\blanknonumber
\chapter{Abstract}\label{abstract}

Magnetic fields generated by a dynamo mechanism due to differential
rotation during stellar mergers are often proposed as an explanation for
the presence of strong fields in certain classes of magnetic stars, including
high field magnetic white dwarfs (HFMWDs). In the case of the HFMWDs,
the site of the differential rotation has been variously proposed to be the
common envelope itself, the massive hot outer regions of a merged
degenerate core or an accretion disc formed by a tidally disrupted
companion that is subsequently incorporated into a degenerate core.
\newline\newline
In the present study I explore the possibility that the origin of HFMWDs
is consistent with stellar interactions during the common envelope evolution (CEE).
In this picture the observed fields are caused by an $\alpha-\Omega$
dynamo driven by differential rotation. The strongest fields would arise when
the differential rotation equals the critical break up velocity and would occur
from the merging of two stars during CEE or double degenerate (DD) mergers
in a post common envelope (CE) stage. Those systems that do not coalesce
but emerge from the CE on a close orbit and about to initiate mass transfer
will evolve into magnetic cataclysmic variables (MCVs),
\newline\newline
The population synthesis calculations carried out in this work have shown
that the origin of high fields in isolated white dwarfs (WDs) and in WDs
in MCVs is consistent with stellar interaction during common envelope evolution.
I compare the calculated field strengths to those observed and test the correlation
between theory and observation by means of the Kolmogorov--Smirnov (K--S) test
and show that the resulting correlation is good for values of the CE energy
efficiency parameter, $\alpha{_{\rm{CE}}}$, in the range 0.1--0.3.

\blanknonumber
\chapter{Acronyms}\label{acronyms}
\normalsize{
AGB         \tabto*{2.5cm}Asymptotic Giant Branch; \\
CE           \tabto*{2.5cm}Common envelope;\\
CEE         \tabto*{2.5cm}Common Envelope Evolution; \\
CS           \tabto*{2.5cm}Fully or deeply Convective MS Star (Mass $< 0.7\Msun$); \\
CV           \tabto*{2.5cm}Cataclysmic Variable star system; \\
DD          \tabto*{2.5cm}Double Degenerate binary star system; \\
HFMWD   \tabto*{2.5cm}High Field Magnetic White Dwarf; \\
HRD        \tabto*{2.5cm}Hertzsprung--Russell diagram of stellar evolution; \\
ISM         \tabto*{2.5cm}Inter-Stellar Medium;\\
IMF         \tabto*{2.5cm}Initial Mass Function;\\
MCV        \tabto*{2.5cm}Magnetic Cataclysmic Variable star system; \\
MS          \tabto*{2.5cm}Main Sequence of the Hertzsprung-Russell diagram; \\
\Msun      \tabto*{2.5cm}One Solar mass; \\
RGB        \tabto*{2.5cm}Red Giant Branch\\
\Rsun      \tabto*{2.5cm}One Solar Radius; \\
SDSS       \tabto*{2.5cm}Sloan Digital Sky Survey; \\
WD          \tabto*{2.5cm}White Dwarf star; \\
}
\blanknonumber
\tableofcontents\blanknonumber
\listoffigures\blanknonumber
\listoftables\blanknonumber

\mainmatter
\chapter{Introduction}\label{Introduction}
\vspace{-30pt}
For decades astrophysicists have been working to develop computer codes that can model the nuclear and hydrodynamic evolution of stars for ranges of masses and metallicities. This led to computer codes such as those described by \citet{SchM1958} and \citet{Iben65} through to codes such as the Kippenhahn code \citep{Kipp67, KippWeig1990} which in turn led to GARSTEC (Garching Stellar Evolution Code) utilised at the Max-Planck Institute in Garching Germany \citep{WeissSchat2008}.  The TYCHO stellar evolution code derives from previous work on supernovae by David \citet{Arnett1996}. At Cambridge University (UK), the STARS code was originally written by \citet{Eggl71} and was developed through to the versions described by \citet{Han1994} and \citet{Pols95}.
\vspace{-12pt}

\subsubsection{Non-Hydrodynamic Methods:}\vspace{-10pt}
These detailed evolution codes can take many hours to run for a single stellar formulation so that for population studies where it is necessary to evolve a large sample of stars a more rapid method of generating the population must be found. One method is to compute detailed stellar models from a number of computer runs of differing input parameters such as stellar masses and metallicities and to present the results in a tabular form that is easy to interpolate as required \citep[e.g.][]{Schal92, Char93, Mow98, Pols98}.

A second method is to construct a set of formulae that represent the results of the stellar evolution codes analytically.  \citet{Tout96} initially fitted analytical functions of mass and metallicity to stars at all stages of evolution and achieved a fit with an error of generally less than 7.5 per cent in mass and 3 per cent in radius over the range of metallicities from $Z=0.0001$ to 0.03.  Thus these analytic formulae are designed to represent the motion of a star in the Hertzsprung-Russell diagram as a function of time.  Follow-up work was carried out by \citet{Hurley2000} achieving fits within 5 per cent of the detailed computer codes.  They present stellar luminosity, radius and core mass as a function of age from the ZAMS to the remnant stages and describe a mass-loss scheme that can be integrated into the formulae.
\vspace{-12pt}

\subsubsection{Binary Star Evolution}\vspace{-10pt}
\citet{Rap83} used a composite polytrope model for the core and envelope of the stars in a binary system while carrying out detailed stellar evolution of the binary for all other relevant aspects.  This technique, faster than detailed models, was used to study the effects of magnetic braking using a range of braking laws.

The method using analytical formulae to represent the time evolution allows a much faster computation of stellar interactions in binary stars and N-body situations such as cluster environments \citep[e.g.][]{HurShar02, Hurley08}.  In a binary star system Roche-lobe overflow, Common Envelope Evolution (CEE) and magnetic braking with tidal friction are facilitated by the compact nature of the formulae over the tabular interpolation.  \citet{Tout97} provide an algorithm for rapid evolution of binary stars applied to the evolution of Algol variables.  They explain how their algorithm can be incorporated into N-body simulations of colliding stars.

\citet{Hurley2002} present a rapid binary star evolution algorithm, \textsc{bse}, that allows modelling mass transfer, mass accretion, CEE, collision, supernovae kicks as well as spin and orbital momentum losses owing to tidal interactions.  By comparing systems with and without tidal evolution they show that tides are  required to draw correct conclusions from population synthesis studies.  Orbit circularisation occurs on a dynamical timescale that is short compared to the nuclear evolution timescale so orbit eccentricity is of minor importance in the evolution of binary systems.  A comprehensive review of the theory of binary star evolution outlining the various factors that contribute to their interactions can be found in \citet{Tout06}.  He sets out the mathematical basis of the factors, viz: orbit, tides, mass transfer, its stability and period evolution.  He also discusses the binary evolution of Algol binaries and their critical mass ratio, cataclysmic variables, CEE and type Ia supernovae. 

In this work I modify the Binary Star Evolution (BSE) code to model the origin of isolated and binary High Field Magnetic White Dwarfs (HFMWDs).

\vspace{-12pt}
\subsubsection{White Dwarfs and Magnetic Fields}\vspace{-10pt}
This work concentrates on the origin of HFMWDs some of which are observed to have fields as high as $10^9$G.

A number of recent reviews give a good overview of the physics of white dwarfs (WD)s. \citet[][and references therein]{Isern02} discuss their evolution and summarise the four stages of neutrino, fluid, crystallisation and Debye cooling.  They also discuss the use of WDs in the determination of the age of the Galaxy.

\citet[][WF]{wickramasinghe2000} deal extensively with magnetism in isolated and binary WDs.  WF give an extensive review of the methods of measuring magnetic fields in WDs followed by the observations, physical properties and theoretical considerations of isolated HFMWDs.  They finish this extensive review by examining HFMWDs in interacting binary systems in particular the AM~Herculis systems.

The most relevant previous work on which this project is based is that of \citet{Reg95} and \citet{Tout95b}.  In these papers they present a model that could be applied to Cataclysmic Variables (CVs) to explain the presence of strong fields.  In particular, they show that the differential velocity between the increasing orbital rate of the shrinking orbit of the binary combined with the decreasing rotation rate of the envelope sets up an $\alpha-\Omega$ dynamo that creates strong magnetic fields.  They also show that the interaction between stellar winds driven by the magnetic fields and the envelope provides a simple explanation for the range of remnant fields observed in WDs.  This work was then used by \citet{Zangrilli97} to show how dynamo generated fields can interact with a CE to create the orbital period gap of CVs.

\citet{WebWick2002} continue the discussion about the period gap in AM~Her binaries.  They find that magnetic braking causes the angular momentum loss in CVs, and that it is its reduction due to trapping of the secondary's wind by the magnetosphere of the primary that causes Magnetic CVs (MCVs) to fill the period gap.

\vspace{-12pt}
\subsubsection{Competing hypotheses for the origin of magnetic fields in white dwarfs}\vspace{-10pt}
The first model of the formation of magnetic fields in WDs was the fossil field theory, first proposed by \citet{Woltjer64} and \citet{landstreet1967}. They predicted the existence of highly magnetic WDs by proposing that the fields are of a fossil origin from before the main sequence (MS) with magnetic flux frozen in from the ISM and conserved in some way during evolution to the WD phase \citep{Mestel2005}.  

\citet{tou2004} discuss the possibility of magnetic fields in WDs being fossil remnants of the fields in Ap and Bp stars and that their magnetic fields are fossil remnants from fields in the pre--MS stars.  \citet{wickramasinghe2005} propose several scenarios for the origin of HFMWDs.  Their first scenario is that only the chemically peculiar Ap and Bp stars on the main sequence evolve into HFMWDs.  In the second scenario they assume that all intermediate--mass MS stars have large scale fields that are below the detectability limit.  Once these stars evolve to WD stage their magnetic flux is conserved and become HFMWDs. The second scenario gives a better match to the observed mass and field distribution of HFMWDs. They also speculate on the possibility of very low--field magnetic white dwarf having progenitors among the F type stars.  This would suggest a bi-modal distribution of magnetic fields with the HFMWDs having fossil fields originating from upper MS stars and low--field magnetic WDs having dynamo
generated fields in lower MS stars.

Many papers have been written on the fossil field model. However none of them solve the duplicity problem.  That is that HFMWDs should occur as often in detached binaries as in single stars whereas no WD in a binary system has been found to be magnetic.  \citet{liebert2005} discuss the results of the Sloan Digital Sky Survey (SDSS) and the discovery that there are no HFMWDs found in the subset of WDs with main sequence companions.  They give possible solutions for these observations but conclude that the sample size of stars may be too small to resolve the issue.  However a much larger and statistically significant sample of binaries studied by \citet{lie2015} led to the same conclusion.

As an alternative to the fossil field model, \citet{tout2008} examine the possibility of magnetic fields being generated during CEE. They propose that the closer the binary pair at the end of the CE phase the stronger the magnetic field.  They then go on to propose that the binaries that merge while in the CE are the progenitors of the isolated HFMWDs. See also \citet{wickramasinghe2014}. 

\vspace{-12pt}
\subsubsection{Goals of the present work}\vspace{-10pt}
The goal of my research is to test the viability of the formation of magnetic fields during CEE.  A CE arises when the radius of the more massive, more evolved, primary star of a binary star system expands during a normal phase of stellar evolution and the orbital radius of the binary is such that the primary overfills its Roche lobe. Mass transfer from the primary star on to the secondary star then occurs.  As the primary expands further the envelope grows in size until it eventually engulfs both stars.

This CE mechanism, first proposed by \citet{Pac1976} and \citet{Ostriker1976}, describes mass transfer becoming unstable if the normal evolutionary process of the primary donor star is affected by loss of mass to the secondary. If the time scale for mass transfer is short compared with the time scale on which the accretor can adjust thermally to the on-flowing material the accreted layer heats up, expands and fills the Roche lobe of the accretor.  Any further mass loss from the donor star is deposited into the CE that now engulfs both stars.

The transfer of orbital energy into the heating of the envelope causes a spiral-in of the binary orbit that accelerates the mass transfer and leads to a run-away process causing the orbit to spiral-in even faster.  If the primary star is ascending the red Red Giant Branch (RGB) or the Asymptotic Giant Branch (AGB) and has developed a deep convective envelope, its radius increases in response to mass loss.  This combined with the shrinking Roche lobe as the orbit spirals in, causes a dynamically unstable mass transfer to occur \citep{Hje1987, deKool1992, Iben1993}.

The resultant drag on the secondary and the transfer of orbital angular momentum from the secondary to the on-flowing material causes the orbit to shrink.  As the orbit shrinks, the kinetic energy of the orbit increases but the potential energy decreases more.  This loss of energy heats and further expands the envelope, which is then ejected into space.

An important quantitative model of CEE is the energy formulism.  In this model the change in orbital energy $E_{\textrm{orb}}$ of the in-spiralling cores is equated to the energy required to heat and eject the envelope to infinity, the binding energy $E_{\textrm{bind}}$. This ratio is represented by the parameter
	\begin{align*}
	\alpha=\frac{\Delta E_{\textrm{orb}} }{\Delta E_{\textrm{bind}}},\qquad0.0 \le \alpha \le 1.0
	\end{align*}
\citet{Rick12} carried out a hydrodynamic evolution of the CE phase of a low-mass binary composed of a 1.05{\,\Msun} red giant and a 0.6{\,\Msun} companion.  They followed the evolution for five orbits and found that only about 25 per cent of the orbital energy loss goes into ejecting the envelope inferring a value for $\alpha$ of 0.25.  In general, the process ends when the envelope has been ejected and the stars are either on a much tighter orbit or have merged.  Circularization and spiral-in begin rapidly after the beginning of CE and the phase is probably short-lived, of the order 10$^3$yr.  In considering the progenitors of HFMWDs, the interest is in the situation where the two stars have merged while considering the progenitors of MCVs, the interest is in the situation when the two stars emerge from CE on a very tight orbit and are about to exchange mass.  

This work is organised as follows.  In chapter 2, I show that population synthesis studies of stars merging during CEE can explain the incidence of magnetism among WDs and the mass distribution of HFMWDs.  In chapter 3, I show that these calculations can also reproduce very well the observed magnetic field distribution.  In chapter 4, I synthesize a population of binary systems to explore the hypothesis that the magnetic fields in the MCVs also originate during stellar interactions in the CEE phase and find that the observed characteristics of the MCVs are consistent with those of a population of binaries that is born already in contact or close to contact, as first proposed by \citet{tout2008}.  This finding is also in agreement with the hypothesis advanced by \citet{sch2009} that the binaries known as PREPs (pre-polars), where a HFMWD accretes matter from the wind of a low-mass companion, are the progenitors of the MCVs.  Finally, in chapter 5, I show that the evolutionary path of the double degenerate super Chandrasekhar system NLTT\,12758 is consistent with that of a binary that underwent two phases of CEE.  The thesis ends in chapter 6 with a summing up and conclusions derived from the research.

\blanknonumber

\chapter{Merging Binaries and Magnetic White Dwarfs}
\label{Chapter 2}
\vspace{-9mm}
This chapter is a reproduction of the paper published in
Monthly Notices of the Royal Astronomical Society, viz:
\vspace{1mm}
\\
{\color{blue}\textit{Briggs, Ferrario, Tout, Wickramasinghe \& Hurley, MNRAS (2015), 447(2): 1713--1723. Merging binary stars and the magnetic white dwarfs}}
\vspace{-12pt}

\section{Abstract}
\label{Paper1abstract}
\vspace{-12pt}
A magnetic dynamo driven by differential rotation generated when 
stars merge can explain strong fields in certain classes of magnetic
stars, including the HFMWDs.  In their case the site of the differential
rotation has been variously proposed to be within a CE, the massive
hot outer regions of a merged degenerate core or an accretion disc
formed by a tidally disrupted companion that is subsequently incorporated
into a degenerate core.  We synthesize a population of binary systems to
investigate the stellar merging hypothesis for observed single
HFMWDs. Our calculations provide mass distribution and the
fractions of WDs that merge during a CE phase or as DD systems in
a post CE phase.  We vary the CE efficiency parameter $\alpha$ and compare
with observations.  We find that this hypothesis can explain both
the observed incidence of magnetism and the mass distribution of
HFMWDs for a wide range of $\alpha$.  In this model, the majority of
the HFMWDs are of the Carbon--Oxygen type and merge within a CE.
Less than about a quarter of a per cent of HFMWDs
originate from DD stars that merge after CE evolution and these
populate the high-mass tail of the HFMWD mass distribution.\newline
\newline\noindent
Keywords: white dwarfs -- magnetic fields -- binaries: general -- stars: evolution

\section{Introduction}\label{Introd}
\vspace{-8pt}
Magnetic fields are seen in main-sequence stars of most spectral types.
They are usually considered to be either of fossil origin, arising from a
conserved primordial field, or generated in a contemporary dynamo
\citep{Mestel2005}.
The latter is the accepted explanation for magnetic
stars with convective envelopes such as the low-mass ($M < 1.5\,$\msun)
main-sequence stars.  The origin of the fields in the higher-mass magnetic
Ap and Bp main-sequence stars with radiative envelopes is less certain.
While a fossil origin remains possible, it has been proposed that
magnetic fields may be generated by a dynamo mechanism driven by
various instabilities, including the magnetorotational instability, in
differentially rotating radiative regions of single stars 
\citep[see e.g.][]{potter2012}.

The origin of the HFMWDs has been the topic of much discussion in
recent years.  The incidence of
magnetism in WDs in the high field group ($B>10^6$\,G) is
estimated to be about 8-16\,per cent \citep{Liebert2003, Kawka2007}.
A traditional explanation has been that the fields are of a fossil
origin from the main sequence with magnetic flux conserved in some way
during evolution to the WD phase \citep{Mestel2005}.
\citet{Kawka2007} pointed out that the strongly magnetic Ap and
Bp~stars could not be their sole progenitors because the birth rate of
these main-sequence stars is insufficient to explain the observed
birth rate of the HFMWDs.  However this turned out not to be a strong
argument against the fossil hypothesis.  In an earlier paper
\citet{wickramasinghe2005} noted that it could be reconciled if about
$40$\,per cent of late B~stars had fields below the observed threshold
for Ap and Bp~stars.  This would be consistent with the observations
of \citet{power2008} who conducted a volume-limited study of the
magnetic Ap and Bp~stars within 100\,pc of the Sun.  Their study has
shown that the incidence of magnetism in intermediate mass stars
increases with the mass of the stars.  At $1.7\,\rm\msun$ the fraction
of magnetic among non-magnetic stars is only $0.1$\,per cent, while at
$3.5\,\rm\msun$ it is 37.5\,per cent.

Some 50\,per cent of stars are in binary systems. As these evolve some
can interact and merge. So one may expect that some stars that appear
single today are the result of the merging of two stars.  The
possibility of generating strong magnetic fields during such merging
events has often been discussed in the literature as an alternative
explanation for magnetic fields in certain classes of stellar object.
Indeed, as an alternative to the fossil field model, \citet{ferrario2009}
proposed that the strong fields in the magnetic
A, B and O~stars are generated as stars merge.

Here we focus on the hypothesis that the entire class of
HFMWDs with fields $10^6<B/{\rm G}<10^9$ owe their magnetic 
fields to merging \citep{tout2008}. This model was first devised
to explain the observation that there are no examples
of HFMWDs in wide binary systems with late-type companions while
a high fraction of non-magnetic WDs are found in such systems
\citep{liebert2005}. 

In the CE scenario, when a giant star fills its Roche
lobe, unstable mass transfer can lead to a state in which the giant's
envelope engulfs both cores.  As the two cores spiral together, energy
and angular momentum are transferred from their orbit to the
differentially rotating CE until it is ejected, leaving
behind a close binary system, or a merged single object.  In the original
model for formation of HFMWDs \citet{tout2008} envisaged that the
fields are generated by a dynamo in the CE and diffuse
into the partially degenerate outer layers of the proto-WD
before the CE is ejected.  If the end product is a single
star it can have a highly magnetic core and if it is a very close
binary, it can become a MCV. \citet{potter2010} attempted to
model this phenomenon and found a potential problem in that
the time-scale for the diffusion of the field into the WD is
generally significantly longer than the expected CE lifetime.

\citet{wickramasinghe2014} suggested that strong magnetic fields in WDs
are generated by a dynamo process that feeds on the differential
rotation in the merged object as it forms.  A weak poloidal seed field
that is already present in the pre-WD core is amplified by the dynamo
to a strong field that is independent of its initial strength but
depends on the amount of the initial differential rotation.  We note
in this context that weak fields of $B\le1\,$kG may be present in
most WDs \citep{landstreet2012}. Presumably these can be
generated in a core--envelope dynamo in the normal course of stellar
evolution.

\citet{Nordhaus2011} proposed an alternative but similar model
(hereinafter the disc field model).  They noted that if the companion
were of sufficiently low mass it would be disrupted while merging
and form a massive accretion disc around the proto-WD.  Fields
generated in the disc via the magnetorotational instability or other
hydrodynamical instabilities could then be advected on to the surface
of the proto-WD and so form a HFMWD.  Such a model could apply to some
merging cores within the CE, depending on component masses, and to
post-CE merging DDs.  It depends on the time-scale for the diffusion of
the field into the WD envelope.

\citet{garcia2012} used the results of a three-dimensional
hydrodynamic simulation of merging DDs to argue that a massive hot and
differentially rotating convective corona forms around the more
massive component and used equipartition arguments to estimate that
fields of about $3\times 10^{10}\,$G could be generated.  They also
presented a population synthesis study of WDs that formed
specifically as merging DDs, assuming a CE energy
efficiency parameter $\alpha=0.25$, and showed that there is general
agreement with the observed properties of high-mass WDs
($M_{\rm WD}>0.8\,\rm{\Msun}$) and HFMWDs.  However
they did not consider merging when the companion is a
non-degenerate star.

We hypothesize that single WDs that demonstrate a strong
magnetic field are the result of merging events, so we carry out a
comprehensive population synthesis study of merging binary systems for
different CE efficiencies $\alpha$.  We consider all
possible routes that could lead to a single WD.  We isolate
the WDs formed by the merging of two degenerate cores, either
as WDs, a red giant plus a WD or two red giants,
from those formed by a giant merging with a main-sequence star and
show that the observed properties of the HFMWDs are generally
consistent with the CE hypothesis for $0.1 \le \alpha \le
0.3$.  Both groups contribute to the observed distribution but
main-sequence companions merging with degenerate cores of giants form
most of the HFMWDs.

\section{Common Envelope Evolution and Formulism}

When one of the stars in a binary system becomes a giant, it expands
and overfills its Roche lobe. Mass transfer soon proceeds typically,
but not always, on a dynamical time-scale \citep{Han2002}. The giant 
envelope rapidly engulfs both the companion star and the core of the
donor to form a CE.  The two dense cores, that of the
giant and the accreting star itself, interact with the envelope,
transferring to it orbital energy and angular momentum.  The envelope
can be partly or wholly ejected and the orbit of the engulfed star
shrinks.  It is not known how long this process takes but it is
generally thought to last for more of a dynamical stellar time-scale
than a thermal or nuclear time-scale.  It probably has never been
observed.  If the companion succeeds in fully ejecting the envelope
the two cores survive in a binary system with a much smaller
separation.  If the envelope is not fully ejected the orbit may
completely decay and the two stars coalesce.  When the envelope of a
giant engulfs a degenerate companion the two cores can merge but if
the companion is non-degenerate it either merges with the envelope or
accretes on to the giant core.  When the initial masses of the two
stars are within a few percent both can expand to giants at the same
time and Roche lobe overflow (RLOF) leads to a double CE.

The CE process was first proposed to explain binary star systems,
such as CVs, whose orbital separations are smaller than the original
radius of the progenitor primary star.  A mechanism was needed
to explain how this could occur.  The possible existence of CEs
was first proposed by \citet{Bisnovatyi1971}.Its qualitative description
is based on evolutionary necessity rather than mathematical physics.
While it is sufficient to explain a variety of exotic stars and binaries that
could not otherwise be explained, a full mathematical model has
yet to be developed to describe the interaction in detail and to test
the various theories.

A simple quantitative model of CEE is the energy or $\alpha$
formulism \citep{vandenHeuvel1976}. For this the change
in orbital energy $\Delta E\rm_{orb}$ of the in-spiralling cores
is equated to the energy required to eject the envelope to infinity,
the binding energy $E\rm{_{bind}}$.  The total orbital energy,
kinetic plus potential, of a binary star with masses ${m_1}$
and~${m_2}$ and separation $a$ is $E_{\rm_{orb}} =-{\rm G}m_{\rm 1}m_{\rm
  2}/2a$.  However the envelope ejection cannot be completely
efficient so \citet{Livio1988} introduced an efficiency parameter
$\alpha$ to allow for the fraction of the orbital energy
actually used to eject the envelope.

\begin{equation}
\Delta E_\textrm{orb} = \alpha E_{\rm{bind}}.
\end{equation}
\vspace{6pt}

 Following \citet{Tauris2001} we use a form of the binding energy
 that depends on the detailed structure of the giant envelope and adopt 

\begin{equation}
E_{\rm bind} = -\frac{{\rm{G}}m_{\rm 1} m_{\rm{1 ,env}}}{{\lambda}R_{1}},
\end{equation}
\vspace{6pt}

where {\it{R}}$_{\rm1}$ is the radius of the primary envelope.  The
constant $\lambda$ was introduced by \citet{deKool1990} to
characterize the envelope structure. Our $\lambda$ depends on the
structure of the particular star under consideration.  It is sensitive
to how the inner boundary between the envelope and the remnant core is
identified \citep{Tauris2001} and includes the contributions from the
thermal energy of the envelope on the assumption that it remains in
equilibrium as it is ejected.

The initial orbital energy is that of the secondary star $m_2$ and
the primary core $m_{\rm 1,c}$ at the orbital separation $a_{\rm i}$
at the beginning of CEE and is given by

\begin{equation}
E_{\rm orb, i}=-\frac12\frac{{\rm G}m_{\rm 1,c} m_{\rm 2}}{a_{\rm i}}
\end{equation}
\vspace{6pt}

and the final orbital energy is

\begin{equation}\label{Ermf}
E_{\rm orb, f}=-\frac12\frac{{\rm G}m_{\rm 1,c} m_{2}}{a_{\rm f}},
\end{equation}
\vspace{6pt}

where $a_{\rm f}$ is the final orbital separation.  Thus we have

\begin{equation}\label{deltaEorb}
\Delta E_{\rm orb}=E_{\rm orb, f}-E_{\rm orb, i}.
\end{equation}

From this we can calculate $a_{\rm f}$ which is the separation of the
new binary if the cores do not merge.  If $a_{\rm f}$ is so small that
either core would overfill its new Roche lobe, then the cores are
considered to merge when $a_{\rm f}$ is such that the core just fills
its Roche lobe. Setting $a_f$ to this separation we calculate $E_{\rm
  orb,f}$ and $\Delta E_{\rm orb}$ with equations \ref{Ermf} and
\ref{deltaEorb}. Then we calculate a final binding energy for the
envelope around the merged core

\begin{equation}
E_{\rm bind,f}=E_{\rm bind, i}+\frac{\Delta E_{\rm orb}}{\alpha}.
\end{equation}
\vspace{3pt}

Assuming this envelope has a normal giant structure $R(m,m_{\rm c})$
we calculate how much mass must be lost.  In the case of a double
CE, the initial orbital energy is that of both cores and
the binding energies of the two envelopes added.

Some difficulties with the energy formulation arise because $\alpha$
can depend on the duration of the CE phase.  If it lasts
longer than a nuclear or thermal time-scale then alterations in the
envelope, owing to adjustments in its thermal equilibrium, can change
its structure and hence $\lambda$.  Changes to the energy output from
the core, owing to the decreasing weight of the diminishing envelope,
can also affect the thermal equilibrium and thence $\lambda$.  We do
not consider these complications in this work. Nor do we include
ionization and dissociation energy, as proposed by \citet{Han1994}
in the envelope binding energy.\vspace{-3mm}

\section{Population synthesis calculations}
\label{sec:calculations}

\begin{table*}[h]
\centering
\captionsetup{width=190mm, labelfont=bf}
\caption{Stellar types distinguished within the {\sc bse} algorithms.}
\begin{tabular}{ r l }
\toprule
Type & Description\\
\midrule
  0.& Deep or fully convective low-mass MS star (CS)\\
  1.& Main-sequence star (MS) \\
  2.& Hertzsprung gap star (HG) \\
  3.& First giant branch (RGB) \\
  4.& Core helium Burning \\
  5.& First asymptotic giant branch (early AGB) \\
  6.& Second asymptotic giant branch (late AGB) \\
  7.& Main-sequence naked helium star \\
  8.& Hertzsprung gap naked helium star \\
  9.& Giant branch naked helium star \\
10.& Helium WD (He CE)\\
11.& Carbon/oxygen WD (CO WD)\\
12.& Oxygen/neon WD (ONe WD)\\
13.& Neutron star \\
14.& Black hole \\
15.& Massless supernova/remnant \\
\bottomrule
\end{tabular}
\label{tab:bsetypes}
\end{table*}

We evolve synthetic populations of binary star systems from the
zero-age main sequence (ZAMS).  Each system requires three initial
parameters, the primary star mass, the secondary star mass and the
orbital period.  The primary masses $M_1$ are allocated between
$0.8$ and $12.0\rm{\,\msun}$ and the secondary star masses
$M_2$ between $0.1$ and $12.0\,\rm{\msun}$.  The binary
orbits are specified by a period $P_0$ at ZAMS between $0.1$ and
$10\,000\,$d and zero eccentricity.  Each parameter was uniformly
sampled on a logarithmic scale for 200\, divisions.  This scheme gives
a synthetic population of some 6 million binary systems. We calculate
the effective number of actual binary systems by assuming that the
primary stars are distributed according to Salpeter's mass function
\citep{Salpeter1955}
$N(M)\,{\rm d}M\propto M^{-2.35}\,{\rm d}M$,
where $N(M)\,{\rm d}M$ is the number of stars with masses between $M$
and $M+{\rm d}M$, and that the secondary stars follow a flat mass
ratio distribution for $q\le 1$ \citep[e.g.][]{Ferrario2012}. The initial
period distribution was taken to be logarithmically uniform in the
range $-1\le \log_{10} P_0/{\rm d} \le 4$.

Each binary system was evolved from the ZAMS to an age of $9.5\,$Gyr,
taken to be the age of the Galactic disc \citep[e.g.][]{Oswalt1996,liu2000},
with the rapid binary star evolution ({\sc bse}) algorithm
developed by \citet{Hurley2002}. This is an extension of their single
star evolution algorithm \citep{Hurley2000} in which they use
analytical formulae to approximate the full numerical hydrodynamic and
nuclear evolution of stars. This includes mass-loss episodes during
various stages of evolution.  The {\sc bse} code adds interactions
between stars, such as mass transfer, RLOF, CEE, supernova kicks
and angular momentum loss by gravitational radiation and magnetic
braking as well as tidal interaction. I summarize the type of stars that
play a role in the {\sc bse} code in Table\,\ref{tab:bsetypes}.

In the {\sc bse} model we use the $\alpha$ (energy) formulism for
CE phases and have taken a fixed $\lambda=0.5$ as
representative of the range expected for our stars.  We take $\alpha$
to be a free parameter between $0.1$ and~$0.9$.  Efficiencies of
$\alpha > 1$ are only possible if additional energy sources are
involved in the process. We do not consider this here. We use the
full suite of mass-loss rates described by \citet{Hurley2000}.
We found that, in order to generate sufficient low-mass WDs,
$\eta=1.0$ for Reimers' mass-loss parameter is necessary so we have
used this throughout.  

\newpage
\begin{landscape}
\begin{table*}
\captionsetup{width=190mm, labelfont=bf}
\caption{Fraction of binary systems that merge during CE for various
values of $\alpha$.  The fraction of WDs born from merged stars in a single
generation of binary systems of age 9.5\,Gyr (the age of the Galactic disc) is $N$.
The remaining six columns give the smallest and the largest parameters on the
search grid for systems that are found to have merged.  The parameters are the
progenitors' ZAMS masses and orbital period.}
\vspace{4mm}
\centering
\begin{tabular}{ c c c c r c c r}
\toprule
$\alpha$ & $N\,$per cent & $M_{1_{\rm min}}/\msun$ & $M_{2_{\rm
    min}}/\msun$ & $P_{0_{\rm min}}/$d &$M_{1_{\rm max}}/\msun
$&$M_{2_{\rm max}}/\msun $ & $P_{0_{\rm max}}/$d \\
\midrule
0.05   & 11.58   &  1.08  &  0.10  &   348.9  & 11.06  &  2.77  &     16.3 \\ 
0.10   & 10.35   &  1.08  &  0.10  &   195.6  & 11.06  &  2.90  &     20.5 \\
0.20   &   8.86   &  1.08  &  0.10  &     97.7  & 11.21  &  2.77  &     20.5 \\
0.25   &   8.17   &  1.08  &  0.10  &     82.1  & 11.21  &  4.06  &   932.9 \\
0.30   &   7.55   &  1.08  &  0.10  &     65.2  & 11.21  &  4.06  &   784.3 \\
0.40   &   6.51   &  1.08  &  0.10  &     48.8  & 11.21  &  4.06  &   587.3 \\
0.50   &   5.70   &  1.08  &  0.10  &     38.7  & 11.21  &  4.06  &   493.7 \\
0.60   &   5.06   &  1.08  &  0.10  &     30.7  & 11.21  &  4.06  &   391.7 \\
0.70   &   4.60   &  1.08  &  0.10  &     25.8  & 11.21  &  3.87  &   195.6 \\
0.80   &   4.18   &  1.08  &  0.10  &     23.7  & 11.21  &  3.12  &     82.1 \\
0.90   &   3.75   &  1.08  &  0.10  &     19.3  & 11.06  &  4.06  &   415.0 \\
\bottomrule
\end{tabular}
\label{tab:statsCE}
\end{table*}

\newpage
\begin{table*}
\captionsetup{width=190mm, labelfont=bf}
\caption{Fraction of merging DD systems, WDs formed by merging of
two degenerate objects outside a CE in a single generation of binary
systems of age 9.5\,Gyr.  Other columns are as in Table~\ref{tab:statsCE}  }
\vspace{4mm}
\centering
\begin{tabular}{ c c c c r c c r }
\toprule
$\alpha$ & $N\,$per cent & $M_{1_{\rm min}}/\msun$ & $M_{2_{\rm
      min}}/\msun$ & $P_{0_{\rm min}}/$d & $M_{1_{\rm max}}/\msun $ &
  $M_{2_{\rm max}}/\msun$ & $P_{0_{\rm max}}/$d \\
\midrule
0.05   & 4.49 x 10$^{-5}$   &  2.41  &  1.79  & 1867.9  &   4.21  &  2.17  &  3331.3  \\
0.10   & 4.89 x 10$^{-4}$   &  2.02  &  1.79  & 1245.9  &  4.21  &  2.28  &  2097.0  \\
0.20   & 1.01 x 10$^{-4}$   &  1.99  &  1.98  &   932.9  &  4.21  &  2.28  &  1867.9  \\
0.25   & 1.29 x 10$^{-4}$   &  1.99  &  1.98  &   784.3  &  4.21  &  2.28  &  1867.9  \\
0.30   & 1.69 x 10$^{-4}$   &  1.52  &  1.52  &   587.3  &  4.27  &  2.23  &  1867.9  \\
0.40   & 2.62 x 10$^{-4}$   &  1.52  &  1.52  &   587.3  &  4.21  &  2.34  &  1663.8  \\
0.50   & 3.42 x 10$^{-4}$   &  1.52  &  1.52  &   587.3  &  4.27  &  2.28  &  1570.3  \\
0.60   & 4.07 x 10$^{-4}$   &  1.52  &  1.52  &   587.3  &  6.24  &  1.59  &      12.2  \\
0.70   & 4.36 x 10$^{-4}$   &  1.52  &  1.52  &   587.3  &  6.33  &  1.59  &      11.5  \\
0.80   & 4.11 x 10$^{-4}$   &  1.54  &  1.52  &   587.3  &  6.59  &  1.71  &      10.2  \\
0.90   & 3.74 x 10$^{-4}$   &  1.54  &  1.52  &   587.3  &  6.42  &  1.71  &        9.7  \\
\bottomrule
\end{tabular}
\label{tab:statsDD}
\end{table*}
\end{landscape}
\newpage

Alternatively sufficient low-mass WDs could be formed with smaller $\eta$
if the Galactic disc were somewhat older. \citet{Meng2008}
produce them with $\eta=0.25$ in populations
of 12\,Gyr in age. The metallicity is taken to be solar ($Z = 0.02$)
in all our calculations.

From all evolved systems we select those that could generate single
HFMWDs. To this end we select all pairs of WDs that merge
outside any CE and leave a single WD
remnant. These are our WD--WD (DD) mergers. Added to
these are WD remnants of systems that underwent at least one
CE phase and merged during the last CE phase
and satisfy two further criteria. Firstly, either one or both of the
stars must have a degenerate core before merging and secondly, there
must be no further core burning before the remnant WD is
exposed. We assume that such a core burning would be convective and
destroy any frozen-in high magnetic field.

\section{Population Synthesis Results}

\begin{figure*}
\begin{center}
\includegraphics[trim={2mm -5mm -2mm -2mm}, width=15cm]{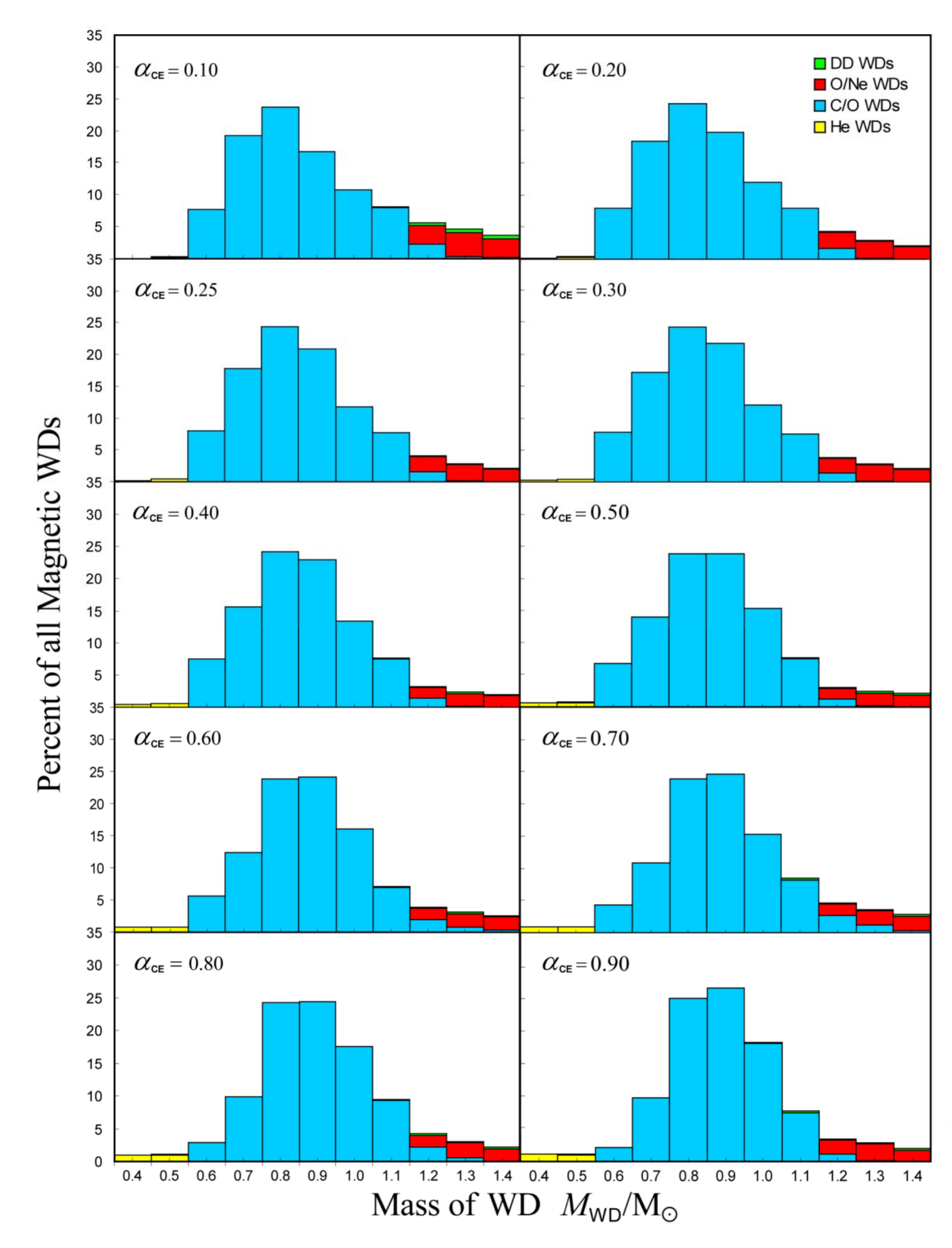}
\end{center}
\captionsetup{width=140mm, labelfont=bf, margin={5mm,-2mm}, justification=justified}
\vspace{-10mm}
\caption{Theoretical mass distribution of remnant WDs formed
  by merging for a range of values $\alpha$ and a Galactic disc age of
  9.5\,Gyr. ''DD WDs'' are WDs resulting from DD mergers, ''ONe WDs''
  are Oxygen--Neon WDs, ''CO WDs'' are Carbon--Oxygen WDs
  and ''He WDs'' are Helium WD remnants after merging.}
\label{fig:WDmasses}
\newpage
\end{figure*} 
\vspace{-3pt}

Assuming a constant star formation rate, each synthetic population was
integrated to the Galactic disc age over the entire parameter space
with $0.05\le\alpha\le0.9$.  Table~\ref{tab:statsCE} lists the
fraction by type of all binary systems that merge in a CE
and Table~\ref{tab:statsDD} those that merge as DDs in a single
generation of stars of age 9.5\,Gyr. The tables also show the limits
of the parameter space within which the cores merge.  The minimum ZAMS
masses of the systems that merged and ended their lives as single
WDs are listed in the columns $M_{1_{\rm min}}$ and
$M_{2_{\rm min}}$ and the minimum initial period in the column
$P_{0_{\rm min}}$.  The maximum ZAMS masses and period are shown in
the columns $M_{1_{\rm max}}$, $M_{2_{\rm max}}$ and $P_{0_{\rm
    max}}$.  For systems that merge during a CE phase the
minimum ZAMS primary mass is determined by the age of the Galactic
disc and thus by the time taken by this star to evolve off the main
sequence.  For the DD route both stars must evolve to WDs.
After the last CE episode, the two stars continue their
evolution to the WD final stage.  The stars are then brought
together by gravitational radiation and eventually coalesce.  This
process takes longer than the CE route.  As a consequence,
the main-sequence evolution lifetime of the primary star
must be shorter and thus the minimum ZAMS mass must be larger than
that required to merge during CEE.  Otherwise such systems would
not be able to coalesce within the age of the Galactic disc.\vspace{-2mm}

\begin{table}
\centering
\captionsetup{width=100mm, labelfont=bf, margin={20mm,19mm}, justification=justified}
\caption{Types and fractions per cent of WDs formed from
 CE and DD by merging binary systems in a population
 aged 9.5\,Gyr.  All DD WDs are of CO type.}
\label{tab:WDtypes}

\begin{tabular}{c r r r r c }
\toprule
$\alpha$  &  \multicolumn{3}{c}{Common}  &  {Double}   \\
&   \multicolumn{3}{c}{envelope}  &   {degenerate} \\
& He &   CO  & ONe & CO  \\
\midrule
0.05  &  0.04 &  88.77 & 11.04 &	0.15  \\
0.10  &  0.16 & 88.73  &   9.79 & 1.32  \\ 
0.20  &  0.43 & 92.14  &   7.08 & 0.36  \\
0.25  &  0.55 & 92.24  &   6.74 & 0.47  \\
0.30  &  0.68 & 92.10  &   6.63 & 0.58  \\
0.40  &  0.94 & 92.89  &   5.42 & 0.75  \\
0.50  &  1.20 & 92.55  &   5.41 & 0.84  \\
0.60  &  1.45 & 91.70  &   5.95 & 0.89  \\
0.70  &  1.68 & 91.20  &   6.27 & 0.85  \\
0.80  &  1.92 & 91.12  &   6.12 & 0.84  \\
0.90  &  2.20 & 90.42  &   6.47 & 0.91  \\
\bottomrule
\end{tabular}
\vspace{10mm}
\end{table}
\vspace{10mm}

\vspace{-8mm}
For low values of $\alpha$ the envelope clearance efficiency is low and the time
for the envelope to exert a drag force on the orbit is largest.
Correspondingly, Table\,\ref{tab:statsCE} shows that, for low
$\alpha$, the number of coalescing stars in the CE is maximal.

As $\alpha$ increases, the time for ejection of the
envelope decreases and the number of systems that merge while still in
the CE also decreases.  WDs formed from merged
stars are of the three types He, CO and ONe.  The small fraction of He
WDs increases with $\alpha$ while that of the ONe WDs falls.  The He WDs
originate when RGB stars coalesce with very low-mass main-sequence stars.
At low $\alpha$ these stars merge when there is very little envelope left
and the resulting giant can lose the rest of its envelope before helium ignition.
As $\alpha$ is increased, more of the envelope remains after coalescence and the
stars pass through core helium burning before being exposed as CO
WDs.  The ONe WDs form when the most evolved AGB
stars coalesce with their companions.  These stars have only rather
weakly bound envelopes so that as $\alpha$ is increased more of them
emerge from the CE phase detached.  For the DD case we
find that only CO WDs are formed in the models.
Table~\ref{tab:WDtypes} sets out the types and fractions of all WDs
that form from CE and DD merging systems as a function of $\alpha$.
The lack of merged He WDs seems to indicate that, while it is true that
very low-mass WDs ($M \lsimeq 0.4\,\rm{\msun})$ must arise from binary
interaction, they do not arise from DD mergers within a Galactic disc age of 9.5\,Gyr.

\subsection{Example Evolutionary Histories}

\vspace{-6pt}
The precise evolutionary history of a binary system depends on
its particular parameters.  For example the number of CE
events that can occur can vary from one to several \citep{Hurley2002}.
Here we give a few examples to illustrate the difference between
CE and DD merging events.\vspace{-12pt}

\subsubsection{Common Envelope Coalescence\vspace{-6pt}}
Table~\ref{tab:evolCE} sets out the evolutionary history of an
example system that merges during a CE with $\alpha=0.2$.
The progenitors are a primary star S1 of $4.44\rm\,\msun$ and a
secondary S2 of sub-solar mass $0.72\rm\,\msun$.  At ZAMS the initial
period is $219.6\,$d and the orbit is circular with a separation of
$264.7\,\rm R _\odot$.  S1 evolves first and reaches the early AGB at 161.77\,Myr
having lost $0.02\,\rm\msun$ on the way.  Roche lobe overflow starts 0.2\,Myr
later with mass flowing from S1 to S2.  At this point the orbital separation
has decreased to 141.4$\,\textrm{R}_\odot$ because orbital angular momentum
has been lost through tidal spin up of S1.  A CE develops and the two cores
coalesce when their separation reaches $0.53\,\textrm{R}_\odot$.
A further $0.6\,\rm\msun$ of the envelope has been lost.  At 162.78\,Myr,
approximately 0.9\,Myr after coalescing, S1 becomes a late stage AGB star.
After a further 0.7\,Myr it becomes a CO WD.\vspace{-9pt}

\subsubsection{DD coalescence\vspace{-6pt}}

In the DD pathway both stars survive the CE without
merging and both continue to evolve to WDs approaching each
other through gravitational radiation to eventually coalesce.
Table~\ref{tab:evolDD} illustrates this for $\alpha=0.1$.  At ZAMS the
progenitors are a $3.7\,\rm\msun$ primary and a $1.9\,\rm\msun$
secondary with an initial period of 3\,444\,d and a separation of
$1603\,\textrm{R}_\odot$, again in a circular orbit.  The primary
evolves through to a late stage AGB star after 270.5\,Myr losing
$0.6\,\rm\msun$ on the way. The separation falls
to~$1509\,\textrm{R}_\odot$.  As a late AGB star S1 loses
$0.9\,\rm\msun$ of which $0.02\,\rm\msun$ is accreted by S2 from the
wind.  Approximately 0.5\,Myr later, at 271\,Myr with S1\, of mass
$2.68\,\rm\msun$ and S2 $1.95\,\rm\msun$, RLOF commences and a CE
develops.  The orbital separation falls to
$374\,\rm\textrm{R}_\odot$ when the envelope is ejected.  S2 continues
to evolve, first as a blue straggler then through the Hertzsprung gap,
red giant and core helium burning stages until it becomes an early AGB
star at 1513.4\,Myr.  At 1517.3\,Myr RLOF begins again and a second
CE forms.  At an orbital separation of only
$2.43\,\rm{R}_\odot$ the envelope is ejected and S2 emerges as a CO
WD of mass $0.54\,\rm\msun$.  A long period of orbital
contraction by gravitational radiation follows until at 9120.8\,Myr
the two WDs are separated by $0.04\,\textrm{R}_\odot$ and
RLOF from S2 to S1 begins followed rapidly by coalescence of the DDs.
The remnant star is still a CO WD but now of mass $1.36\,\rm\msun$.

\clearpage
\begin{landscape}
\begin{table*}
\captionsetup{width=190mm, format=hang, labelfont=bf, margin={30mm,29mm}, justification=justified}
\caption{Evolutionary history of an example binary system that
merges during CE. \\Here $\alpha=0.2$, $P_0= 219.6\,$d,
S1 is the primary star and S2 is the secondary star.}

\label{tab:evolCE}
\centering
\begin{tabular}{ c r c c c l }
\toprule
Stage  & Time/Myr & $M_1/\msun$ & $M_2/\msun$ & $a/{\rm R}_\odot$ & Remarks \\
\midrule
1    &      0.0000    &  4.444   &   0.719   &   264.679   &   ZAMS \\
2    &  138.1295    &  4.444   &   0.719   &   264.679   &   S1 becomes a Hertzsprung gap star \\
3    &  138.7479    &  4.444   &   0.719   &   264.739   &   S1 becomes a red giant  \\
4    &  139.1676    &  4.443   &   0.719   &   179.877   &   S1 starts core helium burning. Some mass loss occurs \\
5    &  161.7637    &  4.402   &   0.719   &   181.495   &   S1 first AGB \\
6    &  161.9691    &  4.402   &   0.719   &   141.380   &   S1 begins RLOF \\
7    &  161.9691    &  4.524   &           - &  { } 0.529   &   CE: S1, S2 coalesce; RLOF ends \\  
8    &  162.8725    &  4.494   &           - &               -  &   S1 becomes late AGB \\
9    &  163.5543    &  0.924   &           - &               -  &   S1 becomes a CO WD \\
\bottomrule
\end{tabular}
\vspace{10mm}
\end{table*}
\end{landscape}

\clearpage
\begin{landscape}
\begin{table*}
\captionsetup{width=195mm, format=hang, labelfont=bf, margin={20mm,20mm}, justification=justified}
\caption{Evolutionary history of an example of WD that formed
in a DD coalescence.\\Here $\alpha=0.1$, $P_0= 3144$
\,days, S1 is the primary star and S2 the secondary star.}
\label{tab:evolDD}
\vspace{4mm}
\centering
\begin{tabular}{ c r c c r l }
\toprule
Stage  & Time/Myr & $M_1/\msun$ & $M_2/\msun$ & $a/{\rm R}_\odot$ & Remarks \\
\midrule
{ }1   &         0.0000   &   3.673   &   1.928   &   1603.362   &   ZAMS \\
{ }2   &     222.4734   &   3.673   &   1.928   &   1603.362   &   S1 becomes a Hertzsprung gap star \\
{ }3   &     223.6164   &   3.673   &   1.928   &   1603.416   &   S1 becomes a Red Giant \\
{ }4   &     224.6021   &   3.672   &   1.928   &   1603.678   &   S1 starts core helium burning \\
{ }5   &     268.5530   &   3.645   &   1.928   &   1611.505   &   S1 becomes early AGB \\
{ }6   &     270.4541   &   3.614   &   1.928   &   1583.219   &   S1 becomes late AGB \\
{ }7   &     270.9681   &   2.682   &   1.947   &   1509.115   &   S1 begins RLOF, mass transfers on to S2, mass loss occurs \\
{ }8   &     270.9681   &   0.821   &   1.947   &     374.233   &   CEE begins, S1 emerges as a CO WD, RLOF ends \\
{ }9   &   1260.0681   &   0.821   &   1.947   &     374.233   &   Begin Blue Straggler phase \\
  10   &   1267.0548   &   0.821   &   1.947   &     374.233   &   S2 becomes a Hertzsprung gap star \\
  11   &   1277.4509   &   0.821   &   1.946   &     374.245   &   S2 becomes a Red Giant \\
  12   &   1306.9423   &   0.821   &   1.943   &     375.353   &   S2 starts core helium burning \\
  13  &  1513.3615     &   0.821   &   1.926   &     377.768   &   S2 becomes early AGB \\
  14  &  1517.2953     &   0.821   &   1.913   &     324.600   &   S2 begins Roche lobe overflow \\
  15  &  1517.2953     &   0.821   &   0.536   &         2.433   &   CEE begins. S2 evolves to a CO WD, RLOF ends \\
  16  &  9120.8467     &   0.821   &   0.536   &         0.040   &   S2 begins RLOF\\
  17  &  9120.8467     &   1.357   &          -  &         0.000   &   S1, S2 coalesce \\
\bottomrule
\end{tabular}
\end{table*}
\end{landscape}
\newpage

\subsection{Mass distribution of the synthetic population}

\begin{figure}
\begin{center}
\includegraphics[width=145mm]{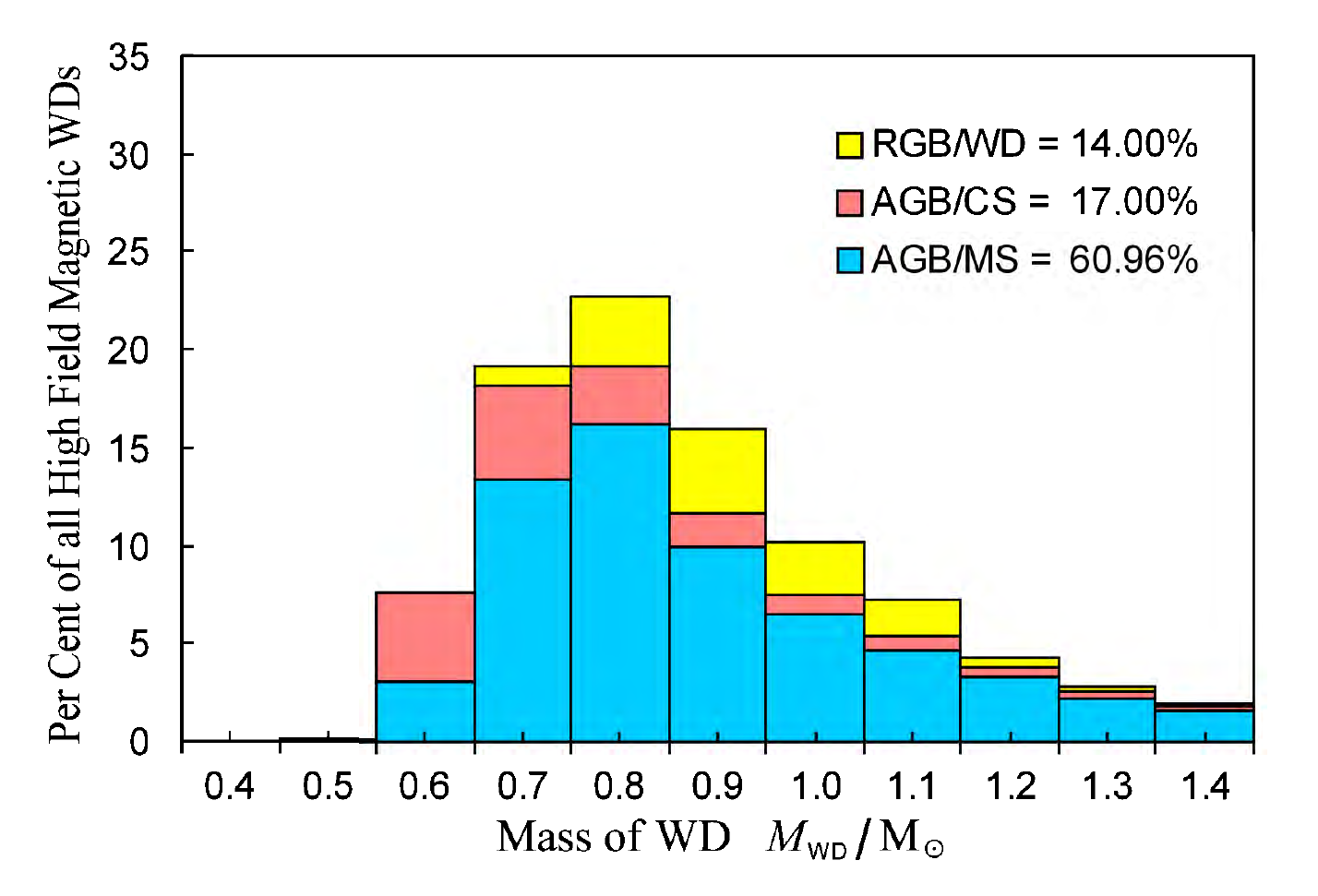}
\vspace{-6mm}
\captionsetup{width=195mm, labelfont=bf, margin={6mm,8mm}, justification=justified}
\caption{Mass distribution of theoretical HFMWDs for $\alpha=0.10$ separated
  according to their pre-CE progenitors.  Other paths also contribute but
  are less than 1\,per cent of the total.  The Galactic disc age is chosen to be 9.5\,Gyr.
  The stellar types are identified in Table~\ref{tab:bsetypes}.}
\label{fig:Path}
\end{center}
\end{figure}

With the selected CE and DD merged systems we generate a
population of putative MWDs by integration over time
from $t=0$ to $9.5\,$Gyr, our chosen age for the Galactic disc.  The
star formation rate is taken to be constant over the lifetime of the
Galactic disc.  Whereas Tables\,\ref{tab:statsCE} and
\ref{tab:statsDD} show the relative numbers of merged WDs
from a single generation of binary stars, continuous star formation
over the lifetime of the Galaxy builds up a population of WDs
that favours higher-mass systems because at lower-mass, especially in
later generations, they do not have enough time to evolve.  Similarly,
the slow orbital contraction by gravitational radiation means that
potential DD coalescence in later generations is not complete and the
fraction of those WDs is further reduced in the present day
population.  Fig.~{\ref{fig:WDmasses}} shows the mass distribution for
CO, ONe and DD WDs in a present day population formed over
the age of the Galactic disc, 9.5\,Gyr.  Fig.~{\ref{fig:Path}} shows
the contributions from the various pre-CE progenitor
pairs that formed the post-CE WDs either through
the CE or DD path when $\alpha$ = 0.1.  Other paths also
contribute but to less than 3\,per cent of the total each.
Table~\ref{tab:Contribs} lists their contributions summed over all
WD masses.

In order to calculate the incidence of HFMWDs we used the same
{\sc{bse}} code to model single star evolution through to the WD stage
also for a Galactic disc age of 9.5\,Gyr under the assumption that all WDs
originating from single star evolution are non-magnetic.. Table\,\ref{tab:MagWDperc}
sets out the incidence of HFMWDs as a percentage of the incidence of field
WDs for a range of $\alpha$.

\begin{table}
\centering
\vspace{6pt}
\captionsetup{width=140mm, labelfont=bf, margin={6mm,8mm}, justification=justified}
\caption{The contributions per cent of pre-CE progenitor pairs
to theoretical HFMWDs when $\alpha = 0.1$. The stellar type `CS' is a deeply
or fully convective low-mass main sequence star (see Table \ref{tab:bsetypes}).}
\label{tab:Contribs}
\vspace{6pt}

\begin{tabular}{c c}
\toprule
Progenitor pairs  & Fraction per cent\\
\midrule
AGB/MS	           & 60.96  \\
AGB/CS	           & 17.00  \\
RGB/CO WD	      & 14.00  \\
AGB/HG                & {  }2.72  \\
AGB/CO WD          & {  }2.21  \\
CO WD/CO WD      & {  }1.32  \\
RGB/RGB	           & {  }0.97  \\ 
RGB/AGB	           & {  }0.46  \\
RGB/CS                & {  }0.16  \\
AGB/AGB              & {  }0.20  \\
\bottomrule
\end{tabular}

\end{table}

\begin{table}
\vspace{15mm}
\captionsetup{width=140mm, labelfont=bf, margin={5mm,7mm}, justification=justified}
\caption{The theoretical incidence of HFMWDs as a fraction of magnetic to non-magnetic field WDs as a function of the CE efficiency parameter $\alpha$.}
\label{tab:MagWDperc}
\centering
\vspace{2mm}
\begin{tabular} { l r c r}

\toprule
$\alpha$ & \multicolumn{3}{c}{HFMWDs per cent} \\
&   {CE} & {DD} & {Total}  \\
\midrule
0.05     & 21.63 & 3.16 x 10$^{-2}$  &   21.67  \\
0.10	& 18.99 & 2.58 x 10$^{-1}$  &  19.25  \\
0.20	& 16.12 & 5.80 x 10$^{-1}$  &  16.18  \\
0.25     & 14.78 & 7.02 x 10$^{-2}$  &  14.85  \\
0.30	& 13.50 & 8.03 x 10$^{-2}$  &  13.58  \\
0.40	& 11.85 & 8.80 x 10$^{-2}$  &  11.67  \\
0.50	& 10.10 & 8.64 x 10$^{-2}$ &  10.18  \\
0.60     &   8.94 &  8.11 x 10$^{-2}$ &    9.02  \\
0.70	&   8.15 &  7.01 x 10$^{-2}$  &    8.22  \\
0.80     & 18.99 &  6.33 x 10$^{-2}$  &    7.50  \\
0.90	& 18.99 &  6.24 x 10$^{-2}$  &    6.78  \\
\bottomrule
\vspace{5mm}

\end{tabular}
\end{table}

\section{Comparison with observations}
\label{sec:Compare}

We compare our theoretical predictions with observations of HFMWDs.
Our comparison includes (i)~the incidence of magnetism among single
WDs and (ii)~the mass distribution of single HFMWDs.  This is
not a simple task because the observational data base of HFMWDs is a
mixed bag of objects from many different ground and space-borne
surveys.  It is plagued by observational biases.  In magnitude-limited
surveys, such as the Palomar-Green (PG) or the Hamburg-Schmidt
surveys, one of the biases against the detection of magnetic WDs
 has been that since these are generally more massive than their
non-magnetic counterparts \citep{Liebert1988}, their radii are smaller
and therefore they are less luminous.  Similar biases would also
apply to UV and X-ray surveys.  However \citet{Liebert2003} have
argued that, in any \emph{explicitly magnitude-limited survey}, it may
be possible to correct for the difference in search volume for the
MWDs.  Thus a re-analysis of the data of the PG survey, that took
into account the different volumes that are sampled
by different mass WDs, gave an estimate for the fraction of
HFMWDs of at least $7.9\pm 3$\,per cent \citep{Liebert2003}.
Volume-limited samples are expected to be less affected by the radius
bias but contain very few MWDs with known masses or
temperatures.  A nearly complete volume-limited sample of nearby WDs
by \citet{Kawka2007} shows that up to 21$\,\pm\,$8\,per cent of
all WDs within 13\,pc have magnetic fields greater than about
3\,kG and 11$\,\pm\,$5\,per cent are HFMWDs with $B\ge 1$\,MG.

\begin{figure}
\begin{center}
\includegraphics[width=150mm]{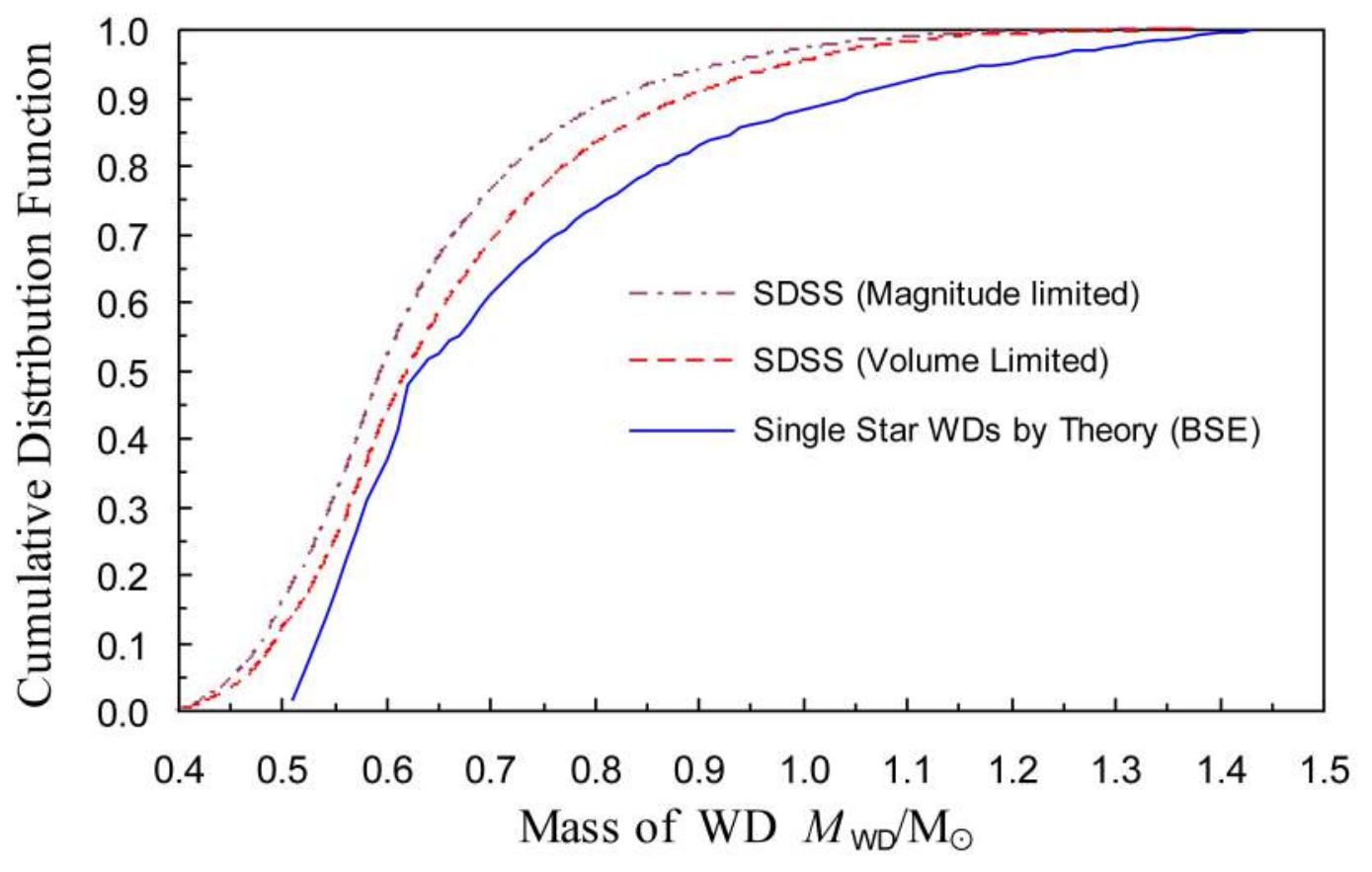}
\vspace{15mm}
\captionsetup{labelfont=bf, margin={5mm,0mm}, justification=justified}
\caption{CDFs of masses of observed SDSS
  DR7  \citep{Kleinman2013}  non-magnetic, magnitude-limited and
  converted-volume-limited field WDs and the theoretical
  ({\sc{bse}}) volume-limited population of non-magnetic WDs
  from single star evolution for a Galactic disc age of 9.5\,Gyr.}
\label{fig:SingleCDF}
\end{center}
\end{figure}

The synthetic population generated by {\sc{bse}} is a volume-limited
sample and so is not directly comparable with a magnitude limited
sample such as the Sloan Digital Sky Survey Data Release\,7 (SDSS DR7)
WD catalogue \citep{Kleinman2013} which has 12\,803 members.
\citet{Liebert2003} estimated that the limiting distance to which a
WD can be found in a magnitude-limited survey is proportional
to its radius $R_{\rm{WD}}$.  Thus the survey volume for a given mass
scales as $R^3_{\rm{WD}}$.  We correct this bias by weighting each
WD found by the SDSS in proportion to $1.0/R^3_{\rm{WD}}$
relative to the radius of a 0.8\,\msun WD.  The cumulative
distribution function (CDF) for the corrected mass distribution along
with the CDF for the uncorrected mass distribution of the SDSS WDs
is shown in Fig.~\ref{fig:SingleCDF}.  The theoretical CDF
obtained with {\sc bse} for the mass distribution of single WDs is shown
for comparison.

We note that the {\sc{bse}} code we use does not produce low-mass
WDs because of the limited age of the Galactic disc. However
\citet{Han1994} and \citet{Meng2008} have constructed single star
models using different assumptions utilizing a superwind that produces
low-mass WDs in older populations. This is also reflected in
the inability of the {\sc{bse}} results to demonstrate the
existence of a significant fraction of low-mass He WDs.

From a theoretical point of view the problem of the determination of
surface gravities and masses from line spectra of HFMWDs has also
proved to be insoluble, except for low-field objects ($B \lsimeq
3\,$MG) for which one can assume that the magnetic field does not
affect the atmospheric structure.  In these objects the field
broadening is negligible and standard zero-field Stark broadening
theories can be used to calculate the line wings \citep[e.g.][]{Ferrario1998}
and thus to determine the mass of the MWD.
In principle it should also be possible to use stationary field components
that are insensitive to field structure to estimate gravities from line
profiles for HFMWDs.  Regrettably this is not yet possible because a
full theory of Stark broadening in the presence of crossed electric
and magnetic fields \citep{Main1998} has not yet been developed.
For now, reliable mass determinations are only available for a few
low-field MWDs, for MWDs which have good trigonometric parallaxes
and MWDs with WD companions whose atmospheric parameters can
be established \citep[e.g. RE\,J0317-853,][]{Barstow1995,Ferrario1997b}.
Currently there are 34~known MWDs with reasonably accurately
determined masses with magnetic fields stronger than $10^5$\,G.  These
are listed in Table~\ref{tab:Obsdb} with their poloidal magnetic field
strengths, effective temperatures, masses and references in the
literature.  If we restrict ourselves to the HFMWDs with $B > 1\,$MG
we end up with 29~objects.  When comparing with our models we exclude
a further two extremely low-mass WDs because it is not
possible to form these within the {\sc bse} formulism.  The most
recent additions to this list are the two common proper motion pairs
from the SDSS reported by \citet{Dobbie2013}. We shall test our
hypothesis on this restricted mass sample with the caveat that we may
well be still neglecting observational biases. We also note that the
observational sample is neither volume nor magnitude limited.

\clearpage
\begin{landscape}
\begin{longtable}{p{5mm} p{22mm} p{55mm} c l l l }

\captionsetup{width=140mm, labelfont=bf, margin={0mm,0mm}, justification=justified}
\caption{Known HFMWDs with poloidal field strength $B_{\rm pol}\ge 10^5$\,G.
In comparison with our models we exclude five of these WDs with
$B_{\rm pol} < 1\,$MG (1, 3, 18, 20 \& 32) and two of extremely low
mass (19 \& 29) that cannot be formed within the {\sc bse} formulism.}
\label{tab:Obsdb}\\
\hline
No. & White Dwarf  & Aliases                        & $B_{\rm pol}/$MG  & $T_{\rm eff}/$K                       & Mass/\msun            & References\\
\hline
\endfirsthead

\hline
No. & White Dwarf  & Aliases                        & $B_{\rm pol}/$MG  & $T_{\rm eff}/$K                       & Mass/\msun            & References\\
\hline
\endhead

\hline
\endfoot

{ }1	& 0009+501     & LHS 1038, G217-037, GR381& $\lesssim  0.2$	& \phantom{0}6540 $\pm$ 150	& 0.74 $\pm$ 0.04		&  1,23 	\\
{ }2	& 0011--134     & LHS 1044, G158-45		& $16.7 \pm 0.6$	& \phantom{0}3010 $\pm$ 120	& 0.71 $\pm$ 0.07		&  2,3	\\
{ }3	& 0257+080     & LHS 5064, GR 476		& $\approx 0.3$	& \phantom{0}6680 $\pm$ 150	& 0.57 $\pm$ 0.09		&  2 \\
{ }4	& 0325--857     & EUVE J0317-855		& $185-450$		& 33000			           & 1.34 $\pm$ 0.03		&  4 \\
{ }5	& 0503--174     & LHS 1734, LP 777-001	& $7.3 \pm 0.2$	& \phantom{0}5300 $\pm$ 120	& 0.37 $\pm$ 0.07		&  2,3 	\\
{ }6	& 0584--001     & G99-37			           & $\approx 10$	& \phantom{0}6070 $\pm$ 100	& 0.69 $\pm$ 0.02		&  5,6,7 \\
{ }7	& 0553+053     & G99-47			           & $20 \pm 3$	& \phantom{0}5790 $\pm$ 110	& 0.71 $\pm$ 0.03		&  2,7,8	\\ 
{ }8	& 0637+477     & GD 77				      & $1.2 \pm 0.2$	& 14870 $\pm$ 120	                     & 0.69				&  9,10 \\
{ }9	&0745+304      & SDSS J074853.07+302543.5	& 11.4		& 21000 $\pm$ 2000	           & 0.81 $\pm$ 0.09		& 44  \\
  10	& 0821--252     & EUVE J0823-254		 & $2.8-3.5$		&  \verb++43200 $\pm$ 1000	& 1.20 $\pm$ 0.04		& 11 \\
  11	& 0837+199     & EG 061$^b$, LB 393	 & $\approx 3$	& 17100 $\pm$ 350	                     & 0.817$\pm$ 0.032	& 12    \\
  12	& 0912+536     & G195-19			            & 100		           & \phantom{0}7160  $\pm$ 190	& 0.75 $\pm$ 0.02		&  2,13,14 \\
  13	&                    & SDSS J092646.88+132134.5	& $210 \pm 25$	& \phantom{0}9500  $\pm$ 500	& 0.62 $\pm$ 0.10		& 15 \\
  14	& 0945+246     & LB11146$^a$			  & 670			& 16000 $\pm$ 2000	           & 0.90 (+0.10, -0.14)	& 16,17 \\
  15	& 1026+117	 & LHS 2273			           & 18			& \phantom{0}7160 $\pm$ 190	& 0.59			& 18 \\
  16	& 1220+234	& PG1220+234			& 3			& 26540			           & 0.81			& 19\\
  17	& 1300+590	& SDSS J13033.48+590407.0	& $\approx 6$	& \phantom{0}6300 $\pm$ 300	& 0.54 $\pm$ 0.06		& 20\\
  18	& 1328+307 	& G165-7		  	           & 0.65		           & \phantom{0}6440 $\pm$ 210	& 0.57 $\pm$ 0.17		& 21\\
  19	& 1300+015	& G62-46 			           & $7.36 \pm 0.11$	& \phantom{0}6040		& 0.25			& 22 \\
  20	& 1350--090	& LP 907-037			     & $\lesssim 0.3$	& \phantom{0}9520 $\pm$ 140	& 0.83 $\pm$ 0.03		& 23,24		\\
  21	& 1440+753	& EUVE J1439+750$^a$ 		& $14-16$		& 20000-50000			& 1.04 (+0.88, -1.19)	& 25 \\
  22	& 1503-070	& GD 175$^a$			& 2.3			& \phantom{0}6990		& 0.70 $\pm$ 0.13	           & 2	 \\
  23	&               & SDSS J150746.80+520958.0	& $65.2 \pm 0.3$	& 18000 $\pm$ 1000	& 0.99 $\pm$ 0.05		           & 15 \\			
  24	&               & SDSS J150813.24+394504.0	& 18.9		& 18000 $\pm$ 2000	& 0.88 $\pm$ 0.06		           & 44 \\
  25	& 1533--057	& PG 1355-057			& $31 \pm 3$	& 20000 $\pm$ 1040	& 0.94 $\pm$ 0.18		           & 26,27,25 \\
  26	& 1639+537	& GD 356, GR 329			& 13			& 7510 $\pm$ 210	& 0.67 $\pm$ 0.07		                      & 2,28,29,45\\
  27	& 1658+440	& 1658+440, FBS 376		& $2.3 \pm 0.2$	& 30510 $\pm$ 200	& 1.31 $\pm$ 0.02		                      & 11,30 \\
  28	& 1748+708	& G240-72			           & $\gtrsim 100$	& \phantom{0}5590 $\pm$ 90	& 0.81 $\pm$ 0.01		& 2,5 \\
  29	& 1818+126	& G141-2$^a$			& $\approx 3$	& \phantom{0}6340 $\pm$ 130	& 0.26 $\pm$ 0.12		& 18,31 \\
  30	& 1829+547	& G227-35			           & $170-180$		& \phantom{0}6280 $\pm$140	& 0.90 $\pm$ 0.07		& 2,8 \\
  31	& 1900+705	& AC +70$^\circ$8247, GW +70$^\circ$8247 & $320 \pm 20$	& 16000		& 0.95 $\pm$ 0.02		& 2,32,33,34,35,36	\\
      	& 		           & EG 129, GL 742, LHS 3424\\	
  32	& 1953--011	& G92-40, LTT 7879, GL 772	& $0.1-0.5$		& \phantom{0}7920 $\pm$ 200	& 0.74 $\pm$ 0.03		& 2,37,38	\\
 	     &		           & LP 634-001, EG 135, LHS 3501 \\
  33	& 2010+310	& GD 229, GR 333		           & $300-700$		& 16000			           &  1.10-1.20			& 33,35,39,40,41,42 \\ 
  34	& 2329+267	& PG 2329+267, EG 161		& $2.31 \pm 0.59$	& \phantom{0}9400 $\pm$ 240	&  0.61 $\pm$ 0.16 & 2,43,24\\
\end{longtable}

\linespread{0.75}\selectfont
\begin{minipage}{216mm}
\vspace{-3mm}
\begin{footnotesize}
$^a$ Unresolved DD, $^b$ Praesepe (M44, NGC 2632)\\
\textbf{References:}
 (1) \citet{Valyavin2005}; 
 (2) \citet{Bergeron2001}; 
 (3) \citet{Bergeron1992}; 
 (4) \citet{Vennes2003}; 
 (5) \citet{Angel1978}; 
 (6) \citet{Dufour2005}; 
 (7) \citet{Pragal1989}; 
 (8) \citet{Putney1995}; 
 (9) \citet{Schmidt92}; 
(10) \citet{Giovannini1998}; 
(11) \citet{Ferrario1998}; 
(12) \citet{Vanlandingham2005}; 
(13) \citet{Angel1977}; 
(14) \citet{Angel1972}; 
(15) \citet{Dobbie2012}; 
(16) \citet{Glenn1994}; 
(17) \citet{Liebert1993}; 
(18) \citet{Bergeron1997}; 
(19) \citet{Liebert2003}; 
(20) \citet{Girven2010}; 
(21) \citet{Dufour2006}; 
(22) \citet{Bergeron1993}; 
(23) \citet{SchmidtSmith1994}; 
(24) \citet{Liebert05}; 
(25) \citet{Vennes1999}; 
(26) \citet{Liebert1985}; 
(27) \citet{Achilleos1989}; 
(28) \citet{Ferrario1997a}; 
(29) \citet{Brinkworth2004}; 
(30) \citet {Schmidtetal92}; 
(31) \citet{Greenstein1986}; 
(32) \citet{Wickramasinghe1988}; 
(33) \citet{wickramasinghe2000};  
(34) \citet{Jordan1992}; 
(35) \citet{Angel1985}; 
(36) \citet{Greenstein1985}; 
(37) \citet{Maxted2000}; 
(38) \citet{Brinkworth2005}; 
(39) \citet{Green1981}; 
(40) \citet{Schmidt1990}; 
(41) \citet{Schmidt1996}; 
(42) \citet{Jordan1998}; 
(43) \citet{Moran1998}; 
(44) \citet{Dobbie2013}. 
(45) \citet{Ferrario1997a}. 
\end{footnotesize}
\end{minipage}
\end{landscape}

\begin{figure}
\begin{center}
\includegraphics[width=140mm]{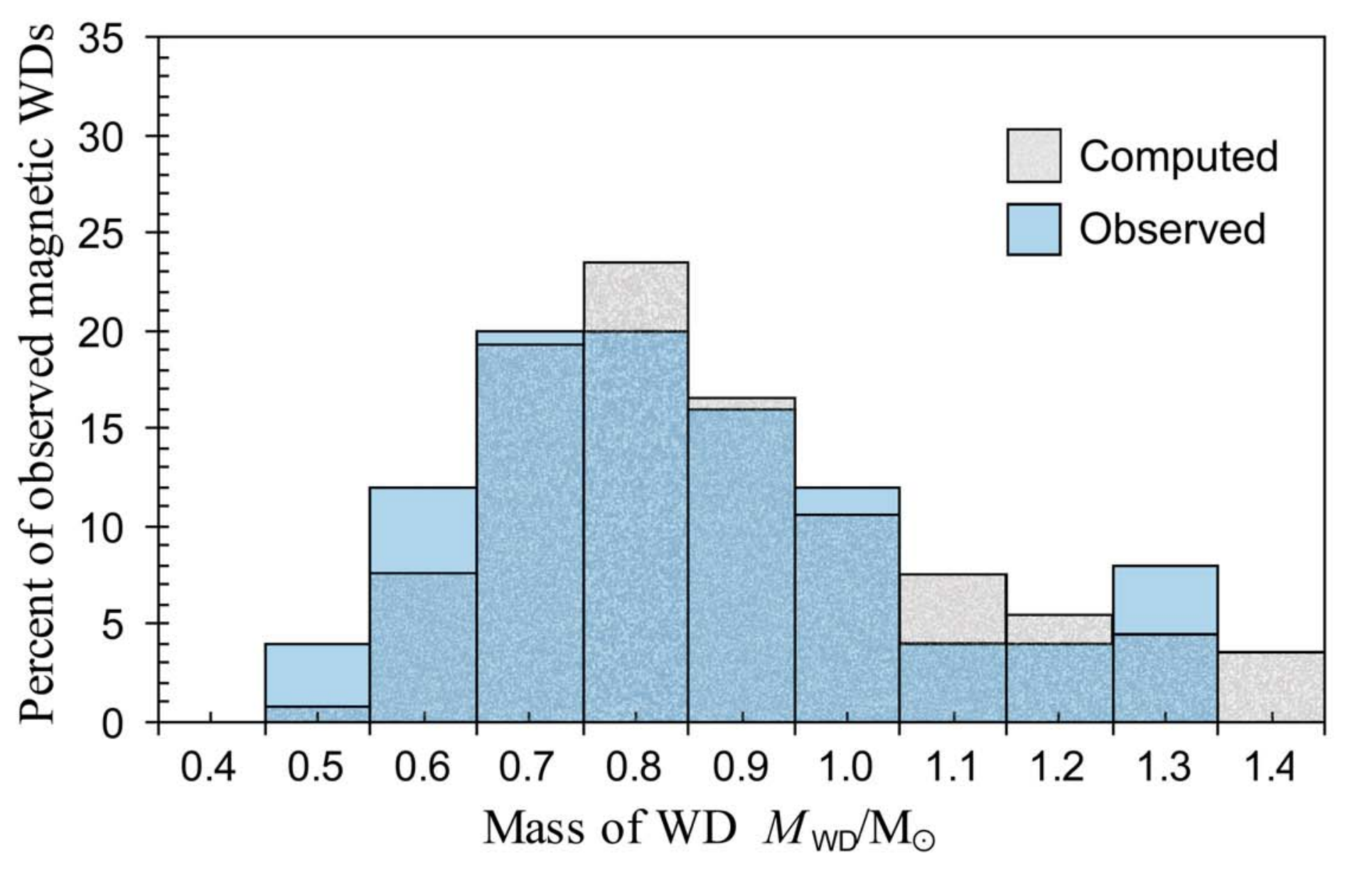}
\vspace{10mm}
\captionsetup{labelfont=bf, margin={7mm,7mm}, justification=justified}
\caption{Mass distribution of 27~observed HFMWDs (objects taken from Table \ref{tab:Obsdb})
compared with the computed sample.}
\label{fig:ObsMagWDs}
\end{center}
\end{figure} 

The comparison of the mass distribution between theory and
observations is shown in Fig.\hphantom{ }\ref{fig:ObsMagWDs}.
Most of our models reproduce the observed peak near $0.8\,\rm\msun$
but are less successful at reproducing the higher and lower mass tails.
Interestingly the peak is dominated by giant cores that merge with
main-sequence stars.  This case was not considered by \citet{garcia2012}
who focused only on merging DDs.  We used a Kolmogorov-Smirnov (K--S)
test \citep{Press1992} to compare the mass distribution of the observed
HFMWDs with our synthetic populations. The K--S test determines the
statistical probability that two sample sets are drawn from the same population.
It uses the CDFs of the two sample sets which naturally agree at the smallest
value of an independent variable where they are both zero and again at its
maximum where they are both unity.  The test then uses the intervening
behaviour to distinguish the populations.  The test gives a statistic
$D$ which is the maximum of the absolute difference between two CDFs
at a given $M_{\rm WD}$ and the probability $P$ that a random selection
from the population would lead to a larger $D$ than that measured.

\begin{figure}
\begin{center}
\includegraphics[width=148mm]{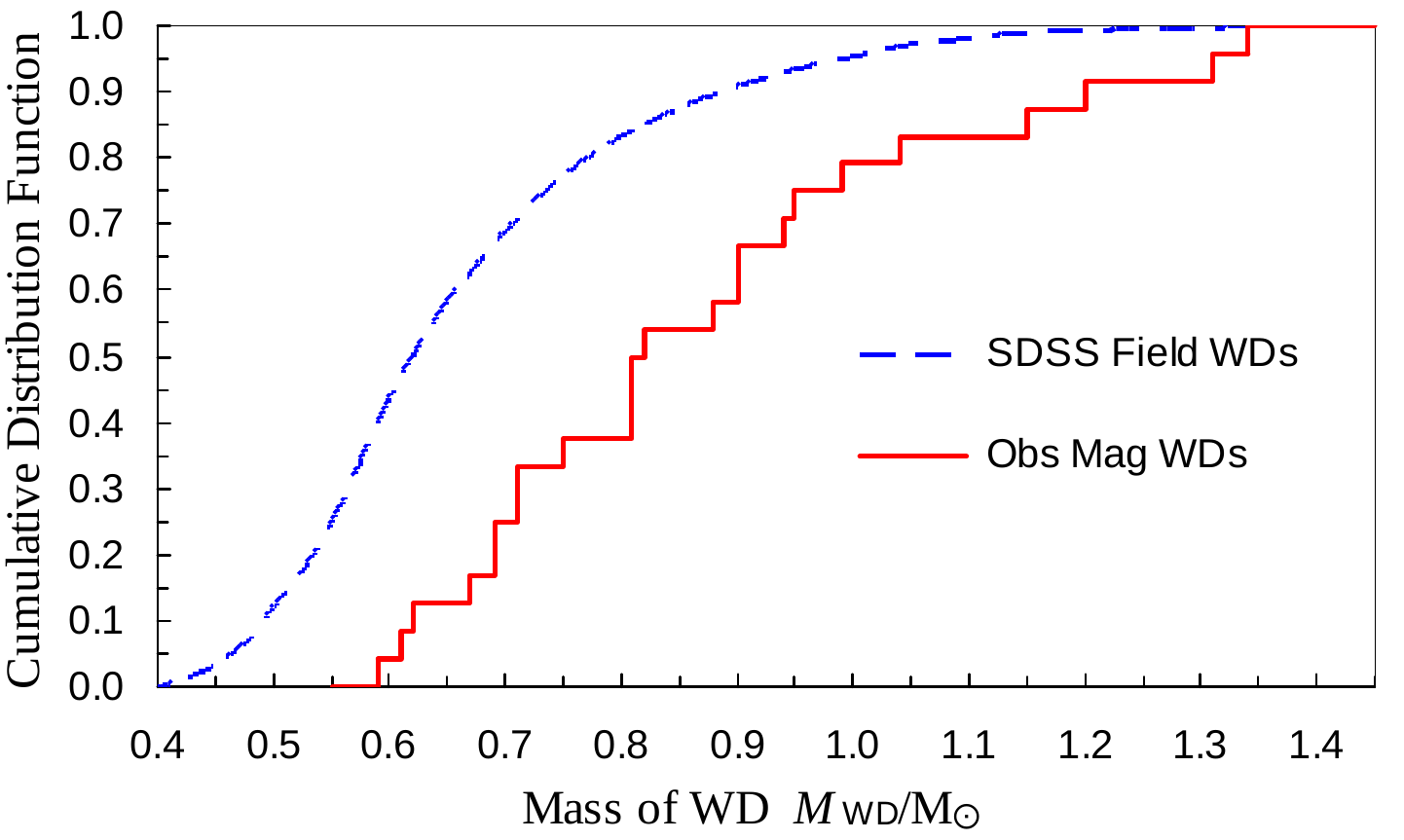}
\vspace{15mm}
\captionsetup{labelfont=bf, margin={2mm,0mm}, justification=justified}
\caption{CDFs of volume-limited-converted masses of observed SDSS DR7
\citep{Kleinman2013} non-magnetic, field WDs and the observed
MWDs.  The population of observed MWDs is not strictly a volume limited
sample since it comes from various surveys as discussed in the text.
A formal application of the K--S test has $D=0.4417$ and  $P = 3\times 10^{-5}$.}
\label{fig:MagNonMagCDF}
\end{center}
\end{figure}

\begin{figure}
\vspace{15mm}
\begin{center}
\includegraphics[trim={0mm 6mm -5mm 0}, width=150mm]{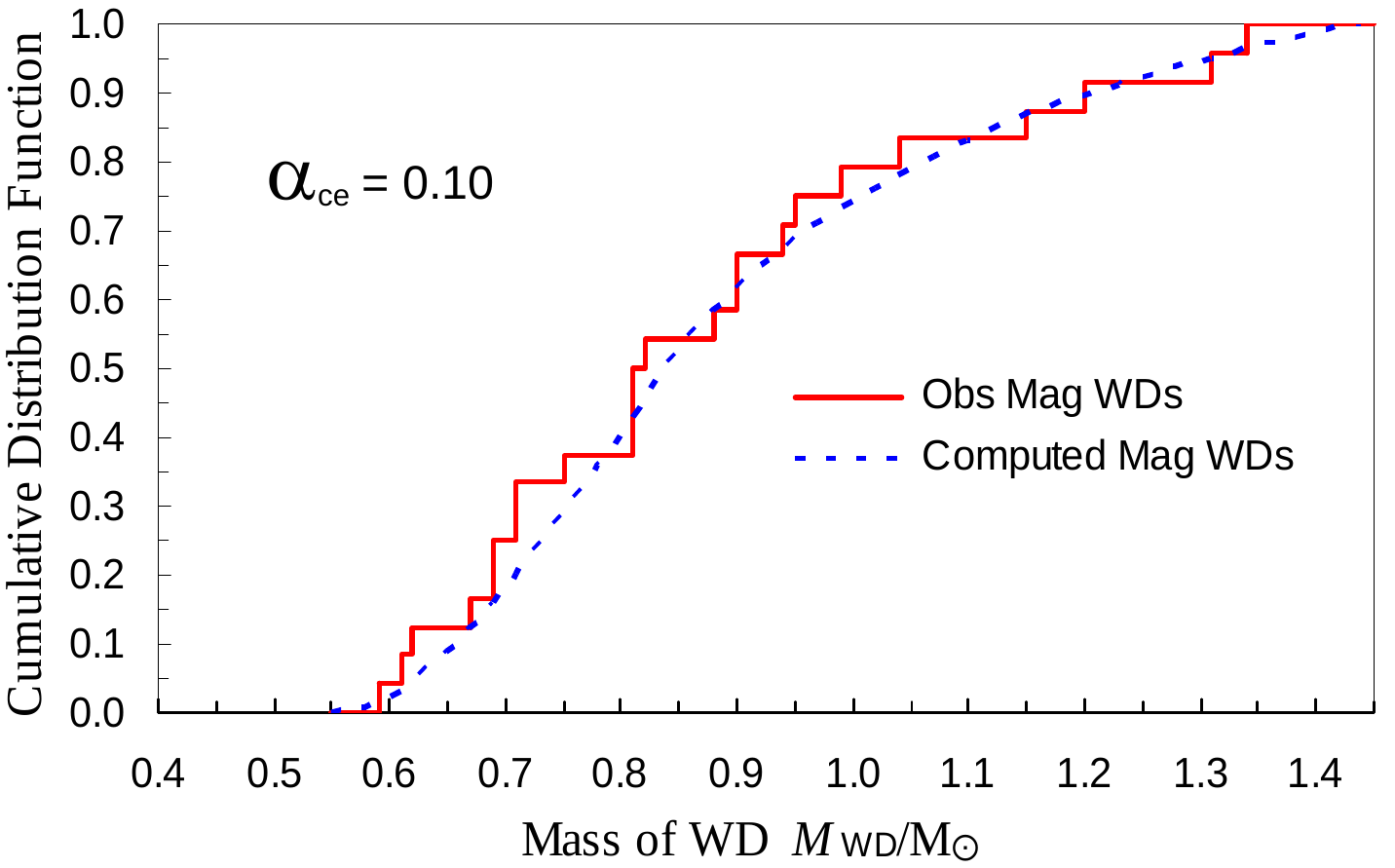}
\captionsetup{labelfont=bf, margin={0mm,0mm}, justification=justified}
\caption{CDF of observed and {\sc{bse}}
  theoretical HFMWD masses for a Galactic disc age of 9.5\,Gyr and
  $\alpha=0.10$.  The K--S test has $D=0.1512$ and $P=0.7095$.
  \label{fig:CDFObsTheory}}
\end{center}
\end{figure}

Fig.~\ref{fig:MagNonMagCDF} shows the mass distribution CDFs for the
27 observed HFMWDs (jagged line) and for the 12\,803 SDSS DR7 field
WDs (smooth curve).  A visual inspection shows the two CDFs
to be distinctly different. The K--S test gives a $D = 0.4417$ and $P
= 3\times 10^{-5}$. So we deduce that HFMWD masses are not distributed
in the same manner as non-magnetic single WDs.  When the CDF
for the observed HFMWD mass distribution is compared to the CDF for
the {\sc{bse}} theoretical mass distribution
(Fig.~\ref{fig:CDFObsTheory}) for $\alpha=0.10$ it can be seen that
the two curves are remarkably similar.  The K--S test gives a smaller
$D$ of 0.1512 with a probability of 0.7095 that indicates success of
our model.  The results of the K--S test for a range of $\alpha$s
(Table~\ref{tab:KSTest}) show that the mass distribution is consistent
over the wide range $0.05 \le \alpha\le 0.7$.  On the other hand,
based on the results in Table~\ref{tab:MagWDperc} the observed
incidence of magnetism, as observed in the \citet{Kawka2007}
volume-limited sample, constrains $\alpha$ to be in the narrower range
$0.1 \le \alpha \le 0.3$.

\clearpage

\begin{table}[htbp]
\vspace{5mm}
\captionsetup{labelfont=bf, margin={0mm,0mm}, justification=justified}
\caption{Kolmogorov-Smirnov $D$ statistic and $P$ of the mass
    distributions of the theoretical ({\sc{bse}}) and observed
    MWD populations being drawn from the same
    distribution for various values of $\alpha$. The theoretical
    population is for a Galactic disc age of 9.5\,Gyr. }
\label{tab:KSTest}
\centering
\begin{tabular} { c c c }
\toprule
$\alpha$  & $D$ & $P$ \\
\midrule
0.05  &  0.1558  &  0.6735  \\
0.10  &  0.1512  &  0.7095  \\
0.20  &  0.1565  &  0.6684  \\
0.25  &  0.1616  &  0.6288  \\
0.30  &  0.1675  &  0.5824  \\
0.40  &  0.1827  &  0.4700  \\
0.50  &  0.2040  &  0.3326  \\
0.60  &  0.2304  &  0.2039  \\
0.70  &  0.2580  &  0.1144  \\
0.80  &  0.2814  &  0.0665  \\
0.90  &  0.2915  &  0.0518  \\
\bottomrule
\end{tabular}
\end{table}
\clearpage

\section{Discussion and Conclusions}
\label{DandC}

Two competing models for the origin of strong magnetic fields in WDs
are broadly the fossil field model and the merging star model.
The proponents of the fossil field model have noted that the maximum
poloidal flux observed in the magnetic Ap and Bp stars is similar to
the maximum poloidal magnetic flux observed in the MWDs.
The two groups of stars could therefore be evolutionarily
linked.  However, to date, there have been no stellar evolution models
that have shown how a strong fossil magnetic flux can survive through
the various stages of stellar evolution through to the WD
phase.  It is also not clear if the similarities in the maximum
magnetic fluxes between two groups of stars is necessarily a reason to
assume a causal link.  The dynamo model of  \citet{wickramasinghe2014}
suggests that similar maximum magnetic fluxes may be expected for
physical reasons if the fields are generated from differential
rotation caused by merging.  Here we have explored the consequences of
such a hypothesis for the origin of the HFMWDs with binary population
synthesis under standard assumptions, discussed in section
\ref{sec:calculations}.  We have found the following.

\begin {itemize}
\item[(i)] While the mass distribution of HFMWDs is not very sensitive
  to $\alpha$, good agreement can be obtained with both the observed
  mass distribution and the observed incidence of magnetism for models
  with {$0.1\,\le\,\alpha\,\le\,0.3$}.  In particular the mean
  predicted mass of HFMWDs is $0.88\,\rm\msun$ compared with
  $0.64\,\rm\msun$ (corrected to include observational biases) for all
  WDs while observations indicate respective mean masses of
  $0.85\,\rm\msun$ \citep[see also][]{Kepler2013} and $0.62\,\rm\msun$
  \citep{Kleinman2013}.
  A K--S test shows that the small number of  reliably measured
  masses of HFMWDs are not distributed in the same way as the
  masses of non-magnetic single WDs.  The probability they are
  is only $3\times 10^{-5}$.  On the other hand our best model fit to the
  observed mass distribution of HFMWDs has a probability of 0.71.
  
\item[(ii)] Stars that merge during CEE and then
  evolve to become WDs outnumber merging post-CE
  DD systems for all $\alpha$.  The CEs yield mainly
  CO~WDs with a few He and ONe~WDs, while the DDs
  yield only CO~WDs.

\item[(iii)] The major contribution to the observed population of
  HFMWDs comes from main-sequence stars merging with degenerate cores
  at the end of CEE.  The resulting giants go on to evolve to HFMWDs.

\item[(iv)]The merging DDs tend mostly to populate the high-mass end of
  the WD mass distribution.

\end{itemize}
We also note that the study by \citet{Zorotovic2010} of the evolution
of a sample of SDSS post-CE binary stars consisting of
a WD and a main-sequence star indicates that the best agreement
with observational data is achieved when $\alpha=0.25$ and this is
consistent with our findings. In summary, available observations of
the mass distribution and incidence of HFMWDs are compatible with
the hypothesis that they arise from stars that merge mostly during CEE
with a few that merge during post-CE as DD systems. Our calculations,
when taken together with the observation that there are no examples of
HFMWDs in wide binary systems, allow us to argue strongly in favour
of this hypothesis.  In the majority of cases the fields may be generated
by a dynamo mechanism of the type recently proposed by
\citet{wickramasinghe2014}. The disc field model of \citet{Nordhaus2011}
or the model proposed by \citet{garcia2012} may be relevant in the case
of merging DD cores depending on mass ratio. The rate of merging of
post-CE DDs alone is too low to account for all observed HFMWDs.

\section*{Acknowledgements}

We would like to thank the Referee, Zhanwen Han, for his suggestions
and comments which have helped me improving the quality of this
work. GPB gratefully acknowledges receipt of an Australian Postgraduate Award. CAT thanks the Australian National University for supporting a visit as a Research Visitor of its Mathematical Sciences Institute, Monash University for support as a KevinWatford distinguished visitor and Churchill College for his fellowship.
\blanknonumber

\chapter{Genesis of the Magnetic Field}
\label{Chapter3}

\vspace{-9mm}
This chapter is a reproduction of the paper accepted for publication in
Monthly Notices of the Royal Astronomical Society, viz:
\vspace{1mm}
\\
{\color{blue}\textit{Briggs, Ferrario, Tout \& Wickramasinghe, MNRAS (2018),
(In publication). Genesis of magnetic fields in isolated white dwarfs}}
\vspace{-12pt}

\section{Abstract}
\label{Paper2abstract}
\vspace{-12pt}
  A dynamo mechanism driven by differential rotation when stars merge
  has been proposed to explain the presence of strong fields in
  certain classes of magnetic stars. In the case of theHFMWDs, the site of
  the differential
  rotation has been variously thought to be the CE, the
  hot outer regions of a merged degenerate core or an accretion disc
  formed by a tidally disrupted companion that is subsequently
  accreted by a degenerate core. We have shown previously that the
  observed incidence of magnetism and the mass distribution in HFMWDs
  are consistent with the hypothesis that they are the result of
  merging binaries during CEE. Here we calculate
  the magnetic field strengths generated by CE
  interactions for synthetic populations using a simple prescription
  for the generation of fields and find that the observed magnetic
  field distribution is also consistent with the stellar merging
  hypothesis. We use the Kolmogorov-Smirnov K--S test to study
  the  correlation between
  the calculated and the observed field strengths
  and find that it is consistent for low envelope ejection
  efficiency. We also suggest that field generation by the plunging of
  a giant gaseous planet on to a WD may explain why magnetism
  among cool WDs (including DZ\,WDs) is higher than
  among hot WDs. In this picture a super-Jupiter residing in
  the outer regions of the WD's planetary system is perturbed
  into a highly eccentric orbit by a close stellar encounter and is
  later accreted by the WDs.
\vspace{4mm}

\section{Introduction}
\label{Introd}
\vspace{-11pt}
The existence of strong magnetic fields in stars at any phase of
their evolution is still largely unexplained and very puzzling
\citep[see][]{fer2015a,wickramasinghe2000}.
HFMWDs have dipolar magnetic field strengths of up to $10^9$\,G.
There are no observed HFMWDs with late-type companions found in
wide binary systems. \citet{liebert2005,lie2015} pointed out that this
contrasts with non-magnetic WDs, a large fraction of which are
found in such systems.  This led \citet{tout2008} to hypothesise that the entire
class of HFMWDs with fields $10^6 < B/{\rm G} < 10^ 9$ owe their
magnetic fields to binary systems which have merged while in a CE
stage of evolution. In this scenario, when one of the two
stars in a binary evolves to become a giant or a super-giant its
expanded outer layers fill its Roche lobe.  At this point unstable
mass transfer leads to a state in which the giant's envelope engulfs
the companion star as well as its own core. This merging idea to
explain the origin of fields in WDs is now favoured over the
fossil field hypothesis first suggested by \citet{Woltjer64} and
\citet{landstreet1967} whereby the the magnetic main-sequence
Ap and Bp stars are the ancestors of the HFMWDs if magnetic
flux is conserved all the way to the compact star phase
\citep[see also][and references therein]{tou2004,wickramasinghe2005}.
  
During CEE, frictional drag forces acting on the
cores and the envelope cause the orbit to decay. The two cores spiral
together losing energy and angular momentum which are transferred to
the differentially revolving CE, part of which at least, is ejected from the
system.  This process is thought to proceed on a dynamical time scale
of less than a few thousand years and hence has never been observed.
The original model of \citet{tout2008}
suggested that high fields were generated by a dynamo between the CE
and the outer layers of the proto-WD before the CE is ejected. If the cores
merge the resulting giant star eventually loses its envelope to reveal a
single HFMWD. If the envelope is ejected when the cores are close but
 have not merged a magnetic CV is formed. \citet{potter2010}
found problems with this scenario in that the time-scale for diffusion
of the field into the WD is significantly longer than the expected CE lifetime.
Instead  \citet{wickramasinghe2014} suggested that a weak seed field
is intensified by the action of a dynamo arising from the differential rotation
in the merged object as it forms. This dynamo predicts a poloidal magnetic
flux that depends only on the initial differential rotation and is independent
of the initial field. \citet{Nordhaus2011} suggested another model where
magnetic fields generated in an accretion disc formed from a tidally disrupted
low-mass companion are advected onto the surface of the proto-WD.
However, this would once again depend on the time-scale for diffusion of
the field into the surface layers of the WD. \citet{garcia2012} found that a
field of about $3\times10^{10}$\,G could be created from a massive, hot
and differentially rotating corona forming around a merged DD. They also
carried out a population synthesis study of merging DDs with a CE efficiency
factor $\alpha=0.25$.  They achieved good agreement in the observed
properties between high--mass WDs
($M_\textrm{WD}\ge$ 0.8\msun) and HFMWDs but their studies did not
include degenerate cores merging with non-degenerate companions as I did
in chapter 2.

The stellar merging hypothesis may only apply to HFMWDs.
\citet{landstreet2012}
point out that weak fields of $B\le1$\,kG may exist in most WDs and
so probably arise in the course of normal stellar evolution from a dynamo action
between the core and envelope.

With population synthesis we showed, in chapter 2 that the origin of
HFMWDs is consistent with the stellar merging hypothesis. The
calculations presented in chapter 2 could explain the observed incidence of
magnetism among WDs and showed that the computed mass
distribution fits the observed mass distribution of the HFMWDs
more closely than it fits the mass distribution of non-magnetic WDs.
This demonstrated that magnetic and non-magnetic WDs
belong to two populations with different progenitors. We now present
the results of calculations of the magnetic field strength expected
from merging binary star systems.

\section{Population synthesis calculations}
\label{sec:calculations}

As described in chapter 2, we create a population of binary systems by
evolving them from the zero-age main sequence (ZAMS) to 9.5\,Gyr, the
age of the Galactic disc
\citep{Kilic2017}.
Often an age of 12\,Gyr is assumed when population synthesis studies
are carried out but an integration age of 12\,Gyr, that encompasses not
only the thin and thick disc but also the inner halo, would be far too large
for our studies of the origin of HFMWDs. The HFMWDs belong to the thin
disc population, according to the kinematic studies of HFMWDs by
\citet{Sion1988} and \citet{Anselowitz1999},
who found that HFMWDs come from a young stellar disc population
characterised by small motions with respect to the Sun and a dearth of
genuine old disc and halo space velocities. The more recent studies of
the WDs within 20\,pc of the Sun by
\citet{Sion2009}
also support the earlier findings and show that the HFMWDs in the local
sample have significantly lower space velocities than non-magnetic WDs.

We use the rapid binary stellar evolution algorithm {\sc bse} developed by
\citet{Hurley2002}
that allows modelling of the most intricate binary evolution.  This algorithm includes not only
all those features that characterise the evolution of single stars
\citep{Hurley2000}
but also all major phenomena pertinent to binary evolution. These comprise Roche
lobe overflow, CEE \citep{Pac1976}, tidal interaction, collisions, gravitational radiation and magnetic braking.

As in chapter 2, we have three initial parameters. The mass of the
primary star $0.8\le M_1/\msun\le 12.0$, the mass of its companion
$0.1\le M_2/\msun\le 12.0$ and the orbital period
$0.1\le P_0/{\rm d}\le 10\,000$. These initial parameters are on a
logarithmic scale of 200\,divisions.  We then compute the real number
of binaries assuming that the initial mass of the primary star is
distributed according to Salpeter's (1955) mass function and the
companion's mass according to a flat mass ratio distribution with $q\le 1$
\citep[e.g.][]{Hurley2002,Ferrario2012}.
The period distribution is taken to be uniform in its logarithm. We use the
efficiency parameter $\alpha$ (energy) formalism for the CE
phases with $\alpha$ taken as a free parameter between $0.1$
and $0.9$.  In our calculations we have used $\eta=1.0$ for the
Reimers' mass-loss parameter and a stellar metallicity $Z = 0.02$.  We
select a sub-population consisting of single WDs that formed
by merging during CEE. Conditions of the selection are that (i) at
the beginning of CEE the primary has a degenerate core to ensure
that any magnetic field formed or amplified during CE persists in a
frozen-in state and (ii) from the end of CE to the final WD stage
there is no further nuclear burning in the core of the pre-WD star
which would otherwise induce convection that would destroy any
frozen-in magnetic field. In addition to stellar merging during CE,
we also select double WD binaries whose components merge to form
a single WD at any time after the last CEE up to the age of the
Galactic disc. This forms the DD merging channel for the formation of
HFMWDs.

\subsection{Theoretical magnetic field strength}\label{TheoryB}

\begin{figure}
\centering
\includegraphics[width=1.00\columnwidth]{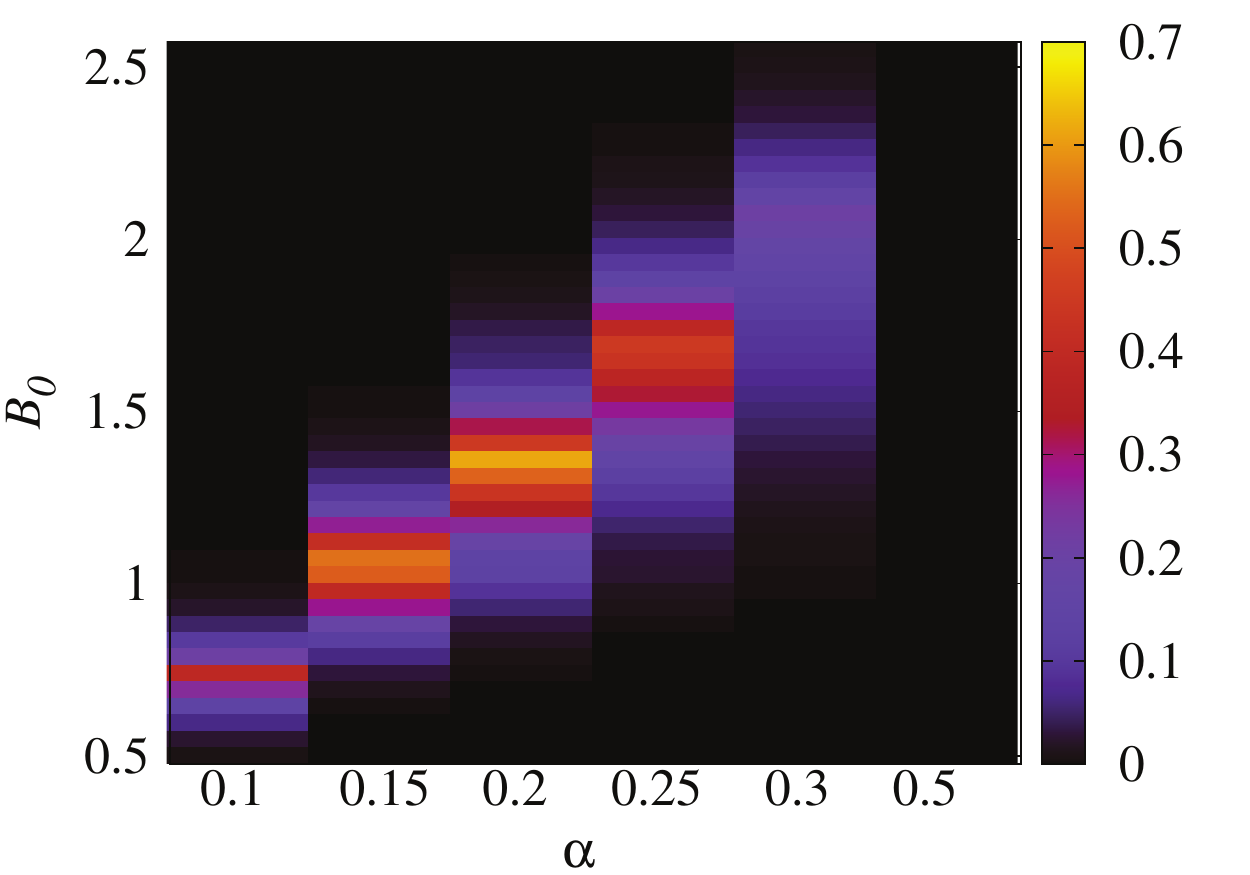}
\caption{Density plot of the probability given by the K--S test that
  the CDFs of the theoretical and observed magnetic field
  distributions are drawn from the same population. This was generated
  for a range of $\alpha$ and $B_0$ (see text). The probability is
  colour-coded according to the palette shown on the right hand side
  of the figure. The sub-structures in this plot are caused by the
  discretisation of $\alpha$ and $B_0$.}
\label{fig:contour}
\end{figure}

The goal of this chapter is to construct the magnetic field distribution
of our synthetic sample of HFMWDs using, as a basis, the results and
ideas set out by \citet{tout2008} and \citet{wickramasinghe2014}.
If the cores of the two stars do not merge during CE,
Our assumption is that a fraction of the maximum angular momentum
available at the point of the ejection of the envelope causes the shear
necessary to generate the magnetic field. The non-merging case, leading
to the formation of MCVs, is presented in chapter 4.
In the case of coalescing cores, a fraction of the break-up angular
momentum of the resulting degenerate core provides the shear
required to give rise to the strongest fields. In the following sections
and in chapter 4 we show that our models indeed show that
the highest fields are generated when two stars merge
and give rise to a HFMWD.

Having obtained the actual number of WDs we then assign a
magnetic field $B$ to each. Our prescription is that the field,
generated and acquired by the WD during CE
evolution or DD merging, is proportional to the orbital angular
velocity
\begin{equation}
 \Omega=\displaystyle{\frac{2\pi}{P_{\rm orb}}}
 \end{equation}
of the binary at the point the envelope is ejected and write
\begin{equation}\label{EqBfield}
B = B_0\left(\frac{\Omega}{\Omega_{\rm crit}}\right)\, \mbox{G}.
\end{equation}
where 
\begin{equation}\label{omega_c}
	\Omega_{\rm crit}= \sqrt{\frac{GM_{\rm WD}}{R_{\rm
              WD}^3}}=0.9\left(\frac{M_{\rm
              WD}}{\Msun}\right)^{1/2}\left(\frac{5.4\times 10^8}{R_{\rm WD}}\right)^{-3/2}
\end{equation}
is the break-up angular velocity of a WD of mass $M_{\rm WD}$
and radius $R_{\rm WD}$.

This model encapsulates the dynamo model of
\citep{wickramasinghe2014}
where a seed poloidal field is amplified to a maximum
that depends \emph{linearly} on the initial differential rotation
imparted to the WD. In view of these results, here we
simply assume a linear relationship between the poloidal field and
the initial rotation and recalibrate the
\citep{wickramasinghe2014}
relation between differential rotation and field using
(i) a more recent set of data and (ii) results from our population
synthesis calculations that provide $\Omega$ in equation
(\ref{EqBfield}). The quantity $B_0$ in equation (\ref{EqBfield}) is
also a parameter to be determined empirically.  Different $B_0$'s
simply shift the field distribution to lower or higher fields with
no changes to the shape of the field distribution which is solely
determined by the CE efficiency parameter $\alpha$.

For the radius of the WD we use the
\citet{Nauenberg1972}
mass-radius formula
\begin{equation}\label{R_WD}
R_{\rm WD}= 0.0112 R_\odot\,\left[\left(\frac{M_{\rm Ch}}{M_{\rm WD}}\right)^{2/3}-\,\,\,\left(\frac{M_{\rm WD}}{M_{\rm Ch}}\right)^{2/3}\right]^{1/2},
\end{equation}
where $M _{\rm{Ch}}=1.44$\,M$_\odot$ is the Chandrasekhar limiting
mass.

\begin{figure}
\centering
\includegraphics[width=1.00\columnwidth]{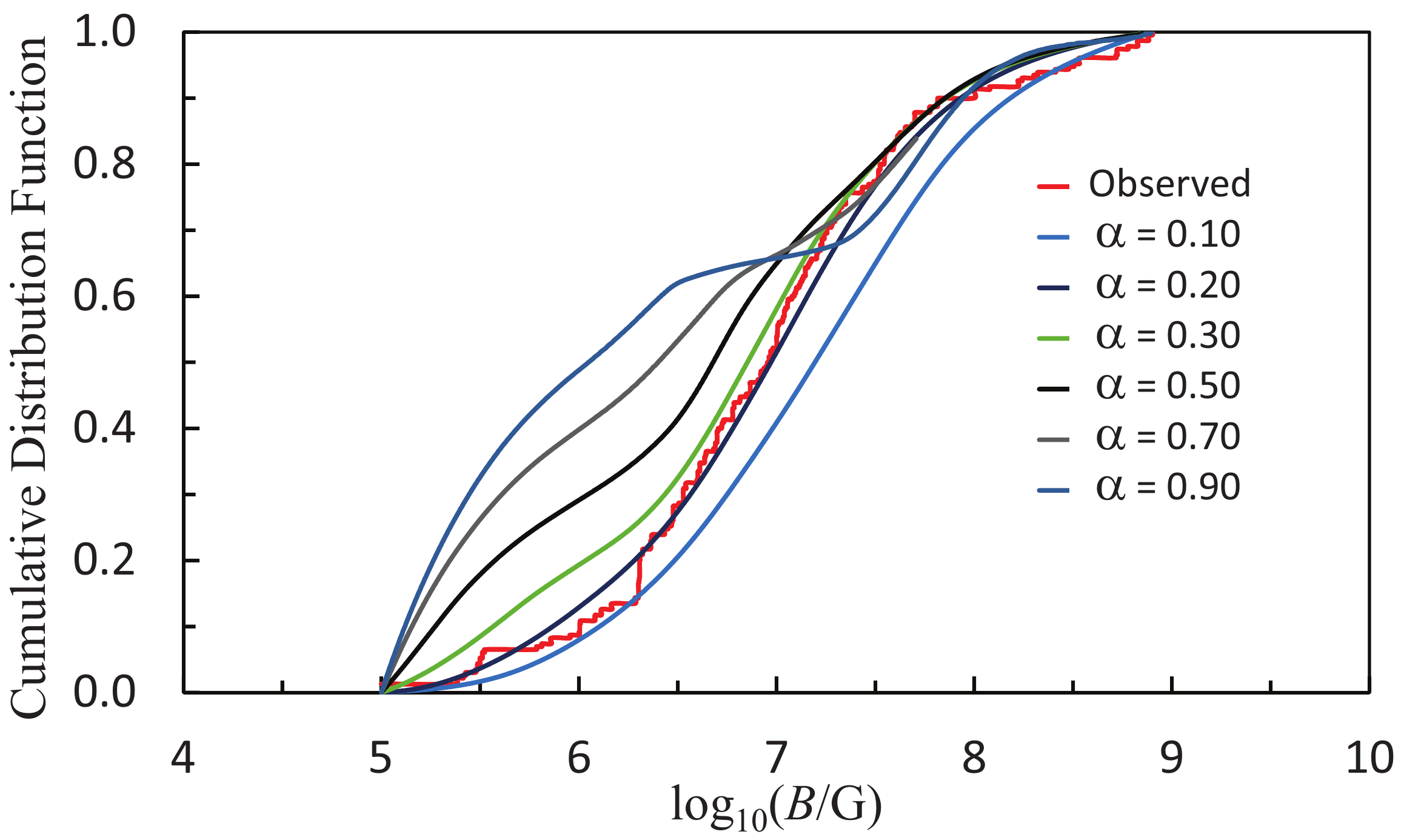}
\caption{CDFs of observed (red) and {\sc{BSE}} theoretical magnetic field
  distributions for a Galactic disc age of 9.5 Gyr and various
  $\alpha$.}
\label{fig:CDF_Magfield}
\end{figure}

\subsection{Parameters calibration}\label{Calibration}

The data set of HFMWDs is affected by many biases, even though
some of the surveys that discovered them were magnitude-limited.
This is because HFMWDs tend to be more massive than their
non-magnetic counterparts, as first noticed by
\citet{Liebert1988},
and therefore their smaller radii, as expected by equation (\ref{R_WD}),
make them dimmer and so less likely to be detected.  Volume-limited samples
are far better, given that our synthetic population mimics a volume-limited sample,
but do not include enough HFMWDs to allow us to conduct any statistically meaningful
study.  In this section we establish the parameter space of relevance to the
observations of HFMWDs by comparing the predictions of the magnetic field
distribution derived from our population synthesis calculations to the fields of
HFMWDs listed in \citet{fer2015b}.

In order to achieve this goal we have employed the K--S test \citep{Press1992}
to establish which combination of $B_0$ and $\alpha$ yield the best fit
to the observed field distribution of HFMWDs.  The K--S test compares
the cumulative distribution functions (CDFs) of two data samples (in
this case the theoretical and observed field distributions) and gives
the probability $P$ that they are drawn randomly from the same
population. We have calculated CDFs for seven different $\alpha$ and
44 different $B_0$s for each $\alpha$. If we discard all combinations
of $\alpha$ and $B_0$ for which $P\le0.01$, we find
$0.5 \times 10^{10} \le B_0/{\rm G} \le 2.5 \times10^{10}$ and
$\alpha<0.5$. We have depicted in Fig.\,\ref{fig:contour} a density
plot of our results.  The highest probability is for
$B_0=1.35\times10^{10}$\,G and $\alpha=0.2$.  We show in
Fig.\,\ref{fig:CDF_Magfield} the theoretical CDFs for
$B_0=1.35\times10^{10}$\,G and various $\alpha$s and the CDF of the
observations of the magnetic field strengths of HFMWDs.

In the following sections we will discuss models with $B_0=1.35\times10^{10}$\,G
and a range of $\alpha$ again noting that a different $B_0$ would simply move
the field distribution to lower or higher fields with no change of shape. Therefore
our discussion in the following sections will focus on the effects of varying $\alpha$.

\begin{figure*}
\centering
\includegraphics[width=1.00\textwidth]{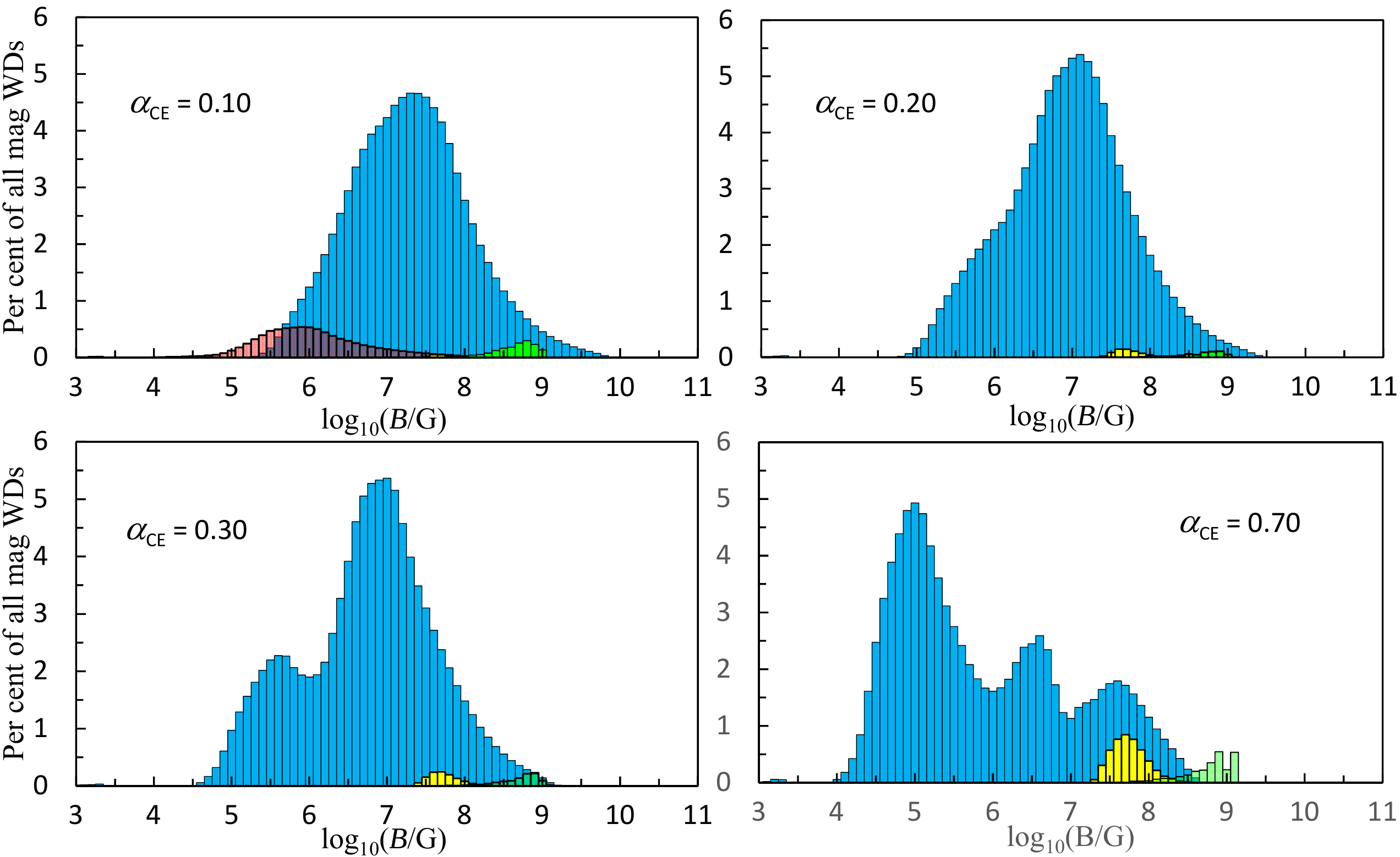}
\caption {Theoretical magnetic field strength for a Galactic disc age
  of 9.5 Gyr and various $\alpha$. The histograms are superimposed,
  not stacked, to highlight the contribution made by each type of
  WD to the overall distribution. The blue, red and yellow
  histograms represent, respectively, CO, ONe, He\,WDs. The
  green histograms depict the merged DD systems.}
\label{fig:MagAlphas}
\end{figure*}

\section{Discussion of results}\label{MagFieldDistr}

Fig. \ref{fig:MagAlphas} shows the calculated magnetic field
distribution and the breakdown of the WD types for $\alpha=0.1$ to
$0.7$. The maximum field strength is a few $10^{9}$\,G and is found
mostly in systems in which the HFMWD forms either via the merging of
two very compact stars on a tight orbit or through the merging of two
WDs after CEE (DD path). The reason for this is that these systems
have very short periods and when they merge produce very strongly
magnetic WDs, as expected from equation \ref{EqBfield}.

\begin{figure*}
\centering
\includegraphics[width=1.00\textwidth]{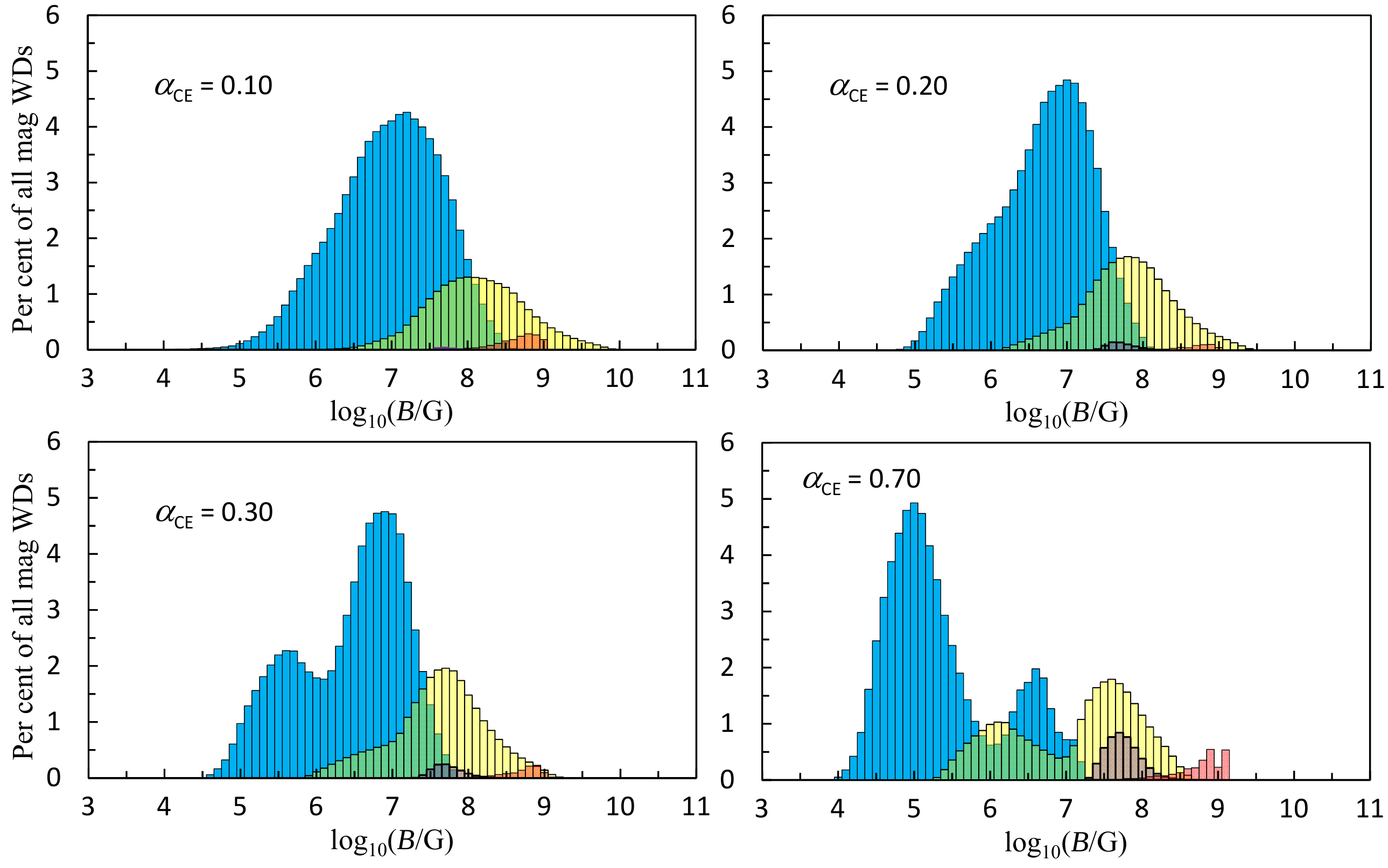}
\caption{Theoretical magnetic field distribution of HFMWDs showing the
  pre-CE progenitors for various $\alpha$. The light blue,
  yellow and purple histograms represent, respectively, the AGB/MS, AGB/CS and RGB/CS
  merging pairs. The red histograms depict the merged DD systems.}
\label{fig:paths}
\end{figure*}

We show in Fig.\,\ref{fig:paths} the theoretical magnetic field
distribution of HFMWDs for $\alpha=0.1$ to $\alpha=0.7$ with the
breakdown of their main formation channels, that is, their pre-CE
progenitors. The overwhelming contributors to the HFMWD
population are AGB stars merging with MS or CS. At low $\alpha$,
systems with initially short orbital periods merge as soon as their
primaries evolve off the main sequence, either whilst in the
Hetzsprung's gap or during their ascent along the RGB. Usually
such merging events produce single stars that continue
their evolution burning helium in their cores and later on, depending
on the total mass of the merged star, heavier elements. Because of
core nuclear burning these stars continue their evolution to
eventually become single non-magnetic WDs. The only
observational characteristic that may distinguish them from other
non-magnetic WDs could be an unusual mass that does not fit
any reasonable initial to final mass function associated to the
stellar cluster to which they belong. On the other hand, if the RGB
star has a degenerate core, as for stars with $M_1\le 2.2$\,M$_\odot$
on the ZAMS, and merges with a low-mass CS, then the resulting object
is a strongly magnetic He\,WD. These RGB/CS merging events do
occur at all $\alpha$ but their fraction is higher at large $\alpha$
owing to fewer overall merging occurrences at high envelope clearance
efficiencies.

When systems do not merge when the primary evolves on the RGB, they
may merge when they undergo CE evolution on the AGB. In
this case those binaries with the shortest orbital periods at the
beginning of the CEE are those that form the
highest magnetic field tail of the distribution. There are two main
types of merging pairs, AGB stars merging with MS stars
($M\ge0.7$\,M$_\odot$) and AGB stars merging with CS
($M<0.7$\,M$_\odot$). Each of these combinations exhibits two peaks as
seen in Fig.\,\ref{fig:paths} for $\alpha>0.2$, although the second peak at
lower fields of the merging AGB/CS pair becomes well defined only when
$\alpha=0.7$.  Because AGB/MS systems have larger orbital periods
at the onset of CEE, their merging gives rise to
generally more massive but less magnetic WDs as expected from
equation (\ref{EqBfield}). This is why the bulk of AGB/MS merging
pairs occupy the lowest and most prominent peak near $B=10^{5.5}$\,G
with the secondary maximum at $B=10^{6.8}$\,G. The AGB/CS merging
pairs form another two peaks, one at $B=10^6$\,G and the other at
$B=10^{7.75}$\,G. RGB stars merging with CS stars also form a maximum
at $B=10^{7.75}$\,G.  The reason for the double peaks in AGB/MS and
AGB/CS merging pairs is because high envelope clearance efficiencies
(high $\alpha$) require more massive primaries to bring the two stars
close enough together to merge during CEE. Thus,
these double peaks are caused by a dearth of AGB/MS merging pairs near
$B=10^6$\,G and of AGB/CS pairs near $B=10^7$\,G. Those systems whose
orbital periods would give rise to magnetic fields in these gaps fail
to merge because their initial periods are large and their primary
stars are not massive enough to bring the two components close enough
to merge. These double peaks are not present at low $\alpha$ 
because low envelope clearance efficiency always leads to tighter
orbits and merging is more likely for a much wider range of
initial masses and orbital periods, more effectively smearing the
contributions made by specific merging pairs.

\section{Comparison to observations}\label{Comparison}

A prediction of our merging hypothesis for the origin of HFMWDs is
that low-mass HFMWDs, mostly arising from AGB/CS merging pairs, should
display fields on average stronger than those of massive HFMWDs which
predominantly result from the merging of AGB/MS pairs. The HFMWDs
formed through the merging of two WDs (DD channel) are
excluded from this prediction.  These are expected to produce objects
that are on average more massive, more strongly magnetic, and may be
spinning much faster than most HFMWDs
\citep[e.g. RE\,J0317-853,][]{Barstow1995,Ferrario1997b,Vennes2003}.
Given the very small number of HFMWDs for which both mass and field are
known, it is not possible to verify whether this trend is present in
observed in HFMWDs. The problem is that it is very difficult to
measure masses of HFMWDs when their field is above a few $10^6$\,G. In
the low field regime one can assume that each Zeeman component is
broadened as in the zero field case. That is, the field does not
influence the structure of the WD's atmosphere. Thus, the
modelling of Zeeman spectra has allowed the determination of masses
and temperatures of lower field WDs such as
1RXS\,J0823.62525 
\citep[$B=2.8-3.5$\,MG and M=1.2\,\msun;][]{Ferrario1998},
PG\,1658+441
\citep[$B=3.5$\,MG and M=1.31\,\msun;][]{Schmidtetal92,Ferrario1998},
and the magnetic component of the double degenerate system NLTT\,12758
\citep[$B=3.1$\,MG and $M=0.69$\,\msun;][]{kaw2017}.
The masses of high field objects can only be determined when their
trigonometric parallax is known
\citep[e.g. Grw\,+70$^\circ$8247 with $B=320\pm20$\,MG and
$M=0.95\pm0.02$\,M$_\odot$,][]{Greenstein1985,Wickramasinghe1988}.
Nevertheless, it is encouraging to see that all the most massive (near
the Chandrasekhar's limit) currently known HFMWDs do indeed possess
low field strengths and that the merged DD RE\,J0317-853 is a strongly
magnetic WD. A test of our prediction of an inverse relation
between field strength and mass will become possible with the release
of the accurate astrometric data of a billion stars by the ESA
satellite {\it Gaia}. This new set of high quality data will not only allow
one to test the (non-magnetic) WD mass--radius relation but
will also provide precise mass and luminosity measurements of
most of the currently known WDs, including the HFMWDs
\citep{Jordan2007}.

\begin{figure}
\centering
\includegraphics[width=1.00\columnwidth]{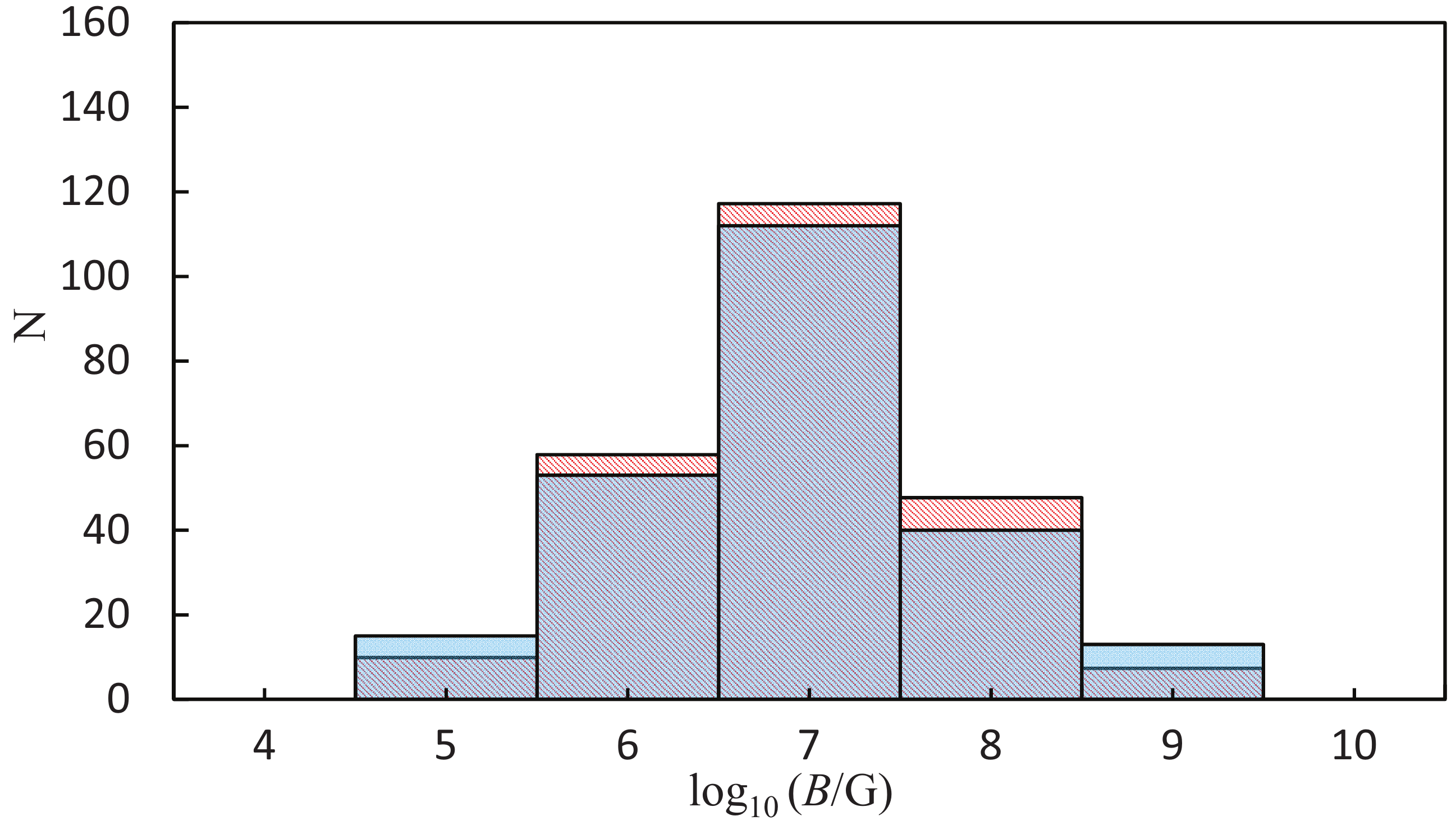}
\caption{Theoretical field distribution for $\alpha=0.2$ of the
  total of the four types of HFMWDs (pink histogram) compared to the
  field distribution of the observed HFMWDs (blue histogram). }
  \label{fig:Theory_OBS_HFMWD} \vspace{-12pt}
\end{figure}

The theoretical distribution for $\alpha=0.2$ overlapped to the
observations of HFMWDs is displayed in
Fig.\,\ref{fig:Theory_OBS_HFMWD}.  This figure shows that the maxima
of the theoretical and observed distributions occur near the same
field strength with the theoretical distribution extending from
$10^5$\,G to $10^9$\,G, as observed. The overwhelming contribution to
the theoretical field distribution is from CO\,WDs (see
Fig.\,\ref{fig:MagAlphas}).  ONe\,WDs are the next most common but at
much lower frequency and with field strengths
$4\le\log_{10} B/{\rm G} \le 8$.  Merged DD WDs present field
strengths $8\le\log_{10}B/{\rm G} \le 9$ at an even lower frequency
than the ONe\,WDs.  Finally, He\,WDs are present in very small numbers
with field strengths centred at $B=10^{7.75}$\,G. This is in contrast to
observations of HFMWDs that show the presence of very low-mass objects
\citep[see table\,1 of][]{fer2015b}
that the {\sc bse} formalism is unable to form. This mismatch between
theory and observations may
be corrected through the use of, e.g., different superwind assumptions
\citep[see][and references therein]{Han1994,Meng2008}.

We note that the models shown in Fig.\,\ref{fig:MagAlphas} with
  $\alpha>0.2$ predict the existence of a large fraction of
  low-field magnetic WDs with a bump appearing near
  $B=10^{5.5}$\,G for $\alpha=0.3$. This bump shifts toward lower
  fields and becomes increasingly more prominent as $\alpha$
  increases. For $\alpha= 0.7$ this low-field hump is the most
  prominent feature of the magnetic field distribution. In the past
  suggestions were made that the incidence of magnetism in white
  dwarfs may be bimodal, sharply rising below $10^5$\,G with an
  incidence that was predicted to be similar to or exceeding that of
HFMWDs \citep{wickramasinghe2000}.
However, recent low-field spectropolarimetric surveys of WDs have
not found anywhere near the number of objects that had been forecast to
exist in this low-field regime \citep{landstreet2012}.
Therefore, there is enough observational evidence to allow us to exclude
the bimodality of the magnetic field distribution that is theoretically predicted
for large $\alpha$'s.
\vspace{-12pt}

\section{Incidence of magnetism among cool white dwarfs}
\vspace{-12pt}
Because WDs have very high gravities, all chemical elements
heavier than hydrogen, helium and dredged-up carbon or oxygen, quickly
sink to the bottom of their atmosphere. Nonetheless, up to 30\,per
cent of WDs exhibit traces of Ca, Si, Mg, Fe, Na and other metals
\citep[DZ\,WDs,][]{Zuckerman2003}.
This metal pollution has been attributed to the steady accretion of debris from
the tidal disruption of large asteroids and rocky planets \citep{Jura2003}
making these WDs important tools for the study of the chemical
composition of exosolar planets.  Interestingly, the incidence of magnetism among cool
($T_{\rm eff}<8\,000$\,K) DZ\,WDs is about thrteen\,per cent
\citep{kaw2014, hol2015}
which is much higher than between two and five\,per cent in the general WD population \citep{fer2015a}.
Although our modelling does not include the merging of sub-stellar companions,
we speculate that the moderately strong magnetic fields observed in metal-polluted WDs \citep[$0.5\le B/10^7{\rm G}\le 1.1$,][]{Hollands2017}
may be caused by giant gaseous planets plunging into the star. The accretion of other
minor rocky bodies would then produce the observed atmospheric pollution.  This
mechanism could be applicable to all WDs, although it is not clear what the
fraction of HFMWDs that may have undergone this process is. Currently only ten
out of about 240 HFMWDs are metal-polluted \citep{Hollands2017}.

Such merging events may occur duringthe latest stages of AGB evolution when the
outer envelope of the star engulfs the innermost planets and the drag forces exerted
on them as they move through the stellar envelope cause them to drift toward the
degenerate stellar core \citep{Li1998}.
Whilst this mechanism is plausible, it does not explain why the incidence of magnetism
is much higher among \emph{cool} DZ\,WDs.  Another possibility involves
close stellar encounters able to significantly disturb the orbits of outer planets and
asteroid belts. Such encounters can trigger dynamical instabilities that cause the inward
migration, and accretion by the WD, of a massive gaseous planet and other
rocky planets and asteroids. Because it takes hydrogen-rich WDs
with $0.5\le M/\Msun\le 1.0$ about $1.5-9$\,billion years to reach effective temperatures
between 5\,000 and 8\,000\,K \citep{Tremblay2011,Kowalski2006},
such stellar encounters are possible, as discussed in detail by \citet{Farihi2011}
to explain the origin of the very cool ($T_{\rm eff}=5310$\,K) and polluted magnetic
WD G77--50.

A similar explanation may be invoked to explain the high incidence of
magnetism among cool WDs of all types, as first reported by
\citet{Liebert1979}. The study of \citet{Fabrika1999}
showed that whilst the incidence of magnetism among hot WDs is only
around 3.5\,per cent, it increases above twenty\,per cent among cool WDs.
The volume-limited sample of \citet{Kawka2007}
also shows a high incidence of magnetism (greater than ten\,per cent) which is
consistent with the fact that volume-limited samples are dominated by cooler objects
\citep{Liebert2003}. Even the Palomar-Green magnitude-limited sample study of
\citet{Liebert05}
shows a higher incidence of magnetism among cooler WDs than hotter
ones. Over the years this topic has been a cause of concern.  It is difficult to think
of how fields could be generated once the star has already evolved into a WD
because, if anything, fields decay over time.  Alternatively, one could argue that the
formation rate of HFMWDs was higher when the Galactic disc was younger, another
hypothesis that is difficult to justify. \citet{wickramasinghe2000} and \citet{fer2015a}
have shown that the field strength is independent of effective temperature as expected 
by the very long ohmic decay time scales of WDs. The cumulative distribution
function of the effective temperatures of the sample of HFMWDs of 
\citet[][see their Figure\,5]{fer2015a} appears to be smooth over the full range of
effective temperatures ($4\,000\le T_{\rm eff}/{\rm K}\le 45\,000$\,K) suggesting
that the birthrate of HMWDs has not altered over the age of the Galactic disc.
However, the sample of HFMWDs at our disposition is neither volume nor
magnitude-limited and biases easily come into play.

Thus, should a future enlarged and less biased sample of HFMWDs confirm
that the incidence of magnetism among cool WDs is indeed
substantially higher than among hot WDs, then the possibility of
field generation by accretion of giant gaseous planets on to an originally
non-magnetic WD may provide a solution to this puzzle.

\citet{Nordhaus2011} found that discs formed from tidally disrupted
companions with masses in the range $0.1-500$ Jupiter masses can explain
the presence of high fields in WDs. Thus, the central issue is, once
again, how the magnetic field can diffuse into the core of a WD over
an appropriate timescale. This is a key question that still needs to be
quantitatively answered.

\noindent The other question concerns the likelihood for an old and presumably stable
planetary system to be sufficiently perturbed to send planets inward to plunge
into the WD.
\citet{Farihi2011} have shown that the number of close stellar encounters that
can have an appreciable effect on the outer regions of a planetary system by
sending objects into highly eccentric orbits is around 0.5\,Gyr$^{-1}$. That is,
the probability is about 50 per cent every 0.5\,Gyr$^{-1}$. Considering typical
cooling times between 1.5 and 9\,Gyr, these close encounters become likely
during the life of a WD. If this hypothesis is correct, we should expect
all WDs hosting a large gaseous planet to develop a magnetic field at
some point in their lifetime.\vspace{-6pt}
  
\section{Conclusions}\label{Conc}
\vspace{-12pt}
In chapter\,2 we discussed the evolution of HFMWDs resulting from two
stellar cores (one of which is degenerate) that merge during a phase
of CEE.  We fitted the observed mass distribution of the HFMWDs and
the incidence of magnetism among Galactic field WD and found that
the HFMWDs are well reproduced by the merging hypothesis for the
origin of magnetic fields if $0.1\le\alpha\le 0.3$.  However in chapter\,2
we did not propose a prescription that would allow me to assign a
magnetic field strength to each WD. This task has been carried out and
the results presented here. We have assumed that the magnetic field
attained by the core of the single coalesced star emerging from CEE
is proportional to the orbital angular velocity of the binary at the
point the envelope is ejected. The break-up angular velocity is
the maximum that can be achieved by a compact core during a
merging process and this can only be reached if the merging stars
are in a very compact binary, such as a merging DD system.

In our model there are two parameters that must be empirically
estimated. These are $B_0$, that is linked to the efficiency with
which the poloidal field is regenerated by the decaying toroidal field
(see \citet{wickramasinghe2014}) and the CE efficiency
parameter $\alpha$.  We carried out K--S test on the CDFs of the
observed and theoretical field distributions for a wide range of $B_0$
and $\alpha$ and we found that the observed field distribution is best
represented by models characterised by $B_0=1.35 \times10^{10}$\,G and
$\alpha=0.2$. Population synthesis studies of MCVs that make use of
the results obtained here and chapter\,2 show that the same
$B_0$ can also explain observations of magnetic binaries.

We have also speculated that close stellar encounters can send a giant
gaseous planet from the outer regions of a WD's planetary
system into a highly eccentric orbit. The plunging of this
super-Jupiter into the WD can generate a magnetic
field and thus provide an answer to why magnetism among cool WDs,
and particularly among cool DZ\, WDs, is higher than
among hot WDs.

\subsubsection{Acknowledgements}
GPB gratefully acknowledges receipt of an Australian Postgraduate Award. CAT
thanks the Australian National University for supporting a visit as a Research Visitor
of its Mathematical Sciences Institute, Monash University for support as a Kevin
Watford distinguished visitor and Churchill College for his fellowship.

\blanknonumber

\chapter{Origin of magnetic fields in cataclysmic variables}
\label{chapter4}

\vspace{-9mm}
This chapter is a reproduction of the paper submitted for publication in
Monthly Notices of the Royal Astronomical Society, viz:
\vspace{1mm}
\\
{\color{blue}\textit{Briggs, Ferrario, Tout \& Wickramasinghe, MNRAS (2018) (Submitted for publication), Origin of Magnetic Fields in Cataclysmic Variables}}
\vspace{-4mm}

\section{\vspace{-3mm}Abstract}
\label{Paper3abstract}
\vspace{-6pt}
  In a series of recent papers it has been proposed that HFMWDs
  are the result of close binary interaction and merging.
  Population synthesis calculations have shown that the
  origin of isolated highly magnetic white dwarfs is consistent with
  the stellar merging hypothesis. In this picture the observed fields
  are caused by an $\alpha-\Omega$ dynamo driven by differential
  rotation. The strongest fields arise when the differential rotation
  equals the critical break up velocity and result from the merging of
  two stars (one of which has a degenerate core) during CEE
  or from the merging of two white dwarfs. We now synthesise a
  population of binary systems to investigate the hypothesis that
  the magnetic fields in the MCVs also originate during stellar interaction
  in the CE phase. Those systems that emerge from CE more tightly bound
  form the CVs with the strongest magnetic fields. We vary the CE efficiency 
  parameter $\alpha$ and compare the results of our population syntheses with
  observations of magnetic cataclysmic variables.  We find that CE
  interaction can explain the observed characteristics of these magnetic
  systems immediately after CE if $\alpha<0.4$.\vspace{-6pt}\newline
\newline\noindent
Keywords: Stars: cataclysmic variables -- stars: white dwarfs --stars: magnetic fields -- stars: binaries.

\section{Introduction}
\label{Introd}
\vspace{-12pt}
Cataclysmic variables (CVs) are close interacting binaries generally
consisting of a low-mass main-sequence (MS) star transferring matter
to the WD primary via Roche lobe overflow \citep{Warner1995}.
In non-magnetic or weakly magnetic systems, which make up
the majority of observed CVs, the hydrogen-rich material
leaving the secondary star from the inner Lagrangian point forms an
accretion disc around the WD. It is estimated that some
$20-25$\,per cent of all CVs host a magnetic WD
\citep[MWDs,][]{wickramasinghe2000,fer2015a}.  These systems are the
MCVs. Among MCVs we have the strongly magnetic
AM Herculis variables or polars. In polars the high magnetic field of
the WD can thread and channel the material from the secondary
star directly from the ballistic stream to form magnetically confined
accretion funnels, so preventing the formation of an accretion
disc. In these systems the two stars are locked in synchronous
rotation at the orbital period. The radiation from the accretion
funnels \citep[e.g.][]{FerWe99} and the cyclotron radiation from
the shocks located at the funnels' footpoints of closed magnetic field
lines dominate the emission of these systems from the X-rays to the
infrared bands \citep[e.g.][]{MeggittWick1982,Wickramasinghe1988}.  Cyclotron
and Zeeman spectroscopy and spectropolarimetry have revealed the
presence of strong fields in the range of a few $10^7-10^8$\,G
\citep[see, e.g.,][]{fer1993,fer1996}.  Weaker fields of about $10^6$ to
$3\times 10^7$\,G are found in the DQ\,Herculis variables or
Intermediate Polars (IPs) where the WD's magnetic field
cannot totally prevent the formation of an accretion disc
\citep[e.g. see][]{fwk1993}. In these systems the material is
magnetically threaded from the inner regions of a truncated accretion
disc and channelled on to the primary forming magnetically confined
accretion curtains \citep{FerrarioWick1993}. In the IPs the white
dwarf is not synchronously locked with the orbital period but is spun
up to a spin period shorter than the orbital period of the system.

\citet{liebert2005} noticed the enigmatic lack of MWDs from the 501
detached binaries consisting of a WD with a non-degenerate
companion found in the DR1 of the Sloan Digital Sky Survey
\citep[SDSS,][]{York2000}. They also noticed that among the 169 MWDs
known at the time, none had a non-degenerate detached companion. This
was puzzling because such a pairing is very common among non-magnetic
WDs \citep[see, e.g.][]{Hurley2002,Ferrario2012}. A similar study
conducted on the much larger DR7 sample of SDSS detached binaries
consisting of a WD with a non-degenerate companion
\citet{Kleinman2013} led to the same conclusion
\citep{Liebert2015}. Over the years, not a single survey conducted to
ascertain the incidence of magnetism among WDs has yielded a
system consisting of a magnetic WD with a non-degenerate
companion \citep[e.g.,][]{sch2001a,Kawka2007}. It is this curious
lack of pairing that led \citet{tout2008} to propose that the
existence of magnetic fields in WDs is intimately connected
to the duplicity of their progenitors and that they are the result of
stellar interaction during CEE.  In this
picture, as the cores of the two stars approach each other, their
orbital period decreases and the differential rotation that takes
place in the convective CE generates a dynamo mechanism
driven by various instabilities.  \citet{Regos1995} argued that it is
this dynamo mechanism that is responsible for the transfer of energy
and angular momentum from the orbit to the envelope which is
eventually, all or in part, ejected.

\citet{wickramasinghe2014} have shown that strong magnetic fields in WDs
can be generated through an $\alpha-\Omega$ dynamo during CEE where
a weak seed poloidal field is wound up by
differential rotation to create a strong toroidal field. However
toroidal and poloidal fields are unstable on their own
\citep{Braithwaite2009}. Once the toroidal field reaches its maximum
strength and differential rotation subsides the decay of toroidal
field leads to the generation of a poloidal field with the two
components stabilising each other and limiting field growth until they
reach a final stable configuration. Thus, a poloidal seed field can be
magnified during CEE by an amount that depends
on the initial differential rotation but is independent of its initial
strength. According to this scenario the closer the cores of the two
stars are dragged at the end of CEE, before the
envelope is ejected, the greater the differential rotation and thus
the stronger the expected frozen-in magnetic field. If CEE
leads to the merging of the cores the result is a highly
magnetic isolated WD. If the two stars do not coalesce they
emerge from the CE as a close binary that evolves into a
MCV.  The viability of such model, in terms of incidence of magnetism
among single WDs and their mass and magnetic field
distribution, have been shown in chapters\,2 and 3 respectively.

In this chapter we continue our study of the origin of fields in MWDs
to explain the parentage of MCVs. To this end we carry out a
comprehensive population synthesis study of binaries for different
CE efficiencies $\alpha$. we examine all paths that lead
to a system consisting of a WD with a low-mass companion
star.  We show that the observed properties of the MCVs are generally
consistent with their fields originating through CEE for $\alpha<0.4$. 
 
\section{Evolution and space density of MCVs}
\vspace{-4mm}
Observed MCVs consist of a WD that accretes matter from a
secondary star that has not gone through any significant nuclear
evolution when the transfer of mass begins. The mass ratio of an MCV
is given by $q=M_{\rm sec}/M_{\rm WD}<1$ where $M_{\rm WD}$ is the
mass of the WD primary and $M_{\rm sec}$ is the mass of the
companion star. The mass accretion process in MCVs is relatively
stable over long periods of time, although the polars suffer from high
and low states of accretion. It is not known what sparks the change in
accretion mode but, because polars do not have a reservoir of matter
in an accretion disc, they can switch very quickly from high to low
states. IPs have never been observed in low states of accretion. 
  Stable mass transfer can be driven by nuclear-timescale expansion of
  the secondary (not generally applicable to MCVs) and/or by loss of
  angular momentum, driven by magnetic braking above the CV period gap
  \citep[caused by the disrupted magnetic braking mechanism,
  see][]{Spruit1983, Rap83, Verbunt1984} and gravitational
  radiation below the gap. Loss of angular momentum shrinks the orbit
  keeping the companion star filling its Roche lobe and so drives mass
  transfer. Therefore, as MCVs evolve, the orbital period diminishes
  until it reaches a minimum when the secondary star becomes a
  substellar-type object whose radius increases as further mass is
  lost. Systems that have reached the minimum period and have turned
back to evolve toward longer periods are often called period bouncers
\citep[e.g.][]{Patterson1998}

The evolution of MCVs is expected to be similar to that of
non-magnetic CVs. However, \citet{LiWuWick1994} have shown that
angular momentum loss may not be as efficient in polars as it is in
non-magnetic or weakly magnetic CVs in bringing the two stars together
because the wind from the secondary star is trapped within the
magnetosphere of the WD. This phenomenon slows down
the loss of angular momentum, reduces the mass transfer rate and leads to
longer evolutionary timescales. This mechanism provides a simple
explanation for the observed high incidence of magnetic systems among
CVs \citep{Araujo2005}.

\begin{figure}
\includegraphics[width=1.05\columnwidth]{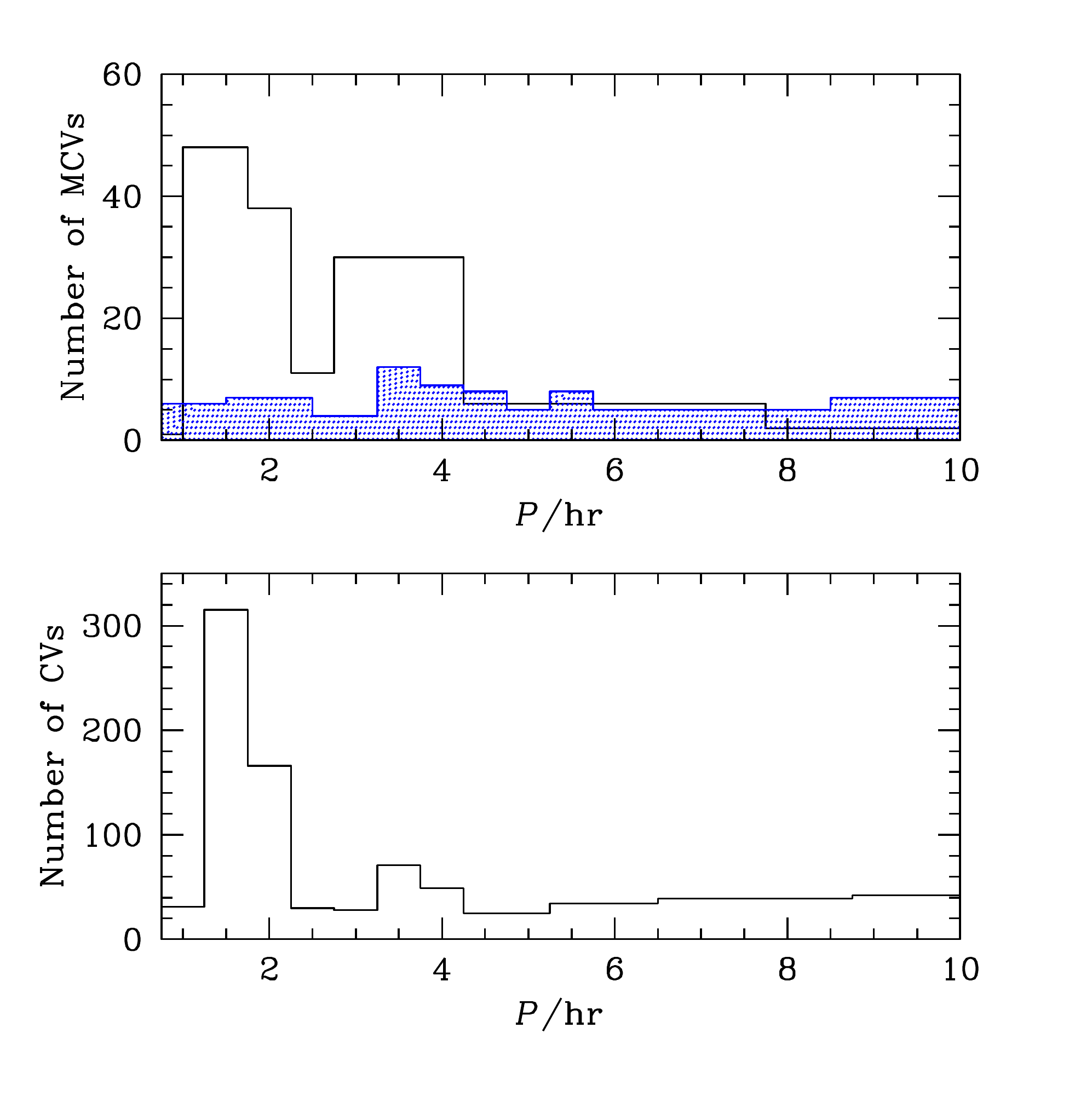}
\caption{The orbital period distribution of MCVs (top) and of the CVs
  (bottom). The MCVs are subdivided into Polars (solid black line histogram) and
  IPs (shaded histogram). We have used the latest version (v7.20) of the
  \citet{RitterKolb2003} CV catalogue to create this figure.}
\label{fig:orbdistr}
\end{figure}
We show in Fig.\,\ref{fig:orbdistr} the period distribution of CVs and
MCVs where the MCVs have been subdivided into polars and intermediate
polars.  The space density of CVs is not well known and, over the
years, there has been some considerable disagreement between
observations and theoretical predictions. The recent study of
\emph{Swift}\,X-ray spectra of an optically selected sample of nearby
CVs conducted by \citet{Reis2013} has unveiled a number of very low
emission X-ray systems.  Hard X-ray surveys of the Galactic ridge have
shown that a substantial amount of diffuse emission can be
resolved into discrete low-luminosity sources.  Because the MCVs are
generally strong X-ray emitters, \citet{Muno2004} and \citet{Hong2012}
have propounded that IPs could be the main components of these
low-luminosity hard X-ray sources.  The SDSS has also revealed a
substantial number of low-accretion rate CVs near the CV period
minimum \citep{Gaensicke2009}.

\citet{Pretorius2013} have conducted a study of the X-ray flux-limited
\emph{ROSAT} Bright Survey (RBS) to determine the space density of
MCVs. They assume that the 30 MCVs in the RBS are representative of
the intrinsic population. They also allow for a 50 per cent high-state
duty cycle for polars under the assumption that polars are below the
RBS detection threshold while they are in low states of accretion.
They find that the total space density of MCVs is
$1.3^{+0.6}_{−0.4}\times 10^{−6}$ pc$^{-3}$ with about one IP per
200\,000 stars in the solar neighbourhood. They conclude that IPs are
indeed a possible explanation for most of the X-ray sources in the
Galactic Centre. These new findings seem to suggest that the space
density of CVs is somewhat larger than initially forecast and thus in
closer agreement with theoretical expectations.

\subsection{Where are the progenitors of the MCVs?}\label{WhereProg}

\begin{figure}
\includegraphics[width=1.03\columnwidth]{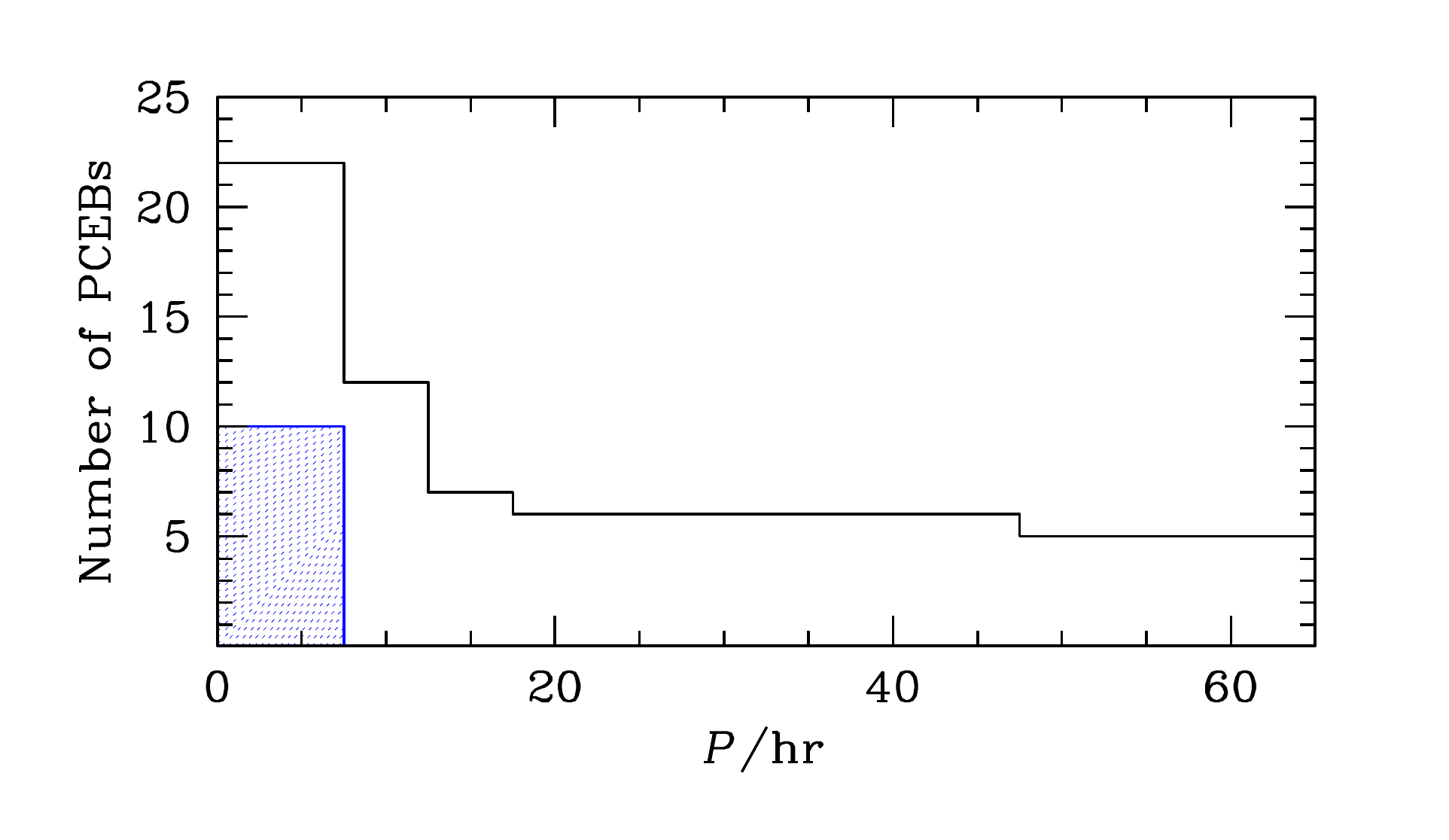}
\caption{The orbital period distribution of PCEBs
  \citep[solid black line histogram,][]{Nebot2011} and PREPs
  \citep[shaded histogram,][]{fer2015a}.}
\label{preps_fig}
\end{figure}

\citet{liebert2005, Liebert2015} asked, ``Where are the magnetic white
dwarfs with detached, non-degenerate companions?''  This question is
awaiting an answer and thus the progenitors of the MCVs still need to
be identified.  As already noted, the proposal by \citet{tout2008},
that the existence of high magnetic fields in all WDs,
isolated and in binaries, is related to their duplicity prior to
CEE is gaining momentum. Observational support
for the binary origin of magnetic fields in MCVs is also
strengthening. \citet{Zorotovic2010} listed about $60$ post CE
 binaries (PCEBs) from the SDSS and other surveys consisting
of a WD with an M-dwarf companion. The periods of these PCEBs
range from about $0.08$ to $20$\,d and the WD effective
temperatures in the range 7\,500 to 60\,000\,K. According to current
binary evolution theory, one third of these systems should lose
angular momentum from their orbits by magnetic braking and
gravitational radiation and are expected to come into contact within a
Hubble time. However none of these 60 binaries contains a MWD, even if
observations indicate that 20 to 25\,per cent of all CVs harbour
one. Furthermore, magnitude limited samples of WD have shown
an incidence of magnetism of about 2 to 3\,per cent and thus some
pre-MCVs should be present among the objects listed by
\citet{Zorotovic2010}.  This finding indicates that magnetic white
dwarf primaries are only present in those binaries that are already
interacting or are close to interaction. The magnetic systems
originally known as Low-Accretion Polars \citep[LARPS,][]{Schwope2002}
have been proposed to be the progenitors of the polars
\citep{sch2009}. The first LARPS were discovered in the
Hamburg/ESO Quasar Survey \citep[HQS,][]{Wisotzki1991} and then by the
SDSS by virtue of their unusual colours arising from the presence of
strong cyclotron harmonic features in the optical band together with a
red excess owing to the presence of a low-mass red companion star. The
MWDs in LARPS are generally quite cool
($T_{\rm eff}\lsimeq10\,000$\,K) and have low-mass MS companions which
underfill their Roche lobes \citep[e.g.][]{Reimers1999, sch2009,
  Parsons2013}. The MWDs in these systems accrete mass from the wind
of their companion at a rate substantially larger than that observed
in detached non-magnetic PCEBs \citep{Parsons2013}. These systems were
thus renamed pre-polars (PREPs) by \citet{sch2009} to avoid
confusion with polars in a low state of accretion. PREPs have orbital
periods which are, on average, only marginally longer than those of
polars. The ages of the WDs in PREPs, as indicated by their
effective temperatures, are generally above a billion years. The
absence of PREPs with hot WDs is puzzling but maybe still not
alarming, if one considers the small number of PREPs currently known
and the initial rapid cooling of WDs. Thus, the hypothesis
that the progenitors of MCVs are expected to emerge from CE
when they are close to transferring mass via Roche Lobe overflow
is well warranted. We show in Fig.\,\ref{preps_fig} the period
distribution of PCEBs and PREPs.

\section{Population synthesis calculations}\label{popsynt}
\label{sec:calculations}
\vspace{-12pt}
Each binary is assigned three initial parameters. These are the mass
$1.0\le M_1/\msun\le 10.0$ of the primary star, the mass
$0.1\le M_2/\msun\le 2.0$ of the secondary star, and the orbital
period $1\le P_0/{\rm d}\le 10\,000$ at the zero-age main sequence
(ZAMS). We set the eccentricity to zero. Each parameter was sampled
uniformly on a logarithmic scale with 200\,divisions.  This sampling
gives a synthetic population of about 70 million binary systems.  The
actual number of binary systems is then calculated on the premise that
$M_1$ follows Salpeter's mass function distribution
\citep{Salpeter1955} and $M_2$ is according to a flat mass ratio
distribution with $q\le 1$.  The initial period distribution is
assumed to be uniform in the logarithm.
\clearpage
\begin{landscape}
\begin{table*}
  \caption{We have indicated with $N$ (second column) the fraction of PREPs that emerge from CE for different efficiency parameters $\alpha$ (first column) in a single generation of binaries. The other columns give the smallest and the largest progenitor masses and initial orbital periods.}
  \centering
\begin{tabular}{ c c c c r c c r}
\hline
\noalign{\smallskip}
$\alpha$ & $N$\,(per cent) & $M_{1_{\rm min}}/\msun$ & $M_{2_{\rm
    min}}/\msun$ &$M_{1_{\rm max}}/\msun
$&$M_{2_{\rm max}}/\msun $ & $P_{0_{\rm min}}/$d & $P_{0_{\rm max}}/$d \\
\noalign{\smallskip}
\hline
\noalign{\smallskip}
0.10   & 1.518  & 1.08 & 0.10 & 8.16 & 1.42   & 369.7 &  3144.0  \\
0.15   & 1.672  & 1.08 & 0.10 & 8.16 & 1.42   & 293.3 &  2800.5  \\
0.20   & 1.663  & 1.08 & 0.10 & 8.16 & 1.42   & 246.6 &  2354.3  \\
0.25   & 1.213  & 1.08 & 0.10 & 8.16 & 1.36   & 207.3 &  2097.0  \\
0.30   & 1.163  & 1.08 & 0.10 & 8.16 & 1.14   & 184.6 &  2221.9  \\ 
0.50   & 0.808  & 1.08 & 0.10 & 8.16 & 0.58   & 123.2 &  2221.9  \\
0.70   & 0.804  & 1.08 & 0.10  & 8.16 & 0.19  &   87.0 &  1867.9  \\
0.90   & 0.859  & 1.08 & 0.10  & 8.16 & 0.13  &   69.1 &  1762.9  \\
\noalign{\smallskip}
\hline
\end{tabular}
\label{tab:statsCE}
\end{table*}
\end{landscape}

We have used the rapid binary star evolution algorithm, {\sc bse},
developed by \citet{Hurley2002}, to evolve each binary system from the
ZAMS to $9.5\,$Gyr \citep[age for the Galactic Disc,][]{Kilic2017}. {\sc bse}
is an extension of the single star evolution code written by \citet{Hurley2000}.
It allows for stellar mass loss, interaction between the two stars as mass transfer,
Roche lobe overflow, CEE \citep{Pac1976}, tidal interaction, supernova
kicks, and angular momentum loss caused by gravitational radiation and
magnetic braking.

We use the $\alpha$ (energy) formalism for CE phases
where $\alpha$ is a free parameter ranging between $0.1$ and $0.9$
(see chapters\,2 and 3 for more details).  Single star mass loss rates are
described by \citet{Hurley2000}. In our calculations we have adopted
a solar metallicity $Z=0.02$ and $\eta=1.0$ for the Reimers' mass-loss
parameter.

Our theoretical sample of PCEBs consists of systems that (i)\,have
undergone CEE, (ii)\,have a primary that evolves
into a WD, (iii)\,have a companion that remains largely
unevolved and (iv)\, have a mass ratio $q\le 1$.  A subset of these
systems come into contact over the age of the Galactic Disc and become
classical CVs. Those systems with a WD that develops a strong
magnetic field become MCVs.

Of our sample of PCEBs, we then select the subset consisting of the
PREPs (the progenitors of the MCVs). PREPs must fulfil two additional
criteria: (i) the primary star must have a degenerate core before
entering the last CE phase and (ii) no further core
burning occurs. The reason for the first criterion is that a
degenerate core is essential for a stellar magnetic field to persist,
in a frozen-in state, after its formation.  The reason for the second
is that nuclear burning in the core would ignite convection that would
destroy any frozen-in magnetic field. Systems that violate either
criterion but come into contact over the age of the Galactic Disc are
expected to evolve into classical non-magnetic CVs. We show in
Table\,\ref{tab:statsCE} the limits of the parameter space within
which PREPs are formed.  The minimum ZAMS masses of the systems that
give rise to PREPs are listed in the columns with headings
$M_{1_{\rm min}}$ and $M_{2_{\rm min}}$ and the maximum masses are
under the headings $M_{1_{\rm max}}$, $M_{2_{\rm max}}$.  Minimum and
maximum initial periods are in the columns under $P_{0_{\rm min}}$ and
$P_{0_{\rm max}}$ respectively.

Once we have obtained our theoretical PREP sample, we assign a magnetic
field $B$ to each of their WD primaries following the prescription
described in chapter 3 to model the field distribution of high field
magnetic WDs (HFMWDs). That is

\begin{equation}\label{EqBfield}
B = B_0\left(\frac{\Omega}{\Omega_{\rm crit}}\right)\, \mbox{G}.
\end{equation}

where $\Omega$ is the orbital angular velocity and
$\Omega_{\rm crit}=\displaystyle{\sqrt{GM_{\rm WD}/R_{\rm WD}^3}}$ is
the break-up angular velocity of a WD of mass $M_{\rm WD}$
and radius $R_{\rm WD}$.  The parameter $B_0$ is a free parameter that
was determined empirically in chapter 3, that is,
$B_0=1.35\times 10^{10}$\,G. The parameter $B_0$ does not influence
the shape of the field distribution which is only determined by
$\alpha$. Lower (or higher) $B_0$ shift the field distribution to
lower (or higher) field strengths. Unlike HFMWDs, both stars emerge
from CEE but on a much tighter orbit that will allow them to come into
contact over a Hubble time and appear as MCVs.
\vspace{12pt}

\section{Synthetic population statistics}\label{SyntPop}
\vspace{-12pt}
We have time integrated each population, characterised by
$\alpha$, to the Galactic Disc age under the assumption that
the star formation rate is constant. We have listed in
Table\,\ref{tab:PCEBS_PREPS_Percent} the percentage by type of all
binaries that emerge from CE over the age of the Galactic
Disc.

\begin{table}
  \caption{The number of PCEBs born, the fraction of PREPs from PCEBs
    and of MCVs (magnetic systems already exchanging mass) from PREP as a
    function of the CE efficiency parameter $\alpha$ over the
    age of the Galactic Disc. The number of PREPs is maximum close to
    $\alpha\,=\,0.15$ while the number of MCVs is maximum at $\alpha\,=\,0.10$.}
\centering
\begin{tabular} {l l l l}
\hline
\noalign{\smallskip}
$\alpha$ & Number of & \underline{PREPs} $\times100$& \underline{MCV~~} $\times100$ \\
              & PCEBs~~~   &PCEBs~~         & PREPS~   \\
\noalign{\smallskip}
\hline
\noalign{\smallskip}
0.10 & 30517472 &20.9 & 61.0 \\ 
0.15 & 36099023 &18.9 & 56.4 \\
0.20 & 38666876 &15.3 & 49.9 \\
0.30 & 41197674 &  8.7 & 45.0 \\
0.40 & 43654871 &  5.6 & 48.0 \\
0.50 & 46289395 &  4.5 & 51.0 \\
0.60 & 49010809 &  4.1 & 52.0 \\
0.70 & 51888317 &  3.8 & 52.4 \\
0.80 & 54664759 &  3.3 & 52.4 \\
\hline
\end{tabular}
\label{tab:PCEBS_PREPS_Percent}
\end{table}

Column 2 in Table\,\ref{tab:PCEBS_PREPS_Percent} shows that while the
number of PCEBs increases when $\alpha$ increases, the percentage of
PREPs (progenitors of the MCVs) decreases. This is because as $\alpha$
increases the envelope's clearance efficiency increases causing the
two stars to emerge from CE at wider separations and thus
less likely to become PREPs and thence MCVs. On the other hand, the
overall number of PCEBs increases because stellar merging events
become rarer at high $\alpha$, as shown in chapter 2.  Fewer merging
events are also responsible for the high incidence of systems with 
low mass He\,WDs (He\,WDs) whose ZAMS progenitors were born at
short orbital periods and entered CEE when the primary star
became a Hertzsprung gap or a RGB star. At larger
initial orbital periods CEE may occur on the AGB. However
as $\alpha$ increases only stars in those systems that harbour
massive enough WDs can come sufficiently close to each other
to allow stable mass transfer to occur within the age of the Galactic
Disc (see section \ref{mass_distribution}). In contrast, at low $\alpha$ the
clearance efficiency is low and so there is a longer time for the
envelope to exert a drag force on the orbit. This results in (i)
more merging events, (ii) tighter final orbits for all WD
masses and (iii) a larger number of systems coming into contact over
the age of the Galactic Disc. Point (i) reduces the overall number of
PCEBs while both (ii) and (iii) increase the number of PREPs.

\clearpage
\begin{landscape}
\begin{table*}
  \caption{Evolutionary history of an example binary system that
    becomes a MCV after CEE with $\alpha\,=\,0.1$.
    Here RLO\,=\,Roche Lobe Overflow. }
  \centering
\begin{tabular}{ r r r r r r c l  }
\hline
\noalign{\smallskip}
Stage  & Time/Myr & $M_1/\msun$ & $M_2/\msun$ & $P/$\,d & $a/{\rm R}_\odot$ &$B/{\rm G}$& Remarks \\
\noalign{\smallskip}
\hline 
\noalign{\smallskip}
  1    &       0.000    &  4.577    & 0.230  & 2244.627 & 1218.030    & 0.000E+00   & ZAMS  \\  
  2    &    128.515   &  4.577    & 0.230  & 2244.627 & 1218.030    & 0.000E+00   & S1 is a Hertzsprung gap star \\
  3    &    129.078   &  4.577    & 0.230  & 2245.210 & 1218.188    & 0.000E+00   & S1 is a RGB star.  Separation increases slightly. \\
  4    &    129.445   &  4.574    & 0.230  & 2247.427 & 1218.790    & 0.000E+00   & S1 starts core He burning. Some mass loss occurs.\\
  5    &    149.930   &  4.466    & 0.230  & 2352.896 & 1247.059    & 0.000E+00   & S1 is an AGB star. Further mass loss occurs. \\  
  6    &    150.947   &  4.390    & 0.230  & 2173.184 & 1176.321    & 0.000E+00   & S1 is a late AGB star. Separation decreases significantly \\
  7    &    150.989   &  4.364    & 0.230  &  861.296  & 633.510      & 0.000E+00   & RLO \& CE start. Separation decreases dramatically. \\
  8    &    150.989   &  0.918    & 0.230  &    0.117    &  1.053        & 1.218E+07   & S1 emerges from CE as a CO\,MWD and RLO ceases.\\
  9    &    326.073   &  0.918    & 0.230  &    0.099    &  0.945        & 1.218E+07   & Separation decreases and MCV phase starts \\
 10   &  9\,500.000 &  0.918    & 0.037  &    0.139    &  1.112        & 1.218E+07   & Separation reaches a minimum between stages 9 and 10 \\
        &                   &               &           &                &                  &                     & and increases again. S2 is a brown dwarf. \\
\hline
\end{tabular}
\label{tab:evolMCV1}
\end{table*}

\begin{table*}
  \caption{Evolutionary history of a second example binary system that
    becomes a MCV after CE with $\alpha=0.4$.}
\centering
\begin{tabular}{ r r r r r r c l }
\hline
\noalign{\smallskip}
Stage  & Time/Myr & $M_1/\msun$ & $M_2/\msun$ & $P$\,(d) & $a/{\rm R}_\odot$ &$B/{\rm G}$& Remarks \\
\noalign{\smallskip}
\hline
\noalign{\smallskip}
  1  &          0.000 & 1.612  &  0.257  &   190.661  &  171.774   &    0.000E+00  & ZAMS  \\ 
  2  &    2197.329 &  1.612  &  0.257  &   190.661  &  171.774   &   0.000E+00  & S1 is a Hertzsprung gap star \\ 
  3  &    2239.430 &  1.611  &  0.257  &   190.743  &  171.811   &   0.000E+00  & S1 is a RGB star, loses mass.  Separation increases slightly. \\
  4  &    2343.048 &  1.580  &  0.257  &   110.351  &  118.629   &   0.000E+00  & S1 loses more mass, separation decreases. \\
  5  &    2343.048 &  0.386  &  0.257  &     0.149    &     1.020   &   3.577E+07  & RLO \& CE start. Separation decreases dramatically. \\
  6  &    2343.048 &  0.386  &  0.257  &     0.149    &     1.020   &   3.577E+07  & S1 emerges from CE as a He\,MWD and RLO ceases.\\
  7  &    3389.278 &  0.386  &  0.257  &     0.102    &     0.792   &   3.577E+07  & Separation decreases and MCV phase starts \\
  8  &  9\,500.000 &  0.386  &  0.052  &     0.100    &     0.687   &   3.577E+07  & Separation reaches a minimum between stages 7 and 8 \\
      &                   &            &            &                  &                &                      & and increases again. S2 is a brown dwarf.\\
\noalign{\smallskip}
\hline
\end{tabular}
\label{tab:evolMCV2}
\end{table*}
\end{landscape}

\subsection{Magnetic CV evolution examples}

The evolutionary history of a binary system depends on the parameters
that characterise it. The number of CE events can vary from one to
several \citep{Hurley2002}.  Whether a classical CV becomes magnetic
or not depends on the evolution before and after the common
envelope. Here we give two typical examples of systems that evolve
into a MCV. In the first the initially rather massive primary
star evolves into a CO\,WD after CEE as a late AGB star. In the second
example the primary evolves into a He\,WD after CE evolution while
ascending the RGB.

\emph{Example 1:} Table~\ref{tab:evolMCV1} illustrates the evolution
of a system that becomes a close binary after CE with
$\alpha=0.1$.  The progenitors are a primary star (S1) of
$4.58\,\msun$ and a secondary star (S2) of sub-solar mass
$0.230\,\msun$.  At ZAMS the initial period is $2\,240\,$d with a
separation of $1\,220\,R _\odot$.

S1 evolves off the ZAMS and reaches the early AGB stage at
$149$\,Myr having lost $0.111$\,$\msun$ on the way.  After a
further $1.02$\,Myr S1 has become a late AGB star. Further evolution
brings the stars closer together at a separation of
$634$\,R$_\odot$. Soon after dynamically unstable Roche lobe
overflow from S1 to S2 takes place and CE begins.  At the
end of the short period of CEE the two stars
emerge with a separation of only $1.05$\,R$_\odot$ because of the
large orbital angular momentum loss during this stage.  The ejection
of the envelope exposes the core of S1 that has now become a magnetic
$0.918$\,$\msun$ CO\,WD.  After a further $175$\,Myr the
separation has further contracted to $0.945$\,R$_\odot$ via magnetic
braking and gravitational radiation.  Roche lobe overflow begins and
the system becomes a bona fide mass-exchange MCV. During the CEE
evolutionary phase the mass of the donor star, separation and orbital
period steadily decrease until the mass of the companion star becomes
too low to maintain hydrogen burning and S2 becomes a degenerate
object. At this point separation and orbital period reach a
minimum. Further evolution sees these two quantities increase again
over time. At an age of $9\,500$\,Myr S2 has lost most of its mass and
has become a $0.037$\,$\msun$ brown dwarf with the separation from its WD
primary increased to $0.112$\,R$_\odot$.

\emph{Example 2:} Table~\ref{tab:evolMCV2} shows the evolution of a
second system that becomes a close binary after CE. This
time we have $\alpha=0.4$.  The progenitors are a MS
primary star (S1) of $1.61$\,$\msun$ and a secondary star (S2) of mass
$0.257$\,$\msun$.  At ZAMS the initial period is $191$\,d and the
separation $172$\,R$_\odot$.

S1 evolves off the ZAMS through the Hertzsprung gap to reach the RGB
after $2\,240$\,Myr having lost $0.001$\,$\msun$ on the way. Still on
the RGB at $2\,340$\,Myr S1 has lost $0.031$\,$\msun$ and the
separation has decreased to $119$\,R$_\odot$. Roche lobe overflow
from S1 to S2 and CEE begin. S1 emerges from CE as a magnetic
He\,WD with a mass of $0.386$\,$\msun$. The orbital separation
has drastically decreased to $1.02$\,R$_\odot$. S2 maintains its
mass and remains an M-dwarf star. From this time onwards magnetic
braking and gravitational radiation cause the orbit to shrink further
until at $3\,390$\,Myr the  separation is $0.792$\,R$_\odot$ and
Roche lobe overflow commences. The system is now a MCV.  Further
evolution leads S2 to lose mass, owing to accretion on to S1, until,
at $9\,500$\,Myr, S2 has become a brown dwarf of mass $0.052$\,$\msun$
and the separation is $0.687$\,R$_\odot$.

\subsection{Property distributions of the synthetic population}

We create our population of putative PREPs by integration over time
from $t=0$ to $t=9.5\,$Gyr.  The star formation rate is taken to be
constant over the age of the Galactic Disc.  Whereas
Table\,\ref{tab:statsCE} shows the relative numbers of PREPs obtained
from a single generation of binaries, continuous star formation
over the age of the Galactic Disc builds up a population of PCEBs,
PREPs, CVs, and MCVs that, as birth time increases, favours systems with
progressively higher mass primaries because lower mass primaries,
especially in later generations, do not have enough time to evolve to
the WD stage.

\begin{figure*}
\begin{center}
\includegraphics[width=1.03\textwidth]{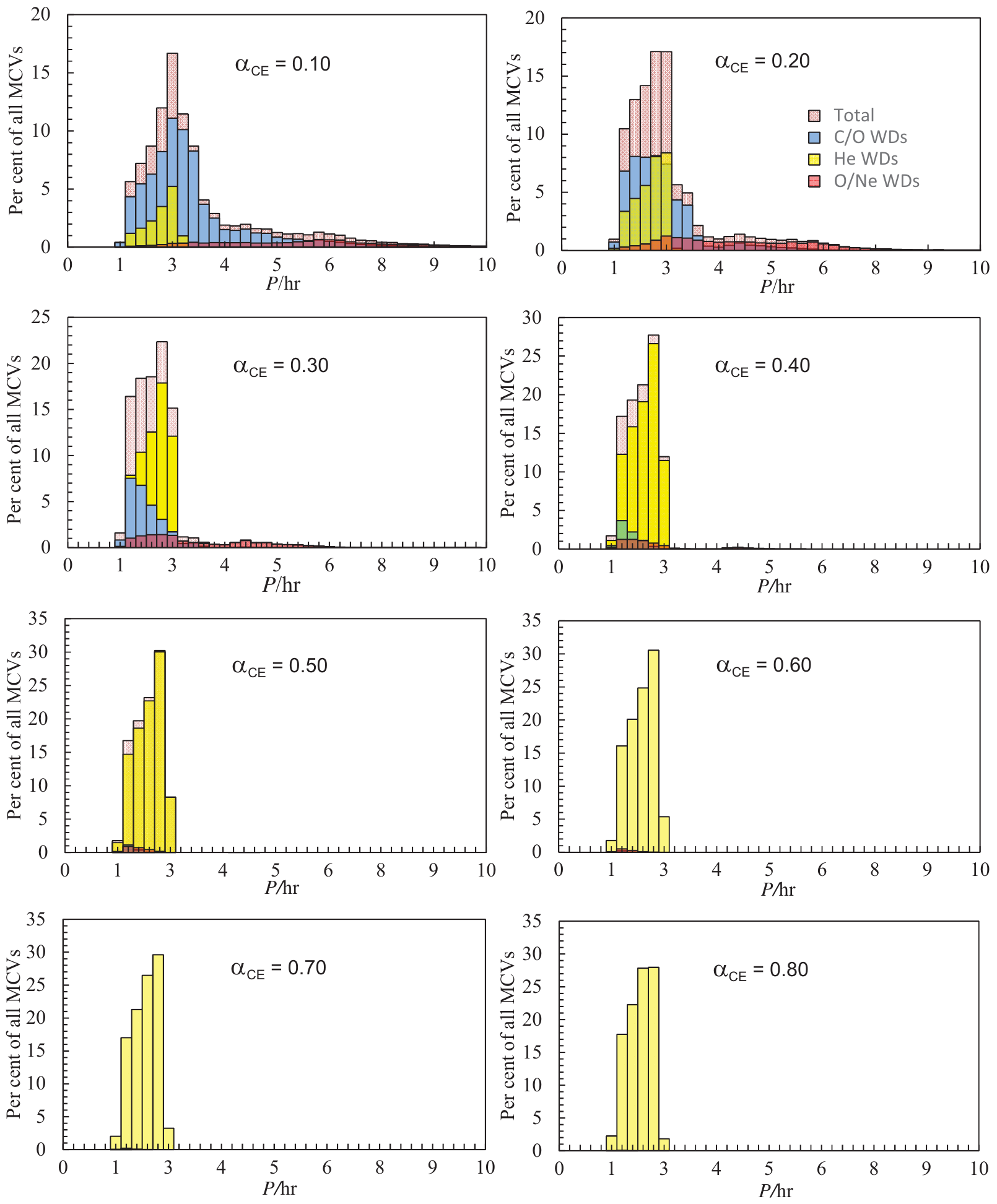}
\caption{Theoretical period distribution of magnetic systems just
  before they start RLOF for various $\alpha$'s.
  The period distribution of the primary WD types is shown as the
  superimposed coloured categories. The total of the distribution is
  shown as the pink background histogram peaking around 2.8 
  to 3.0\,hrs. This is to be compared with the observed distribution
  for PREPs in Fig.\,\ref{preps_fig}}
\label{fig:RLOFPeriodP}
\end{center}
\end{figure*}

\begin{figure*}
\begin{center}
\includegraphics[width=1.03\textwidth]{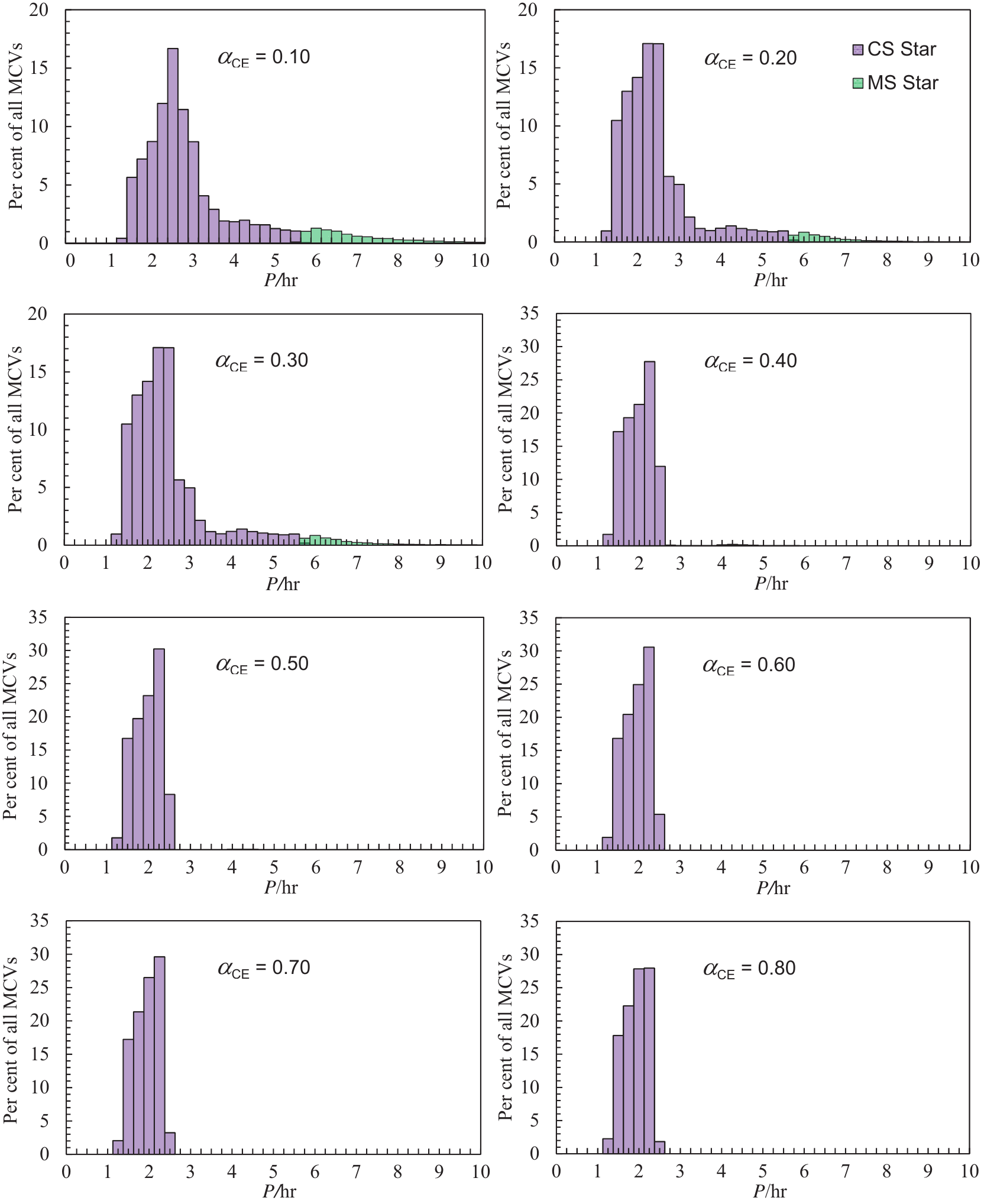}
\caption{Same as Fig.\ref{fig:RLOFPeriodP} but with the secondary star
  types shown as the superimposed coloured categories.  Both secondary
  star types are MS stars.  The CS type is a deeply or fully
  convective MS star with $M<0.7$\,M$_\odot$.}
\label{fig:RLOFPeriodS}
\end{center}
\end{figure*}

\subsubsection{Period distribution}\label{PerDist}

Figs\,\ref{fig:RLOFPeriodP} and \ref{fig:RLOFPeriodS} show the
theoretical period distribution of the PREPs just before the beginning
of Roche lobe overflow (RLOF) in a present day population formed over
the age of the Galactic Disc for various $\alpha$. The contributions
to the period distribution by WD primaries of a certain
type are depicted in Fig.\,\ref{fig:RLOFPeriodP} while the
contributions to the period distribution by the secondaries of a given
type are displayed in Fig.\,\ref{fig:RLOFPeriodS}.

The period distribution peaks around $3\,$hr with a long tail
extending to about $10\,$hr for low $\alpha$. We note that at low
$\alpha$ our synthetic population tends to have orbital periods
clustering around the 1 to 4\,hr region while at higher $\alpha$ they
are confined to the 1 to 3\,hr region.

Fig.\,\ref{fig:RLOFPeriodP} shows that when $\alpha=0.1$ the main
contributors to the whole range of periods are systems with CO\,WD
primaries characterised by orbital periods from about 1 to 7\,hr and a
peak near 3\,hr. Systems with He\,WDs are also present but are fewer
and their periods are below 3\,hr.  Massive ONe\,WD primaries form
a much smaller fraction of the distribution,
as expected from a Salpeter initial mass function, but make some
contribution to the full range of periods when $\alpha< 0.4$.

As $\alpha$ increases the fraction of CO\,WD systems decreases until
these all but disappear for $\alpha>0.6$ while the percentage of
He\,WDs increases dramatically. For $\alpha\ge 0.4$, the orbital
periods are always below 3\,hr and He\,WD systems well and truly
dominate the period distribution.  For $\alpha>0.6$ the only systems
that are predicted to exist are those with He\,WDs. The fraction of
ONe\,WD systems reaches a maximum near $\alpha=0.2$ and then
decreases.

We note that systems with He\,WDs tend to populate the lowest period
range at all $\alpha$. These systems are generally characterised by
initially lower-mass primaries, and thus lower-mass companions
because $q\le 1$, and shorter orbital periods and initiate common
envelope evolution before helium ignition. Usually systems
characterised by short initial periods are unlikely to survive at low
$\alpha$ because the stronger drag force exerted on the two stars
during WD evolution causes them to merge.

Fig.\,\ref{fig:RLOFPeriodS} shows that most companions, particularly
at shorter orbital periods, are low-mass deeply convective stars.
More massive secondaries are generally found at longer periods for
three reasons. First, longer orbital periods require high-mass white
dwarfs to initiate stable mass transfer over the age of the Galactic
Disc and these massive WDs can have secondaries with
masses all the way up to $1.44$\,\msun, provided $q\le 1$. Second,
during CEE for a fixed primary initial mass and orbital period, systems
with more massive secondaries have more orbital energy and so a
smaller portion of this energy is necessary to eject the envelope.
This leads to longer orbital periods. Third, for
a fixed WD mass, more massive secondaries fill their Roche
lobes at longer orbital periods and so systems with more massive
companions are more likely to evolve into PREPS.

\subsubsection{Stellar pair distribution}\label{stellarpairs}

Table~\,\ref{tab:Contribs} lists fractions of the various combinations
of types of WD primaries and secondary types just before RLOF
commences. At low $\alpha$ the predominant combination is a CO\,WD
primary with a low-mass M-dwarf secondary.  Second in abundance are
systems comprised of a He\,WD with a low-mass M-dwarf secondary.
Other combinations are also found but in much smaller numbers.  At
high $\alpha$ the two major categories are swapped and those systems
with He\,WD primaries become the predominant type. The observed
fraction of He\,WDs ($f_{\rm He}$) is generally low among classical
CVs ($f_{\rm He}\lsimeq 10$\,per cent) and pre-CVs
($f_{\rm He}\lsimeq 17\pm8$\,per\,cent as shown by
\citet{Zorotovic2011}. The results in Table\,\ref{tab:Contribs}
indicate that in order to reproduce the observed low fraction of
He\,WDs our models need to be restricted to $\alpha<0.3$.

\clearpage
\begin{landscape}
\begin{table*}
  \caption{The fraction of the
    combinations of types of WD primaries and
    secondary types just before RLOF commences for various $\alpha$.
    The stellar type CS is a deeply or fully convective low-mass MS star with $M<0.7$\,M$_\odot$.}
  \centering
 \label{tab:Contribs}
\begin{tabular}{c c c c c c c }
\hline
\noalign{\smallskip}
 \multicolumn{7}{c}{MCV progenitor pairs, fraction per cent} \\
$\alpha$  &  He\,WD/CS & CO\,WD/CS & ONe\,WD/CS & He\,WD/MS & CO\,WD/MS & ONe\,WD/MS \\
\noalign{\smallskip}
\hline
\noalign{\smallskip}
0.10  &14.86  &69.63  & 5.72   &0.00  &3.77  &6.03 \\
0.20  &30.27  &52.27  &12.99  &0.00  &0.38  &4.10 \\
0.30  &61.36  &25.69  &12.49  &0.00  &0.00  &0.46 \\ 
0.40  &96.44  & 7.78  & 5.78   &0.00  &0.00  &0.00 \\ 
0.50  &95.85  & 1.72  & 2.44   &0.00  &0.00  &0.00 \\ 
0.60  &98.75  & 0.28  & 0.98   &0.00  &0.00  &0.00 \\ 
0.70  &99.67  & 0.01  & 0.32   &0.00  &0.00  &0.00 \\ 
0.80  &99.92  & 0.00  & 0.00   &0.00  &0.00  &0.00 \\ 
\hline
\end{tabular}
\end{table*}
\end{landscape}

\vspace{-6pt}
\subsubsection{Mass distribution}\label{mass_distribution}
\begin{figure*}
\begin{center}
\includegraphics[width=1.02\textwidth]{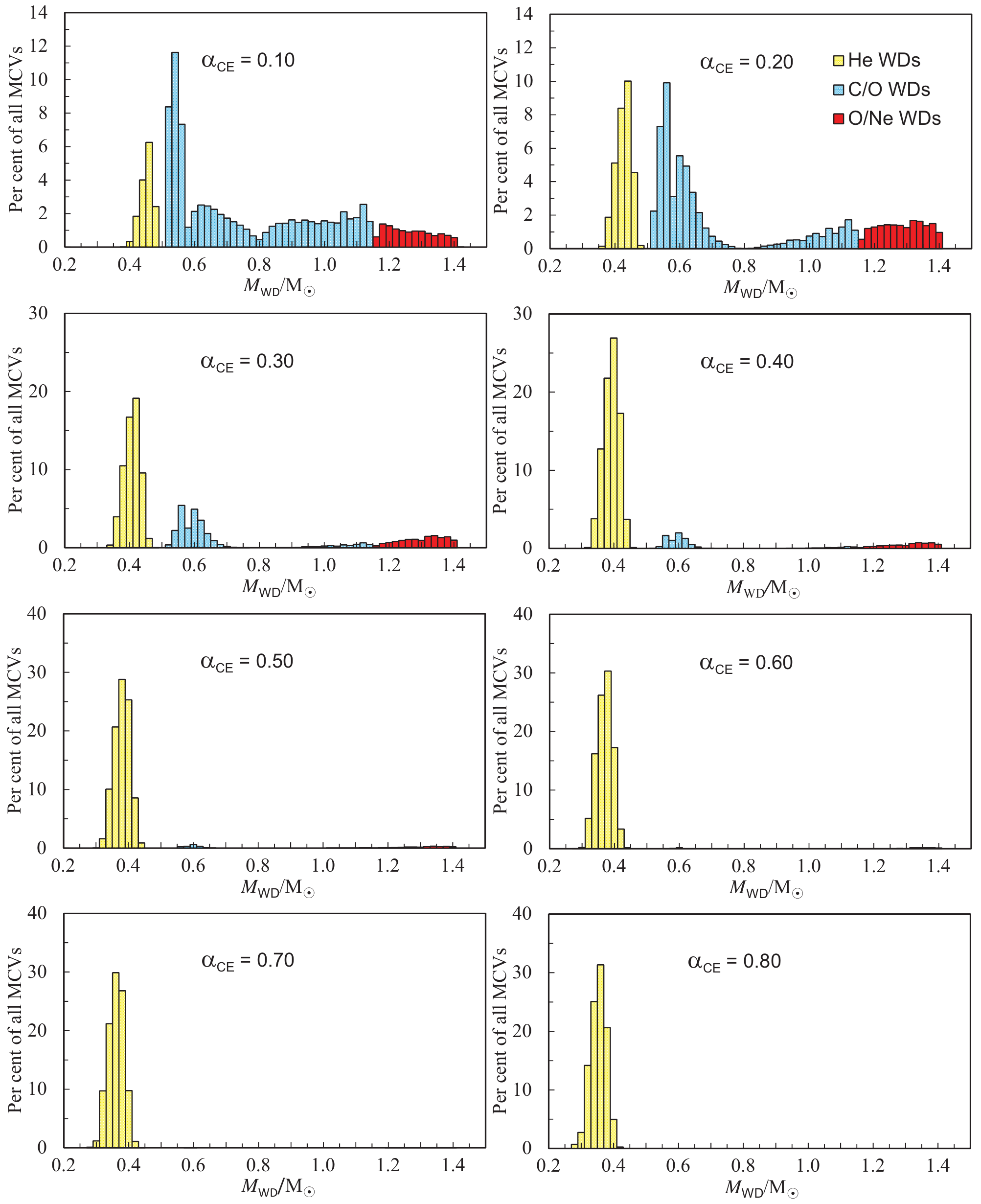}
\caption{Theoretical mass distribution of the WD primary star
  of magnetic systems just before they start RLOF for various
  $\alpha$. The distributions of the three WD types are shown
  as three superimposed coloured categories.}
\label{fig:RLOFMassP}
\end{center}
\end{figure*}

\vspace{-12pt}
Fig.\,{\ref{fig:RLOFMassP}} shows that all our models predict that, on
average, longer orbital period systems contain CO\,WDs while
shorter-period systems tend to have He\,WD primaries. At low $\alpha$
the primaries are predominantly CO\,WDs with masses in the range 0.5
to 1.1\,M$_\odot$ followed in lesser numbers by He\,WDs with masses in
the range 0.4 to 0.5\,M$_\odot$ while ONe\,WDs, with masses in the
range 1.2 to 1.4\,M$_\odot$, are rarer with their incidence reaching a
maximum near $\alpha=0.2$.

We note that there is a curious dip in the WD mass
distribution near $M_{\rm WD}=0.8$\,M$_\odot$ which widens as
$\alpha$ increases until all CO and ONe\,WDs disappear for
$\alpha> 0.5$.  This is because as $\alpha$ increases,
systems emerge from CE at progressively longer periods,
because large $\alpha$ means a high envelope clearance efficiency
which leads to larger stellar separation at the end of the common
envelope stage. However the longer the orbital period, the higher the
WD mass needs to be for stable mass transfer to
commence. Thus the gap in the WD mass distribution is caused
by those systems that emerge from CE at large separations
but with WD primaries that are not massive enough to allow
RLOF to take place. Another, albeit much narrower, gap occurs near
0.5\,M$_\odot$ for all $\alpha$ but becomes wider for
$\alpha\ge0.2$. This gap also persists until all CO and ONe\,WDs
disappear at $\alpha > 0.5$. It divides systems with He\,WDs primaries
from those with CO\,WDs and is linked to whether the stars enter
CEE on the RGB, and so produce a He\,WD primary
with $M_{\rm WD}\lsimeq 0.5$\,M$_\odot$, or on the AGB, and so produce
a CO\,WD primary with $M_{\rm WD}> 0.5$\,M$_\odot$.

\begin{figure*}
\begin{center}
\includegraphics[width=1.03\textwidth]{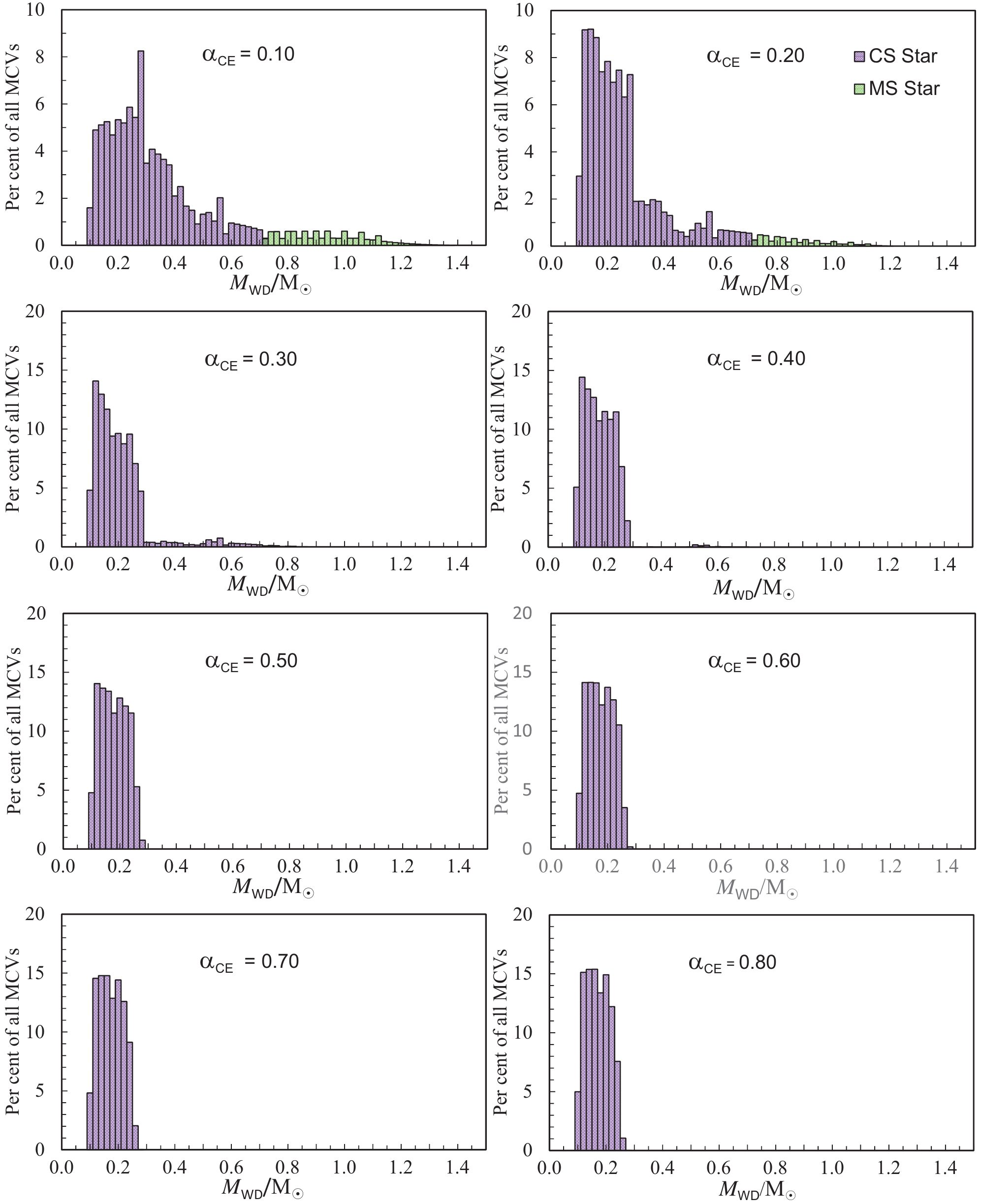}
\caption{As in Fig\,\ref{fig:RLOFMassP} but for the secondary star
  types shown as the coloured categories. Both secondary star types
  are MS stars.  The CS type is a deeply or fully
  convective MS star with $M<0.7$\,M$_\odot$.}
\label{fig:RLOFMassS}
\end{center}
\end{figure*}

Once again Fig.\,\ref{fig:RLOFMassS} shows that the secondaries are
predominantly low-mass deeply or fully convective M-dwarf stars. The
distribution has a broad peak around 0.1 to 0.3\,M$_\odot$ at
$\alpha=0.1$ to 0.2 with a long tail extending to $1.2$\,M$_\odot$. As
$\alpha$ increases, the peak in the secondary mass distribution shifts
to slightly lower masses (around 0.1 to 0.25\,M$_\odot$) but the
high-mass tail shrinks quite dramatically. At $\alpha\ge0.4$ the
distribution is confined to secondary masses of less than about
$0.3$\,M$_\odot$. As already noted in section \ref{PerDist}, the
majority of these very low-mass donor stars belong to systems that
underwent CEE during the Hertzsprung gap or RGB
phases and thus have He\,WD primaries with
$M_{\rm WD}\lsimeq 0.5$\,M$_\odot$.  We also note that systems with
low-mass secondaries ($M_{\rm sec}\lsimeq 0.35$\,\msun) remain detached
for longer because magnetic braking is inefficient in these stars and
gravitational radiation is the main source of loss of angular
momentum.

\vspace{-8pt}
\subsubsection{Magnetic field distribution}\label{MagFieldDistr}
\vspace{-12pt}
Fig. \ref{fig:RLOFOvTBmag} shows the theoretical magnetic field
distribution and the breakdown of the primary WD types for our
range of $\alpha$.  The maximum field strength is a few $10^8$\,G
and is found mostly in systems whose primary is a He\,WD. The reason
for this is that systems that undergo CEE during
the RGB evolution have shorter initial orbital periods and create very
short period binaries with a highly magnetic WD, as expected
from equation \ref{EqBfield}.

The magnetic field distribution is dominated by systems with CO\,WD
primaries when $\alpha\le 0.2$. When $\alpha\ge 0.4$ the field
distribution becomes narrower and its peak shifts to higher field
strengths. For $\alpha\ge0.6$ the field distribution only contains very
highly magnetic He\,WD primaries with a peak near $3.2\times 10^8$\,G
at $\alpha=0.8$. This shift to high fields is because those systems
that go through CEE while their primaries are on
the RGB merge for low $\alpha$ but can survive for high $\alpha$
giving rise to very short orbital period systems with strongly
magnetic, low-mass WDs.

We note that the magnetic field distribution has a dip near
$8\times10^6$\,G appearing at $\alpha\ge0.2$ and persisting until
all CO and ONe\,WDs disappear from the distribution. This is
reminiscent of the dip we encountered in the WD mass
distribution (see \ref{mass_distribution}) and has the same
explanation. The similar behaviour is because the magnetic field
strength is a function of WD mass (by virtue of equations
\ref{EqBfield}). The field dip is thus
caused by the dearth of systems with WD masses around
$0.8\,\msun$ (see Fig.\,\ref{fig:RLOFMassP}).

\section{Comparison to observations}\label{Comparison}

The optimal observational sample with which to compare our results
would be that formed by the known magnetic PREPs. However, this sample
is exceedingly small and observationally biased. To make things worse,
not all PREPS have well determined parameters, such as masses and
magnetic field strengths. Hence, for some of these studies we use the
observed sample of MCVs, noting the following important points (i) the
MCV sample is magnitude-limited, (ii) MCVs suffer from prolonged high
and low states of accretion and (iii) MCVs include systems at all
phases of evolution. Some of them began Roche lobe overflow billions
of years ago while others have only recently begun mass
exchange. Therefore, one should take such a comparison with some
degree of caution particularly when it comes to compare quantities
that change over time, such as orbital periods and masses. When
comparing masses we will also use the observed sample of non-magnetic
Pre-CVs \citep{Zorotovic2011}.

\begin{figure*}
\begin{center}
\includegraphics[width=1.0\textwidth]{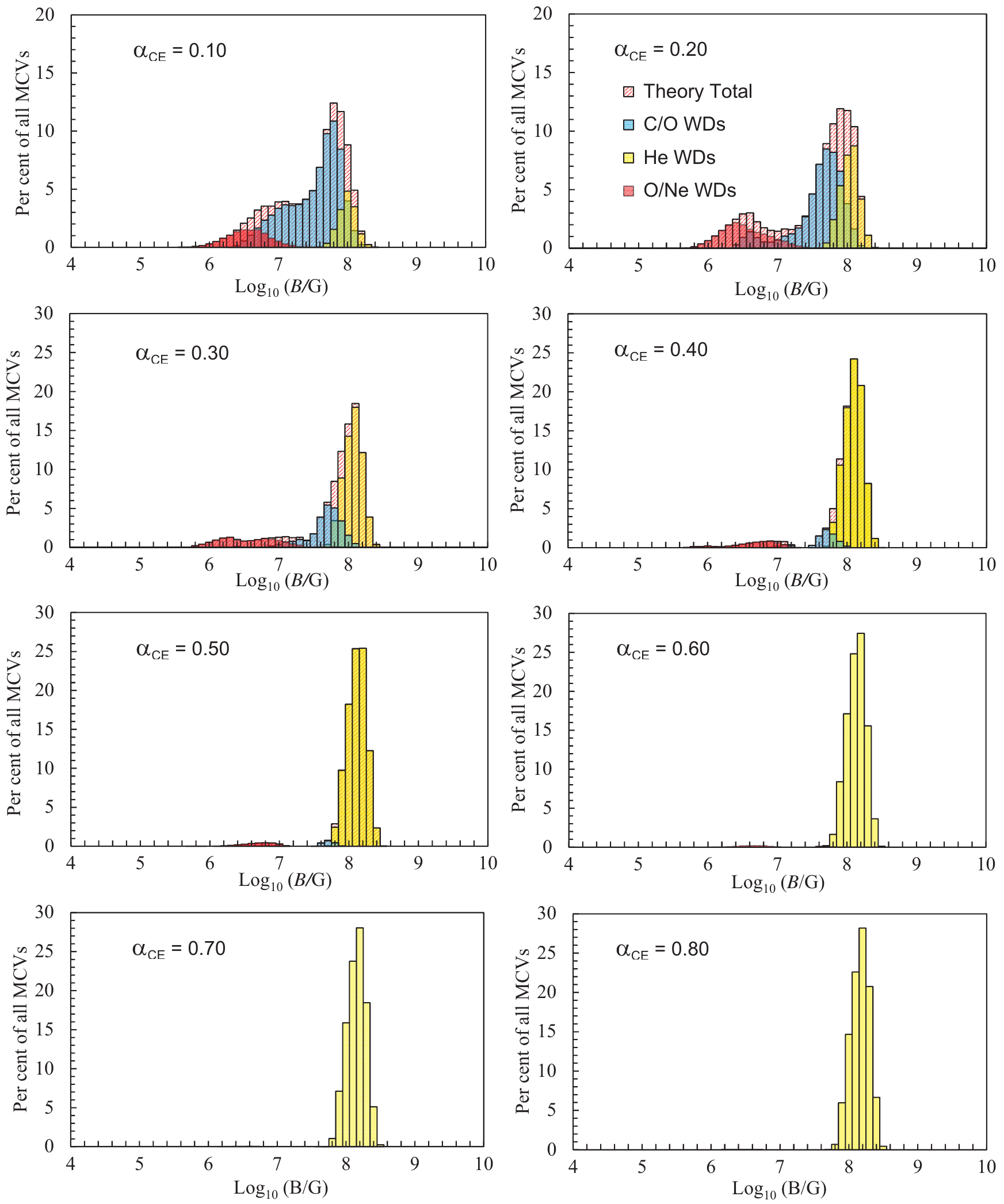}
\caption{Pink shaded histogram: Total theoretical magnetic field
  distribution of the WD primary stars in magnetic systems just before they
  start RLOF for the indicated $\alpha$. The histograms of the three
  types of WDs making up the total theoretical magnetic field distribution are
  shown as the foreground coloured histograms.  These three are made partially
  transparent so that details of the other histograms can be seen through them.}
\label{fig:RLOFOvTBmag}
\end{center}
\end{figure*}

The tables of \citet{fer2015a} show that the observed orbital periods
of MCVs are in the range 1 to 10\,hr, masses are between about $0.4$
and $1.1$\,$\msun$ and that the magnetic field distribution is relatively
broad with a peak near $3.2\times10^7$\,G.  A quick glance at
Figs\,\ref{fig:RLOFPeriodP}, \ref{fig:RLOFMassP} and
\ref{fig:RLOFOvTBmag} immediately reveals that models with
$\alpha>0.3$ are all unable to reproduce the general
characteristics expected from the progenitors of the observed
population of MCVs and we elaborate on this in more detail
below. Generally we see that generally models with $\alpha>0.3$ are
not realistic and evolutionary effects cannot account for the large
degree of discrepancy between theory and observations.

We begn our analysis with the magnetic field distribution. There is
no evidence for field decay among MCVs \citep{fer2015a,Zhang2009}
so we can assume that the magnetic field strength remains unchanged
over the entire life of the magnetic binary.

\begin{figure}
\begin{center}
\includegraphics[width=1.00\columnwidth]{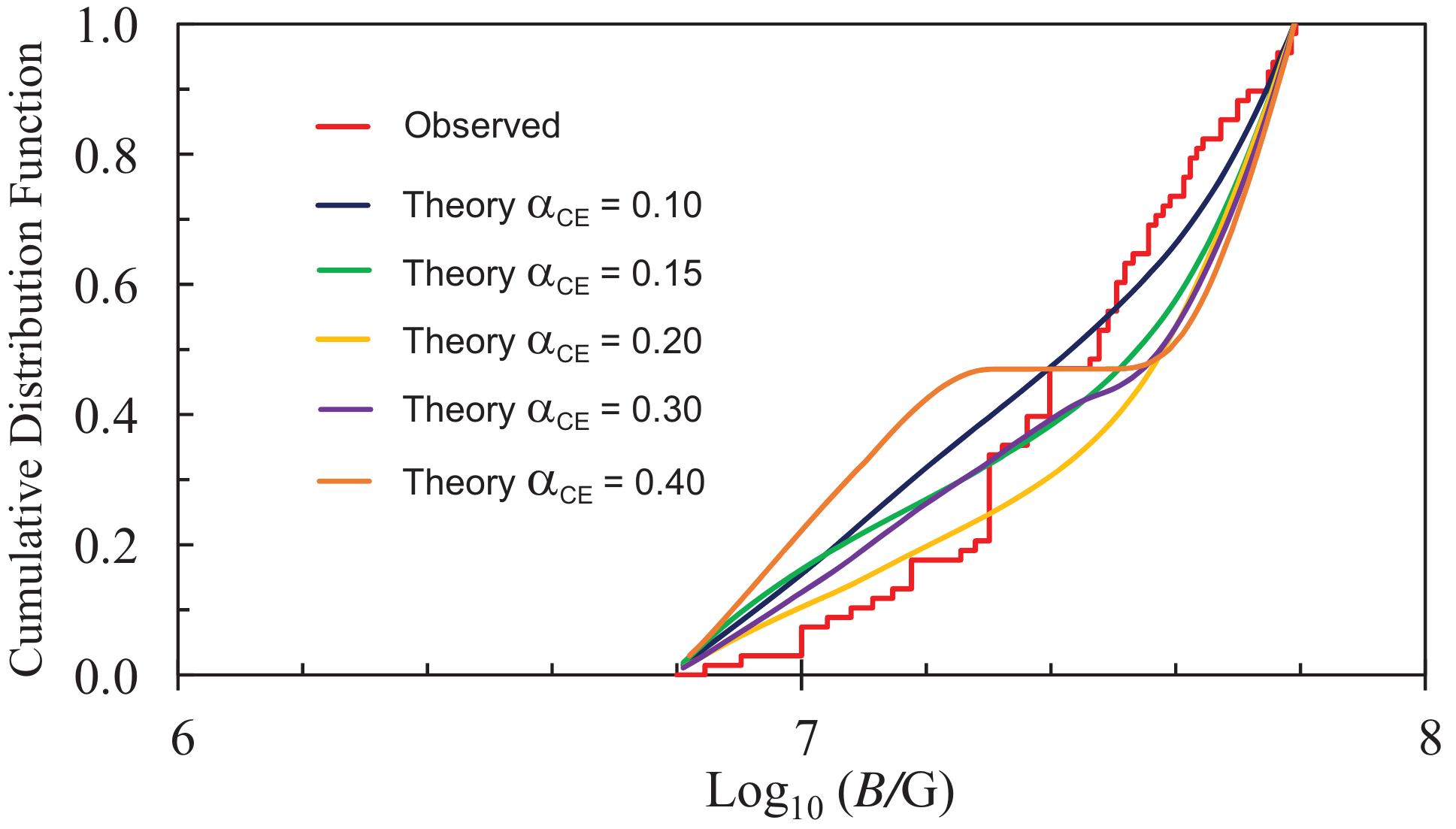}
\caption{Theoretical cumulative distribution functions for the
  magnetic fields of MCV WDs at RLOF for 
  $\alpha=0.10, 0.15, 0.20, 0.30$ and $0.40$ and the CDF of the
observed magnetic field of 81 systems taken from \citet{fer2015a} }
\label{fig:MCVWDBMagCDF}
\end{center}
\end{figure}

We have used a K--S test \citep{Press1992} to compare the magnetic
field distribution of the observed population with the theoretical results.
This test establishes the likelihood that two samples are drawn from the
same population by comparison of the cumulative distribution functions
(CDFs) of the two data samples. The CDFs of the two distributions vary
between zero and one and the test is on the maximum of the absolute
difference $D$ between the two CDFs. It gives the probability $P$ that
a random selection would produce a larger $D$. Five model CDFs for five
different $\alpha$'s and the CDF for the known observed magnetic fields
of 81 MCV systems are compared in Fig.\,\ref{fig:MCVWDBMagCDF}.

\begin{table}

\caption{Kolmogorov--Smirnov $D$ statistic and probability $P$ of the magnetic field
distributions of the  observed and synthetic populations of MCVs for a range of $\alpha$.
\vspace{2mm}}
\centering
\begin{tabular} { c c c}
\hline
\noalign{\smallskip}
$\alpha$ & $D$  & $P$ \\
\noalign{\smallskip}
\hline
\noalign{\smallskip}
0.10 & 0.17476 & 0.36069 \\
0.15 & 0.19349 & 0.24632 \\
0.20 & 0.25141 & 0.05845 \\
0.30 & 0.22962 & 0.10500 \\
0.40 & 0.26939 & 0.04298 \\
0.50 & 0.35186 & 0.00429 \\
0.60 & 0.38035 & 0.00006 \\
0.70 & 0.61987 & 0.00000 \\ 
0.80 & 0.94366 & 0.00000 \\
\noalign{\smallskip}
\hline
\end{tabular}
\label{tab:KS_Bfield}
\end{table}

The observed samples of MCVs and magnetic PREPs are very biased,
particularly at the low and high ends of the magnetic field
distribution. At low fields ($B\lsimeq 10$\,MG) the observed radiation
is dominated by the truncated accretion disc. In these low-field
systems the photosphere of the WD is never visible and Zeeman
splitting cannot be used to determine field strengths. Nor can
cyclotron lines be used to measure fields because they are too weak
and invisible in the observed spectra. In the high field regime
($B\gsimeq 100$\,MG) mass accretion from the companion star is
inhibited \citep{fer1989, LiWuWick1994} and so high field MCVs
are very dim wind accretors often below the detection limits of most
surveys \citep[AR\,UMa,][]{Hoard2004}. Because of these biases the
observed samples in these regimes are far from complete and
theoretical fits are unreliable.  We therefore restrict our comparison
between theory and observations to field strengths in the range 10 to 70\,MG.

\begin{figure}
\begin{center}
\includegraphics[width=1.0\columnwidth]{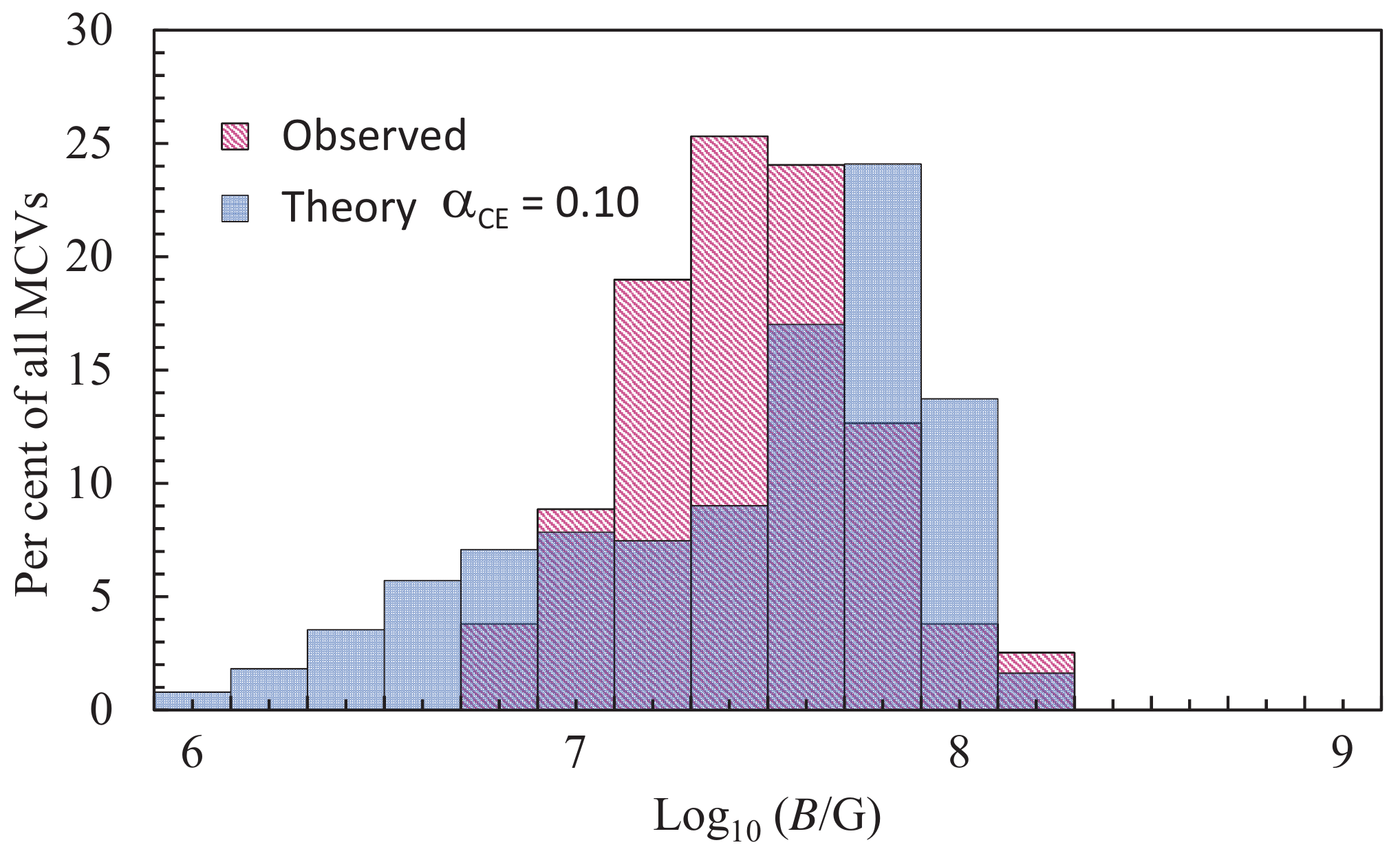}
\caption{Comparison of the theoretical magnetic field strength for
  $\alpha= 0.1$ and the observed magnetic field strength of the 81
  MCVs taken from \citet{fer2015a}}
\label{fig:CompareBmag}
\end{center}
\end{figure}

The results of the K--S test for our range of $\alpha$ are displayed in
Table \ref{tab:KS_Bfield} and show that although no $\alpha$ can be
ruled out the field distribution is a better match to the observations
at low $\alpha$. The comparison of the magnetic field
distribution between theory and observations is shown in
Fig.\,\ref{fig:CompareBmag} for $\alpha=0.1$.

We stress that for $\alpha>0.3$ all the theoretical magnetic
  field distributions shown in Fig.\,\ref{fig:RLOFOvTBmag} are very
  unrealistic because only very high field ($B>60$\,MG) He\,WDs
  ($M\lsimeq0.5$\,\Msun) are predicted to exist.  This is contrary to
  observations that show that fields cover a much wider range of
  strengths ($\mbox{a few}10^6$ to a few $\,10^8$\,G) and WD
  masses (0.4 to 1.2\,\Msun) as seen in Tables\,2 and 3 of
  \citet{fer2015a}.

We have performed a K--S study between the synthetic WD mass
distribution and that of WD masses in MCVs taken from
\citet{fer2015a}. In principle such a comparison can be
justified if we make the usual assumption that the mass of the white
dwarf does not grow in CVs because nova eruptions tend to expel all
material that is accreted. However, we found that the
K--S test applied to the WD masses of the theoretical and
observed population of MCVs yields poor results, as shown in the
second and third columns of Table \ref{tab:KS_WDmass}.  However, such
a conflict is not surprising because our assumption that the mass of
the WD does not grow because of nova eruptions may not be
correct.

In this context, I note that \citet{Zorotovic2011} noticed a curious
discrepancy in their observational data of CVs and Pre-CVs. That is,
they found that the mean WD mass in CVs
($0.83\pm 0.23$\,M$_\odot$) significantly exceeds that of pre-CVs
($0.67\pm0.21$\,\,M$_\odot$) and they excluded that this difference
could be caused by selection effects.  The two possible solutions
advanced by \citet{Zorotovic2011} were that either the mass of the
WD increases during CV evolution, or a short phase of thermal
time-scale mass transfer comes before the formation of CVs during
which the WD acquires a substantial amount of mass via stable
hydrogen burning on the surface of the WD \citep[as first
suggested by][]{Schenker2002}. During this phase the
system may appear as a super-soft X-ray source
\citep[][]{Kahabka1997}. Using this assumption \citet{Wijnen2015}
could build a large number of massive WDs. However their
model still created too many low-mass He\,WDs and too many evolved
companion stars contrary to observations. Another possibility has
recently been advanced by \citet{Zorotovic2017}.  In order to achieve
a better agreement between their binary population synthesis models
and observations of CVs they adopted the ad-hoc mechanism proposed by
\citet{Schreiber2016} which surmises the existence of additional
angular momentum losses generated by mass transfer during the CV
phase. Such losses are assumed to increase with decreasing WD
mass and would cause CVs with low-mass WDs to merge and
create an isolated WD. By removing these merged systems from
the synthetic CV sample the average WD mass
increases. Furthermore such a mechanism would explain the existence of
isolated low-mass WDs ($M_{\rm WD}<0.5\,\msun$) that
constitute around 10\,per cent of all single WDs observed in
the solar neighbourhood \citep[e.g.][]{Kepler2007}.

\begin{figure}
\begin{center}
\includegraphics[width=1.0\columnwidth]{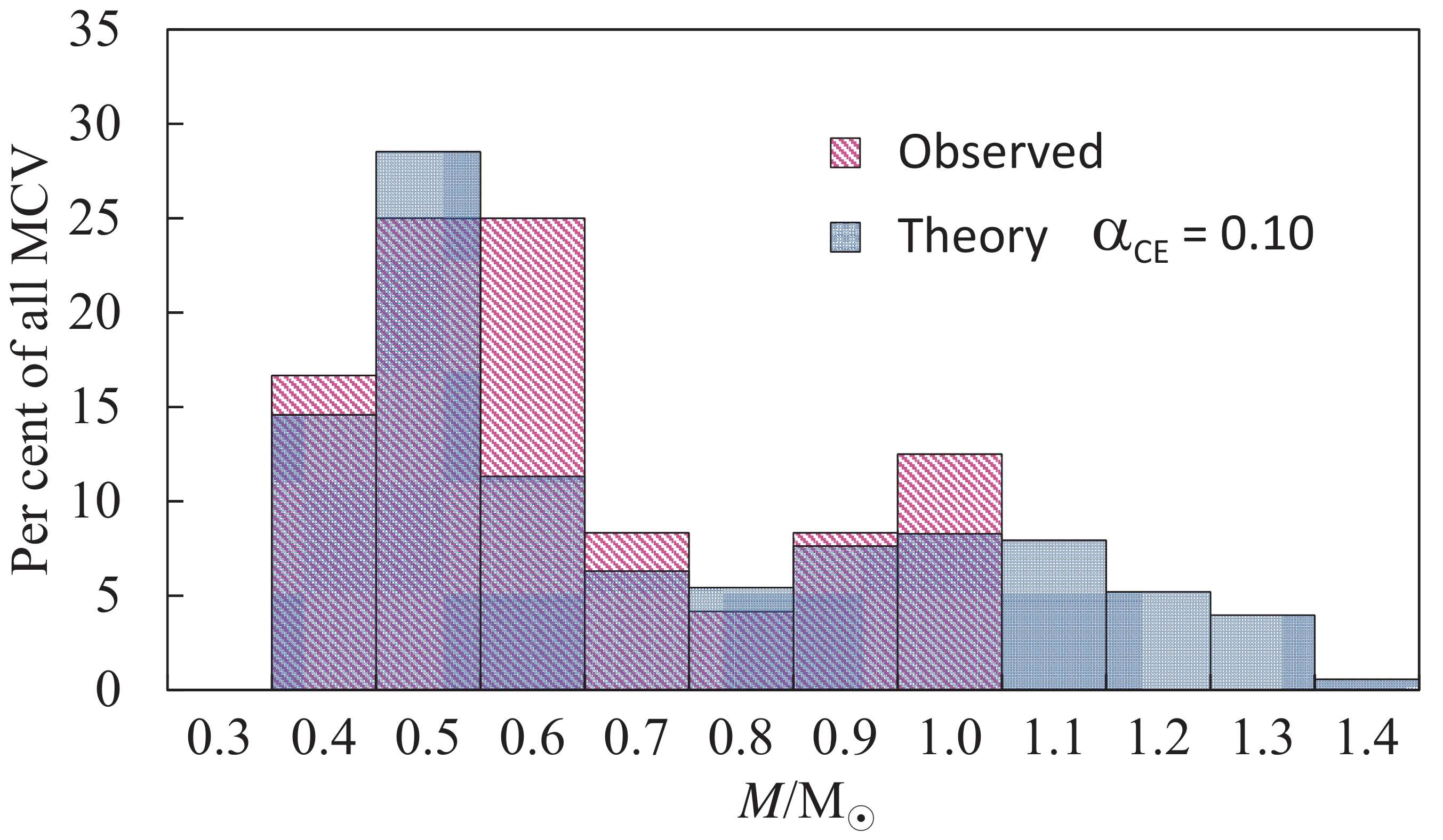}
\caption{Comparison of the mass distributions for the observed pre-CV white
dwarf masses taken from \citet{Zorotovic2011} and the theoretical mass
distribution of the WDs as the systems start RLOF for
$\alpha= 0.10$.\vspace{-5mm}}
\label{fig:CompareMasses}
\end{center}
\end{figure}

Going back to our studies, if a comparison between WD masses
in MCVs and our synthetic population may not be meaningful, the next
best sample to use for our K--S test is the observed WD
masses of pre-CVs \citep{Zorotovic2011}.  The results are reported in
the fourth and fifth columns of Table \ref{tab:KS_WDmass} and show
that the agreement between observations and theory is greatly
improved.  The comparison of the synthetic and observed Pre-CV\,WD
mass distribution is shown in Fig.~\ref{fig:CompareMasses} for the
largest K--S probability when $\alpha = 0.10$.

\begin{figure}
\begin{center}
\includegraphics[width=1.0\columnwidth]{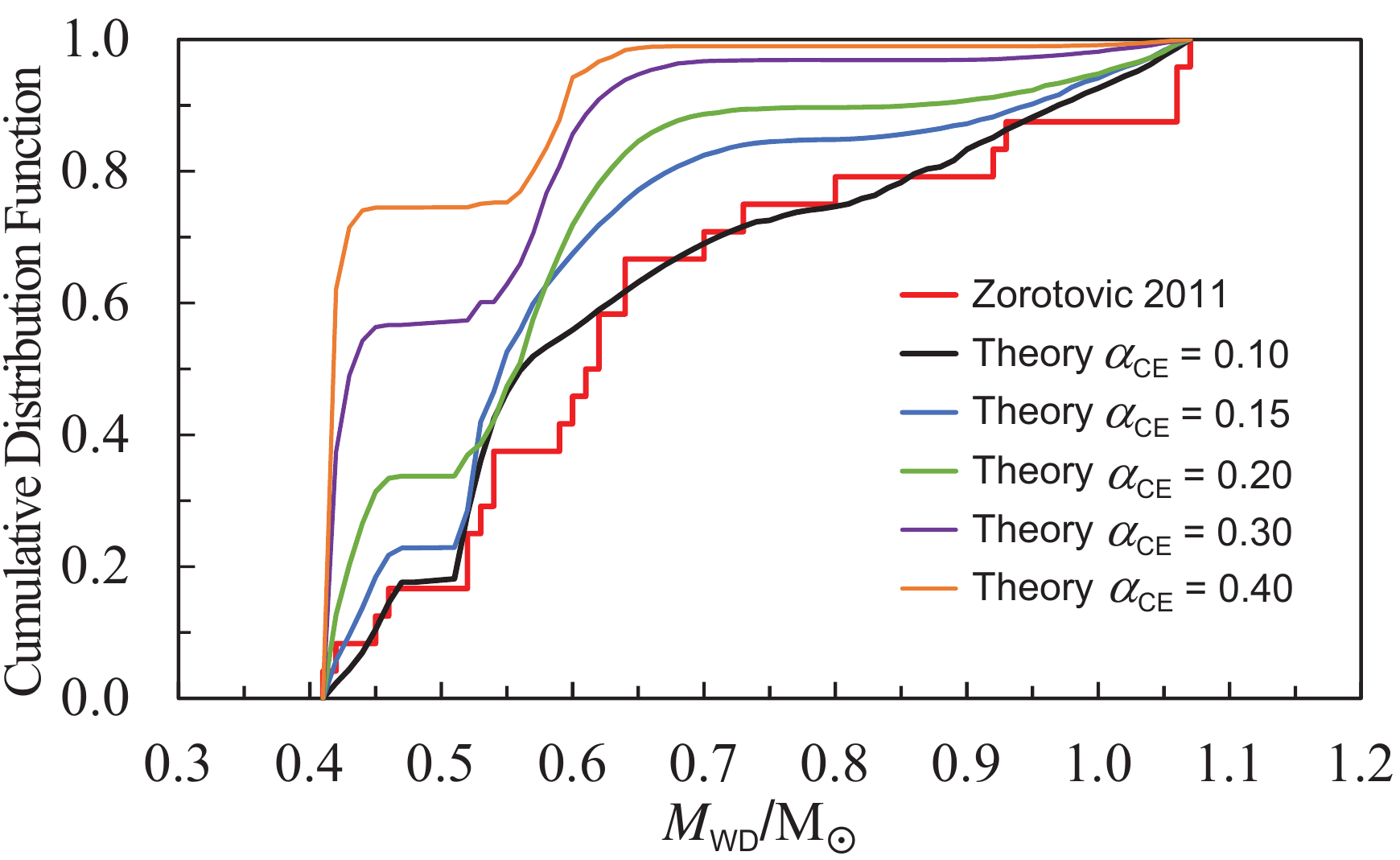}
\caption{Cumulative Distribution Functions of the mass distributions
  for the observed pre-CV WD masses taken from
  \citet{Zorotovic2011} and the theoretical distribution of the
  WDs as the systems start RLOF for $\alpha=0.10,
  0.15, 0.20, 0.30$ and $0.40$.  The K-S statistics for this plot are
  shown in the fourth and fifth columns of table~\ref{tab:KS_WDmass}}
\label{fig:MCV-WD-Mass-CDF}
\end{center}
\end{figure}

\begin{table}
  \caption{K--S $D$ statistic and probability $P$ of the
    WD mass distributions of the observed MCVs listed by
    Ferrario et al. (2015a, second and third columns) and
    our synthetic populations for $\alpha$ given in the first
    column. We show the K--S results of the observed Pre-CV
    masses of \citet{Zorotovic2011} and our synthetic
    populations at the start of  RLOF (fourth and fifth columns).
    \vspace{3mm}}
  \centering
\begin{tabular} {c c c c c}
\hline
\noalign{\smallskip}
$\alpha$&$D$&$P$&$D$&$P$\\
\noalign{\smallskip}
\hline
\noalign{\smallskip}
0.10 & 0.37687  & 0.02023088 & 0.12954  & 0.95281557 \\
0.15 & 0.49861  & 0.00064407 & 0.23478  & 0.34844783 \\
0.20 & 0.56677  & 0.00006150 & 0.26010  & 0.23507547 \\
0.30 & 0.62615  & 0.00000622 & 0.48014  & 0.18713800 \\ 
0.40 & 0.69590  & 0.00000032 & 0.66148  & 0.00106500 \\ 
\noalign{\smallskip}
\hline
\end{tabular}
\label{tab:KS_WDmass}
\end{table}

We note that he Pre-CV observational sample shows a dearth of systems in the WD
mass distribution centred around $0.8\,\msun$. This mass gap was already
noted in the theoretical {\sc{bse}} models and the reasons for its existence
were explained in section \ref{mass_distribution}. The smaller size of this gap
for models with $\alpha\le0.2$ explains whywe achieve a better fit with
observations for $\alpha=0.1$, as indicated by the K--S test.

We stress that if $\alpha>0.3$ the theoretical WD mass distribution
shown in Fig.\,\ref{fig:RLOFMassP} is very unrealistic because only
He\,WDs ($M<0.5$\,\Msun) are predicted to exist by these models.
This is contrary to observations that show that masses cover the
much wider range 0.4 to 1.2\Msun~\citep[see Tables\,2 and 3 in][]{fer2015a}.

Next, we look at the secondary mass distribution, keeping in mind that
a comparison between our synthetic {\sc{bse}} mass sample and the
observed secondary masses in MCVs is definitely not appropriate
because secondary masses decrease over time as mass is transferred to
the WD. Nonetheless it may still be pertinent to use the
observed Pre-CV sample to study and compare the overall
characteristics of these samples so that we can, at the very least,
discard some of the most extreme theoretical models.

Fig. \ref{fig:RLOFMassS} shows that if $\alpha>0.3$ then
$M_{\rm sec}< 0.3\,\msun$, which is inconsistent with observations of
pre-CVs \citep[see][]{Zorotovic2011}. Furthermore,we can see that
when $\alpha>0.2$, the decline towards higher masses becomes far too
steep. This straightforward comparison seems again to indicate that
models with $\alpha>0.3$ are very unrealistic and therefore low $\alpha$
is a better fit.

\section{Discussion and Conclusions}\label{DandC}
\vspace{-6pt}
The origin of large-scale magnetic fields in stars is still a puzzling
question \citep[see][]{fer2015b}. However, the results from
recent surveys such as the SDSS \citep{Kepler2013}, BinaMIcS
\citep{ale2015} and MiMes \citep{wad2016} have provided us with a much
enlarged sample of magnetic stars that have allowed investigators to
conduct studies like this one. There are two main competing scenarios
to explain the existence of magnetic fields in WDs. In 1981,
\citeauthor{ang1981} first proposed that the magnetic Ap and Bp stars
are the most likely progenitors of the highly MWDs under the
assumption of magnetic flux conservation \citep[see
also][]{tou2004,wickramasinghe2005}. According to this scenario the
weak fields recently observed in a number of WDs
\citep[e.g.][]{Landstreet2016} could be either dynamo generated in
late stages of stellar evolution or these WDs may be the
descendants of MS stars with fields below the current
detection limit. However, not much is known about the weak magnetic
field regime. For instance, it is still not known whether all white
dwarfs are magnetic at some level because the weakest measured fields
are just about at the limit of detectability.

The best clue so far on the origin of fields in WDs (isolated
and in binaries) has come from the study of their binary properties
\citep{Liebert05,Liebert2015}, as outlined in section
\ref{WhereProg}. This is why the proposal by \citet{tout2008}, that
the origin of magnetic fields in WDs is related to their duplicity and
stellar interaction during CEE, is becoming more and more appealing.

We have extended our population synthesis study of binary systems
carried out in chapters 2 and 3 for the HFMWDs to explain
the origin of fields in the accreting WDs in MCVs. Similarly to
the investigations conducted in chapters 2 and 3, we varied the CE
efficiency parameter $\alpha$ to investigate its effects on
the resulting synthetic population of MCVs.  We have shown that
models with $\alpha\ge 0.4$ are not able to reproduce the large
range of WD masses, field strengths, and secondary types and
masses that are observed in MCVs and therefore models with
$\alpha< 0.4$ best represent the observed data.  K--S tests
conducted to compare our synthetic WD mass and magnetic field
distributions with the observed populations have given us some
quantitative support in favour of models with $\alpha<0.4$.

However, we need to stress again some of the shortcomings
  of our work and in particular those that arise from our comparison
  to observations. Many of the parameters (e.g., WD mass,
  magnetic field, secondary star mass and type, orbital period) that
  characterise the Galactic populations of MCVs and PREPs and are
  needed for comparison studies are often hard to ascertain owing to
  evolutionary effects and observational biases that are difficult to
  disentangle. For instance, we mentioned in section \ref{Comparison}
  magnetic WDs in PREPs would be the best objects with
  which to compare my theoretical results and in particular the mass
  distribution, because mass is not then contaminated by accretion
  processes. On the other hand there are far too few members of this
  population. The WD mass distribution provided by the much larger
  sample of MCVs cannot be used either for comparison purposes
  because masses vary over time, owing to accretion and nova explosions.
  So instead we have used the sample provided by the non-magnetic Pre-CVs of
  \citet{Zorotovic2011}.
  
  The situation is somewhat ameliorated when we consider the magnetic
  field distribution because fields are not expected to change over
  time \citep[see][]{fer2015a}. However, the true magnetic field
  distribution of MCVs is not well known because it is plagued by
  observational biases. For example, at field strengths below a
  few\,$10^7$\,G most systems (the intermediate polars) have an
  accretion disc from which continuum emission and broad emission lines
  swamp the Zeeman and cyclotron features arising from the WD
  surface \citep{fer1992} and so hide those spectral signatures
  that are crucial to determine their field strengths. Very high field
  polars are also likely to be under-represented in the observational
  set because mass accretion from the companion star is impeded by
  the presence of strong fields \citep{fer1989, LiWuWick1994}
  making these systems very dim wind accretors.

  Despite the limitations highlighted above, we have shown that the
  characteristics of the MCVs are generally consistent with those of a
  population of binaries that is born already in contact (exchanging
  mass) or close to contact, as first proposed by \citet{tout2008}.
  This finding is also in general agreement with the hypothesis first
  advanced by \citet{sch2009} that the binaries known as PREPs,
  where a MWD accretes matter from the wind of a low-mass
  companion, are the progenitors of the MCVs.
  
\subsubsection{Acknowledgements}
GPB gratefully acknowledges receipt of an Australian Postgraduate Award. CAT
thanks the Australian National University for supporting a visit as a Research Visitor
of its Mathematical Sciences Institute, Monash University for support as a Kevin
Watford distinguished visitor and Churchill College for his fellowship.

\blanknonumber


\chapter{A Double Degenerate White Dwarf System}
\label{Chapter 5}
This chapter is a reproduction of the paper published in
Monthly Notices of the Royal Astronomical Society, viz:
\vspace{2mm}
\\
{\color{blue}\textit{Kawka, Briggs, Vennes, Ferrario,  Paunzen \& Wickramasinghe, MNRAS (2017), 466(1): 1127--1139. A fast spinning magnetic white dwarf in the double-degenerate, super--Chandrasekhar system NLTT 12758\footnote{Based on observations made with ESO
telescopes at the La Silla Paranal Observatory under programmes
083.D-0540, 084.D-0862, 089.D-0864 and 090.D-0473. Based in part
on data collected with the Danish 1.54-m telescope at the ESO La
Silla Observatory.}}}
\vspace{-6pt}

\label{firstpage}

\vspace{-12pt}
\section{Abstract}
\label{Paper4abstract}
\vspace{-12pt}
We present an analysis of the close double degenerate NLTT\,12758,
which is comprised of a magnetic white dwarf with a field of about
3.1 MG and an apparently non-magnetic white dwarf. We measured an orbital period
of 1.154 days and found that the magnetic white dwarf is spinning
around its axis with a period of 23 minutes. An analysis of the
atmospheric parameters has revealed that the cooling ages of the two
white dwarfs are comparable, suggesting that they formed within a
short period of time from each other. Our modelling indicates that
the non-magnetic white dwarf is more massive ($M=0.83\, M_\odot$) than
its magnetic companion ($M=0.69\,M_\odot$) and that the total
mass of the system is higher than the Chandrasekhar mass. Although the 
stars will not come into contact over a Hubble time, when they
do come into contact, dynamically unstable mass
transfer will take place leading to either an accretion induced
collapse into a rapidly spinning neutron star or a Type Ia
supernova.

\section{Introduction}
\label{sec:Intro4}
\vspace{-5mm}
The majority of stars will evolve into a white dwarf and a significant
fraction of white dwarfs harbours a magnetic field that ranges from a
few kG to about 1000 MG \citep{Liebert2003, Kawka2007}. Spectroscopic and
spectropolarimetric surveys
\citep[e.g.,][]{sch1995,sch2001a,azn2004,Kawka2007,kaw2012,landstreet2012,Kepler2013}
of white dwarfs have been able to place constraints on the incidence
of magnetism among white dwarfs. The incidence of magnetic white
dwarfs in the local neighbourhood has been estimated by
\citet{Kawka2007} to be around 20\,\%. The local sample, as well as
various surveys, have shown that the incidence of magnetism as a
function of field strength is constant, although \citet{landstreet2012}
suggested a possible field resurgence at the extremely low-field ($<1$
kG) end of the distribution. A higher incidence of magnetism is also
observed in cool polluted white dwarfs. \citet{kaw2014} found an
incidence of $\approx 40$\,\% in cool ($T_{\rm eff} < 6000$ K)
DAZ\footnote{DAZ type white dwarfs show photospheric hydrogen (DA) and metal lines.}
white dwarfs. A higher incidence of magnetism was also observed among
cool DZ\footnote{DZ type white dwarfs show metal lines only.} white dwarfs
\citep{hol2015}. A recent review on the properties of magnetic white
dwarfs can be found in \citet{fer2015a}.

The origin of large scale magnetic fields in stars is still one of the
main unanswered questions in astrophysics, although recent data,
particularly from surveys such as the Sloan Digital Sky Survey
\citep[SDSS, ][]{York2000}, the Magnetism in Massive Stars
\citep[MiMes, ][]{wad2016} and the Binarity and Magnetic Interactions
in various classes of stars \citep[BinaMIcS, ][]{ale2015} may have
finally thrown some light into this matter \citep{fer2015b}. Magnetism
in white dwarfs has been explained with two main evolutionary
scenarios. For a long time the leading theory was that the progenitors
of magnetic white dwarfs are magnetic Ap and Bp stars
\citep{ang1981}. Under the assumption of magnetic flux conservation,
the magnetic field strengths observed in Ap stars would correspond to
magnetic fields in white dwarfs in excess of 10 MG
\citep{kaw2004,tou2004,wickramasinghe2005}. The progenitors of white dwarfs
with weaker fields may be other main-sequence stars whose magnetic
fields are below our current detection limits or could be
dynamo-generated in later stages of stellar evolution.

More recently, proposals that strong magnetic fields are created in
evolving interacting binaries via a dynamo mechanism during a 
common envelope (CE) phase \citep{tout2008,potter2010,Nordhaus2011,garcia2012,wickramasinghe2014}
have gained momentum as a possible origin for strong magnetic fields
in white dwarfs. The main reason for this proposal is that all
magnetic white dwarfs appear to be either single or in interacting
binaries (the magnetic cataclysmic variables). That is,
magnetic white dwarfs are never found paired with a non-interacting,
non-degenerate star, which is at odds with the fact that approximately
30\% of all non-magnetic white dwarfs are found in non-interacting
binaries with a non-degenerate companion (usually an M-dwarf)
\citep{liebert2005,Ferrario2012,lie2015}. This result is hard to explain and
leaves the magnetic cataclysmic variables without obvious
progenitors. Because of this observational peculiarity, the existence
of magnetic fields in white dwarfs has been linked to fields generated
during CE binary interactions or mergers. The merger scenario during
the CE also successfully explains the higher than average mass of
isolated magnetic white dwarfs \citep{bri2015}. The complex magnetic
field structure usually observed in rotating high field magnetic white dwarfs
would also be in support of a merging hypothesis.

However, a few common-proper motion (CPM) magnetic plus non-magnetic
double degenerate systems are now known
\citep{Ferrario1997b,Girven2010,Dobbie2012,Dobbie2013}. In some of these cases, the
more massive magnetic white dwarf is hotter and hence younger than its
non-magnetic companion, which seems to imply that the more massive
star evolved later.  This apparent paradox can be resolved by
postulating that systems of this kind were initially triple
systems and that the magnetic white dwarf resulted from the merger of
two of the three stars \citep[e.g., EUVE\,J0317$-$855, ][]{Ferrario1997b}.

The study of the magnetic field structure in white dwarfs may also
give us important clues on how they formed. Normally, a simple dipole
is assumed for the field structure, but the study of rotating magnetic
white dwarfs have all shown variability, hence revealing much more
complex structures. One of the most extreme examples of a rotating
magnetic white dwarf is the hot ($T \approx 34\,000$ K) and massive
($M\approx 1.35\ M_\odot$) EUVE\,J0317-855, which has a rotation
period of 12 minutes \citep{Barstow1995,Ferrario1997b}. The rotation of the
white dwarf reveals a two component magnetic field structure: A high
field magnetic spot ($B \ge 425$ MG) with an underlying lower field
\citep{Vennes2003}. Another example of a rotating white dwarf with a
complex magnetic field structure is WD\,1953-011
\citep{Maxted2000,val2008}.  In this case, the rotation is slower
\citep[$P_{\rm rot}\,=\,1.448$ days][]{Brinkworth2005} and the magnetic field
strength is much weaker (180 kG - 520 kG) than that of
EUVE\,J0317-855.

NLTT\,12758 was discovered to be a magnetic white dwarf by
\citet{kaw2012}.  They showed that the circular polarization spectra
are variable and that there is also variability in the H$\alpha$ core
suggesting that NLTT\,12758 is a close double degenerate system. Here,
we present our analysis of spectroscopic, spectropolarimetric and
photometric data of NLTT\,12758. The observations are presented in
Section~\ref{sec:Observe}.  The orbital and rotation period analyses are described in
Sections~\ref{subsec:BinParm} and \ref{subsec:Rotation}, respectively. The stellar and atmospheric
parameters are presented in \ref{subsec:SandAParms}, and we discuss the
evolutionary scenarios in subsection~\ref{subsec:Evolution}. We discuss the case of
NLTT\,12758 in comparison to other known double degenerate systems
containing a magnetic white dwarf in section~\ref{sec:Discuss} and
we conclude in section~\ref{sec:Conclude}.

\vspace{-4mm}
\section{Observations}
\label{sec:Observe}
\vspace{-4mm}
\subsection{Spectroscopy and Spectropolarimetry}
\label{subsec:SandS}
\begin{center}
\begin{figure}
\includegraphics[width=14cm]{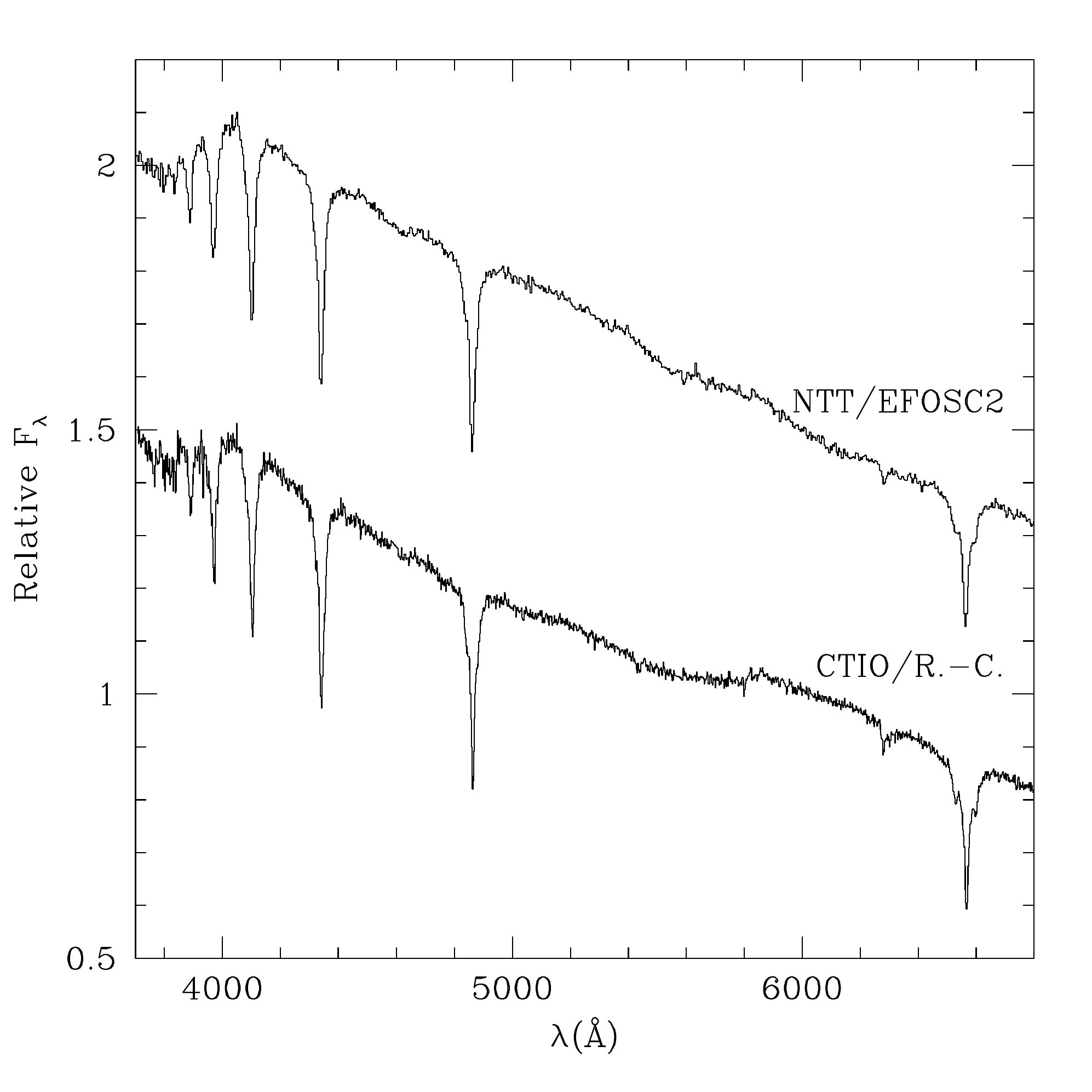}
\vspace{15mm}
\captionsetup{labelfont=bf,justification=justified}
\caption{Low dispersion CTIO/R.-C. and NTT/EFOSC2 spectra of NLTT\,12758 revealing
Zeeman splitted Balmer lines.}
\label{fig:efosc2}
\end{figure}
\end{center}

\vspace{-2mm}
NLTT\,12758 was first observed with the R.-C. spectrograph attached to
the 4m telescope at Cerro Tololo Inter-American Observatory (CTIO) on
UT 2008 February 24. We used the KPGL2 (316 lines per mm) grating with
the slit-width set to 1.5 arcsec providing a resolution of about 8
\AA. We obtained a second set of low-dispersion spectra with the
EFOSC2 spectrograph attached to the New Technology Telescope (NTT) at
La Silla. Two consecutive spectra were obtained on UT 2009 08 27. We
used grism number 11 and set the slit-width to 1.0 arcsec providing a
resolution of about 14 \AA. Both sets of spectra revealed Zeeman
splitting in the Balmer lines. Figure~\ref{fig:efosc2} shows the low
dispersion spectra.

We obtained a first set of spectropolarimetric observations using the
FOcal Reducer and low dispersion Spectrograph (FORS2) attached to the 8m
telescope (UT1) of the European Southern Observatory (ESO) in 2009. We
obtained another set of observations using the same set-up in 2013. We used
the 1200 lines mm$^{-1}$ grism (1200R+93) centred on H$\alpha$ providing a
spectral dispersion of 0.73 \AA\ pixel$^{-1}$. We set the slit-width to
1 arcsec providing a spectral resolution of 3.0 \AA. Each spectropolarimetric
observation consisted of two individual exposures, the first having the
Wollaston prism rotated to $-45^\circ$ immediately followed by the second
exposure with the prism rotated to $+45^\circ$.

We also obtained five spectra of NLTT\,12758 with the EFOSC2 spectrograph
in September 2012. These spectra were obtained with grism number 20 which
provides a spectral dispersion of 1.09 \AA\ per binned pixel. The slit-width
was set to 0.7 arcsec providing a resolution of 3.0 \AA.

Finally we obtained a set of five consecutive spectra of NLTT\,12758 with the
X-shooter spectrograph \citep{ver2011} attached to the VLT at Paranal 
Observatory on UT 2014 August 26. The spectra were obtained with the slit width
set to 0.5, 0.9 and 0.6 arcsec for the UVB, VIS and NIR arms, respectively. 
This setup provided a resolution of $R=9000$, 7450 and 7780 for the UVB,
VIS and NIR arms, respectively.

The log of the spectroscopic observations is presented in Table~\ref{tab:Log} below.

\begin{longtable}{p{2.5cm}p{1.5cm}r l l}

\captionsetup{labelfont=bf, justification=justified}
\caption{Spectroscopic observation log.}
\label{tab:Log}\\
\toprule
UT date & UT start & Exposure && Instrument\\
            &             & time (s)   && \& Telescope\\
\midrule
\endfirsthead

\captionsetup{labelfont=bf, margin={0mm,22mm}, justification=raggedright, singlelinecheck=off}
\caption{Spectroscopic observation log - continued}
\\
\toprule
UT date & UT start & Exposure && Instrument\\
            &             &  time (s)   && \& Telescope\\
\midrule
\endhead

\midrule
\begin{minipage}{100mm}
\footnotesize
\footnotemark{$^1$ Exposure times for the VIS/UVB arms, respectively.}\\
{\bf Table~\ref{tab:Log}:} continues on the next page.
\normalsize
\end{minipage}
\endfoot
\bottomrule

\begin{minipage}{100mm}
\footnotesize
\footnotemark{$^1$ Exposure times for the VIS/UVB arms, respectively.}
\normalsize
\end{minipage}
\endlastfoot

24 Feb 2008 & 02:04:42 & 1200 & & RC/CTI04m \\
24 Feb 2008 & 02:26:56 & 1200 && RC/CTI04m \\
27 Aug 2009 & 09:39:21 & 600 && EFOSC2/NTT \\
27 Aug 2009 & 09:49:57 & 600 && EFOSC2/NTT \\
23 Oct 2009 & 06:35:09 & 900 &&FORS2/UT1 \\
23 Oct 2009 & 06:51:18 & 900 && FORS2/UT1 \\
23 Oct 2009 & 07:16:20 & 900 && FORS2/UT1 \\
23 Oct 2009 & 07:32:28 & 900 && FORS2/UT1 \\
23 Oct 2009 & 07:55:18 & 900 && FORS2/UT1 \\
23 Oct 2009 & 08:11:35 & 900 && FORS2/UT1 \\
24 Nov 2009 & 02:40:00 & 900 && FORS2/UT1 \\
24 Nov 2009 & 02:56:08 & 900 && FORS2/UT1 \\
24 Nov 2009 & 03:24:14 & 900 && FORS2/UT1 \\
24 Nov 2009 & 03:40:22 & 900 && FORS2/UT1 \\
02 Sep 2012 & 08:41:00 & 900 && EFOSC2/NTT \\
02 Sep 2012 & 09:04:14 & 900 && EFOSC2/NTT \\
03 Sep 2012 & 08:09:10 & 900 && EFOSC2/NTT \\
03 Sep 2012 & 08:33:13 & 900 && EFOSC2/NTT \\
03 Sep 2012 & 09:18:43 & 900 && EFOSC2/NTT \\
04 Jan 2013 & 03:49:14 & 700 && FORS2/UT1 \\
04 Jan 2013 & 04:02:02 & 700 && FORS2/UT1 \\
04 Jan 2013 & 04:15:06 & 700 && FORS2/UT1 \\
04 Jan 2013 & 04:27:55 & 700 && FORS2/UT1 \\
04 Jan 2013 & 04:41:01 & 700 && FORS2/UT1 \\
04 Jan 2013 & 04:53:49 & 700 && FORS2/UT1 \\
07 Jan 2013 & 02:39:34 & 700 && FORS2/UT1 \\
07 Jan 2013 & 02:52:22 & 700 && FORS2/UT1 \\
07 Jan 2013 & 03:05:26 & 700 && FORS2/UT1 \\
07 Jan 2013 & 03:18:14 & 700 && FORS2/UT1 \\
07 Jan 2013 & 03:31:18 & 700 && FORS2/UT1 \\
07 Jan 2013 & 03:44:06 & 700 && FORS2/UT1 \\
07 Jan 2013 & 03:57:28 & 700 && FORS2/UT1 \\
07 Jan 2013 & 04:10:17 & 700 && FORS2/UT1 \\
07 Jan 2013 & 04:33:02 & 700 && FORS2/UT1 \\
07 Jan 2013 & 04:46:02 & 700 && FORS2/UT1 \\
03 Feb 2013 & 03:01:49 & 700 && FORS2/UT1 \\
03 Feb 2013 & 03:14:38 & 700 && FORS2/UT1 \\
03 Feb 2013 & 03:27:35 & 700 && FORS2/UT1 \\
03 Feb 2013 & 03:40:24 & 700 && FORS2/UT1 \\
03 Feb 2013 & 03:53:22 & 700 && FORS2/UT1 \\
03 Feb 2013 & 04:06:10 & 700 && FORS2/UT1 \\
26 Aug 2014 & 08:18:30 & 450/540$^1$ && Xshooter/UT3 \\
26 Aug 2014 & 08:28:43 & 450/540$^1$ && Xshooter/UT3 \\
26 Aug 2014 & 08:37:51 & 450/540$^1$ && Xshooter/UT3 \\
26 Aug 2014 & 08:47:00 & 450/540$^1$ && Xshooter/UT3 \\
26 Aug 2014 & 08:56:08 & 450/540$^1$ && Xshooter/UT3 \\
\end{longtable}
\vspace{-6mm}

\subsection{Photometry}
\label{subsec:Photometry}

We collected available photometric measurements from the Galaxy 
Evolutionary Explorer ($GALEX$) sky survey, optical photometry from 
\citet{egg1968} and the AAVSO Photometric All-Sky Survey,{ }Deep Near Infrared 
Survey (DENIS) of the southern sky, 
the Two Micron All Sky Survey (2MASS) and the Wide-field Infrared Survey
Explorer ($WISE$). These measurements are listed in Table~\ref{tab:Phot}.

\clearpage
\begin{table}[htbp]
\captionsetup{labelfont=bf, margin={1mm,1mm}, justification=justified}
\caption{Photometric measurements of NLTT\,12758{\vspace{0pt}}}
\label{tab:Phot}
\centering
\begin{tabular}{lcc}
\toprule
Band & Magnitude & Reference\\
\midrule
$GALEX$ $FUV$    & not detected                       & 1 \\
$GALEX$ $NUV$   & $17.401\pm0.016$               & 1 \\
$V$                     & 15.46, $15.483\pm0.071$     & 2,3 \\
$B-V$                  & $+0.31$                             & 2 \\
$U-B$                  & $-0.71$                              & 2 \\
$B$                     & $15.855\pm0.094$               & 3 \\
$g$                     & $15.607\pm0.037$               & 3 \\
$r$                      & $15.417\pm0.074$               & 3 \\
$i$                       & $15.443\pm0.132$              & 3 \\
DENIS$I$             & $14.976\pm0.07$                & 4 \\
DENIS$J$            & $14.713\pm0.15$                 & 4 \\
2MASS$J$            & $14.809\pm0.032$               & 5 \\
2MASS$H$           & $14.723\pm0.071$               & 5 \\
2MASS$K$            & $14.683\pm0.096$              & 5 \\
$WISE$ $W1$       & $14.703\pm0.034$              & 6 \\
$WISE$ $W2$       & $14.781\pm0.069$              & 6 \\
\bottomrule
\end{tabular}
\begin{minipage}{90mm}
\vspace{2mm}
\footnotesize
{\bf References:} (1) \citet{mor2007}; (2) \citet{egg1968}; (3) \citet{hen2016}; 
(4) \citet{fou2000}; (5) \citet{skr2006}; (6) \citet{cut2012}
\end{minipage}
\end{table}

We obtained new CCD photometric measurements with the 1.54-m Danish telescope
at the La Silla Observatory in Chile on UT 26th December 2014, 30th January
2015 and 11th March 2016. On 26th December 2014, we alternated between the 
$V$ and $R$ filter and on 30th January 2015 11th March 2016 we observed 
NLTT\,12758 with the $R$ filter only.  The integration time was set to 40 seconds
for all observations. The data reduction and differential photometry were
performed using the C-Munipack package\footnote{http://c-munipack.sourceforge.net/}.
Since several comparison stars were available, and these were checked individually to
exclude variable objects.  We compared the results of the final differential light curves
using the aperture photometry routine from IRAF \citep{Stet87}. We found no
differences above the photon noise.

\clearpage
\begin{figure*}
\includegraphics[width=14cm]{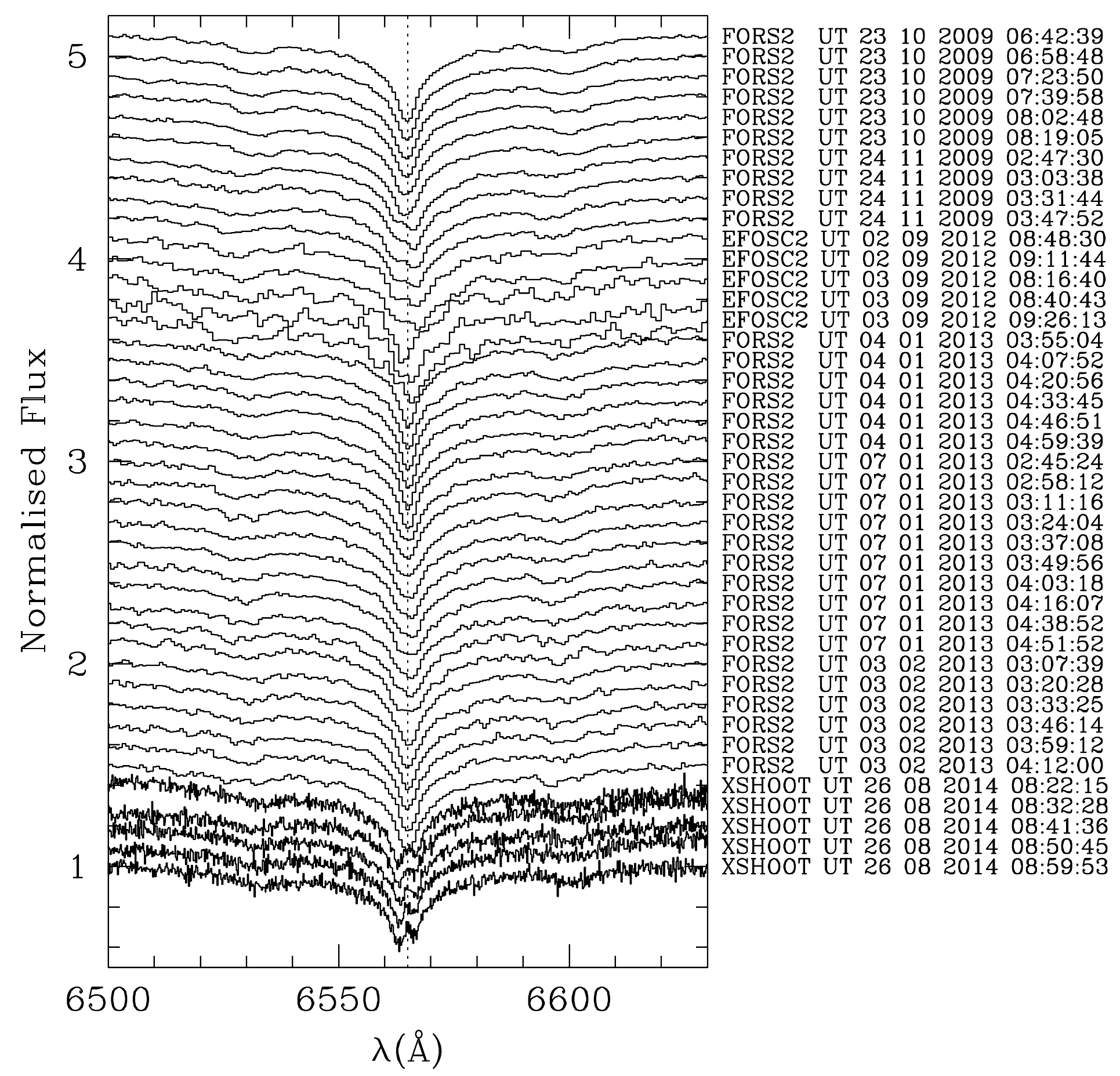}
\vspace{10mm}
\captionsetup{labelfont=bf, margin={10mm,10mm}, justification=justified}
\caption{\normalfont EFOSC2, FORS2 and X-shooter spectra of NLTT\,12758 showing
variations  in the H$\alpha$ core. The mid-exposure UT time is listed for each spectrum.}
\label{fig:Rad}
\end{figure*}
\clearpage

\section{Analysis}
\label{sec:Analysis}

During the first spectropolarimetric observations of NLTT\,12758, we
found that the $\sigma$ components of H$\alpha$ varied with a reversal
in the polarisation spectra, thus revealing itself as a new member of
the DAP white dwarf class\footnote{DAP white dwarfs show hydrogen
lines with detectable polarisation. The DAH classification is reserved for
Zeeman splitted line spectra, but without confirmed  polarization.}.
We also found that the width of the core of the $\pi$ component is
structured and variable, thus suggesting the presence of a close
companion. The FORS2, EFOSC2 and X-shooter spectra displayed
in Figure~\ref{fig:Rad} clearly show the variations in the central
H$\alpha$ core. The resolution of the X-shooter spectra and timing
of the observations allowed us to discern the individual cores of the
two components.

\subsection{Binary parameters}
\label{subsec:BinParm}

We measured the radial velocity of the magnetic white dwarf by first
subtracting a template representing the DA white dwarf and then
cross-correlating the DAP white dwarf FORS/EFOSC2 spectra ($\sigma$
components only) with the X-shooter spectrum.  The DA radial velocity
could only be measured at quadrature, i.e, at maximum line core
separation, and with a sufficient signal-to-noise ratio. Only three
sets of spectra met these criteria.  Consecutive exposures (2 to 4)
were co-added to increase the signal-to-noise and improve the
reliability of the velocity measurements while minimizing orbital
smearing.  Table~\ref{tab:RadVel} lists the barycentric julian date
(BJD) with the measured radial velocities of the magnetic and
non-magnetic white dwarfs in NLTT\,12758. All velocities are
barycentric corrected.

\begin{table}
\centering
\captionsetup{labelfont=bf, margin={1mm,1mm}, justification=justified}
\caption{Radial velocity measurements{\vspace{0pt}}}
\label{tab:RadVel}
\begin{tabular}{c c c}
\toprule
BJD ($2450000+$) & $\varv_{DAP}$ (km~s$^{-1}$) & $\varv_{DA}$ (km~s$^{-1}$) \\
\midrule
5127.78952 &  $157\pm6$ & ... \\
5127.81811 &  $165\pm5$ & ... \\
5127.84523 &  $196\pm5$ & ... \\
5159.62676 &    $2\pm8$ & ... \\
5159.64212 &   ...      & $156\pm20$ \\
5159.65748 &   $18\pm8$ & ... \\
6172.87606 &   $18\pm7$ & $196\pm20$ \\
6173.86764 &   $35\pm8$ & ... \\
6296.67107 &  $112\pm6$ & ... \\
6296.68904 &  $120\pm7$ & ... \\
6296.70703 &  $124\pm6$ & ... \\
6299.62701 &   $47\pm7$ & ... \\
6299.65848 &   $38\pm8$ & ... \\
6299.69457 &   $29\pm9$ & ... \\
6326.64042 &  $100\pm6$ & ... \\
6326.66725 &   $79\pm6$ & ... \\
6895.86223 &  $193\pm5$ & $12\pm5$ \\
\bottomrule
\end{tabular}\\
\end{table}

We searched for a period in the measurements using $\chi^2$
minimization techniques by fitting the sinusoidal function
$v=\gamma + K\times \sin{(2\pi(t-T_0)/P)}$ to the measured radial
velocities where $t$ is time (BJD).  The initial epoch ($T_0$), period
($P$), mean velocity ($\gamma$) and velocity semi-amplitude ($K$) were
determined simultaneously and we normalized the $\chi^2$ function by
setting the minimum reduced $\chi^2$ to 1.

\begin{figure}
\vspace{10mm}
\includegraphics[width=140mm]{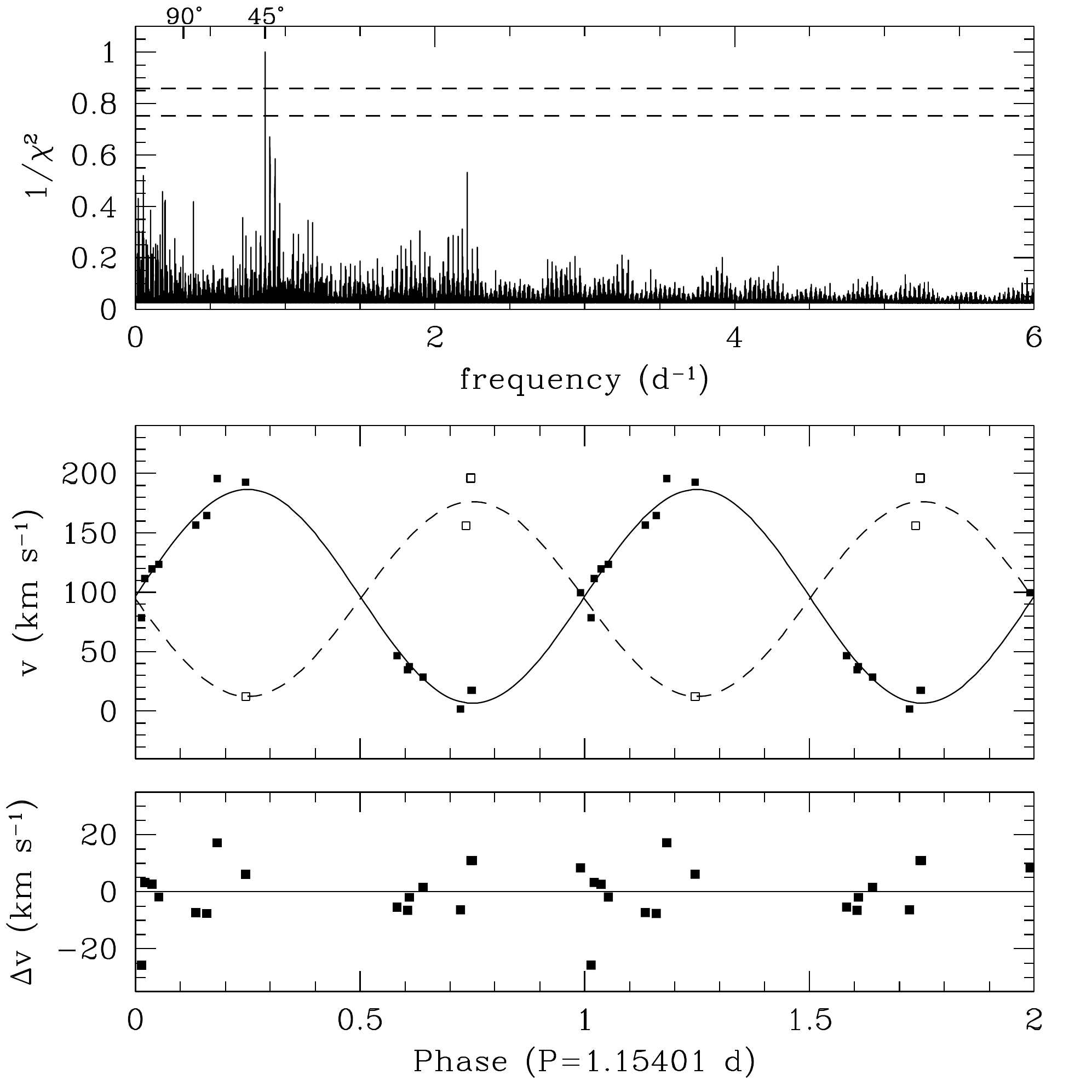}
\vspace{10mm}
\captionsetup{labelfont=bf, margin={5mm,5mm}, justification=justified}
\caption{\normalfont (Top panel) period analysis of the FORS2, EFOSC2 and
X-shooter data with 66 and 90\% confidence level (dashed
lines). (Middle panel) radial velocity measurements
(Table~\ref{tab:RadVel}) of the DA (open squares) and DAP stars
(full squares) phased on the orbital period and the best-fitting
sine curves (Table~\ref{tab:Param}) and (bottom panel) velocity
residuals for the DAP star. The longest period is marked at $90^\circ$ on
the top horizontal axis along with the actual period at $45^\circ$.}
\label{fig:Period}
\end{figure}

Figure~\ref{fig:Period} shows the period analysis of the FORS2, EFOSC2
and X-shooter data sets and Table~\ref{tab:Param} lists the new binary 
parameters. Using the FORS2 and EFOSC2 data combined
with the X-shooter data we determined a period of $1.15401\pm0.00005$
days and a velocity semi-amplitude for the DAP star of
$89.7\pm3.8$\,km\,s$^{-1}$ with an average residual of only
7.7\,km\,s$^{-1}$ and commensurate with measurement errors
(Table~\ref{tab:RadVel}). The corresponding mass function is
$f(M_{\rm DA}) = 0.0863\pm0.0110\ M_\odot$. Since the X-shooter
spectra were taken near quadrature and clearly show the cores of both
components, we were able to estimate a semi-amplitude of
$81.9\pm17.3$\,km\,s$^{-1}$ for the non-magnetic white dwarf. The orbital mass
ratio $M_{\rm DA}/M_{\rm DAP}=0.85-1.35$ is not sufficiently accurate
to constrain the evolution of the system, and additional constraints will
be provided by the spectroscopic analysis (Section~\ref{subsubsec:AtmParam}).

\subsection{Rotation}
\label{subsec:Rotation}

The spectropolarimetric data have revealed a modulation that we
attribute to the rotation of the magnetic white dwarf.

\begin{figure}
\includegraphics[width=1.0\columnwidth]{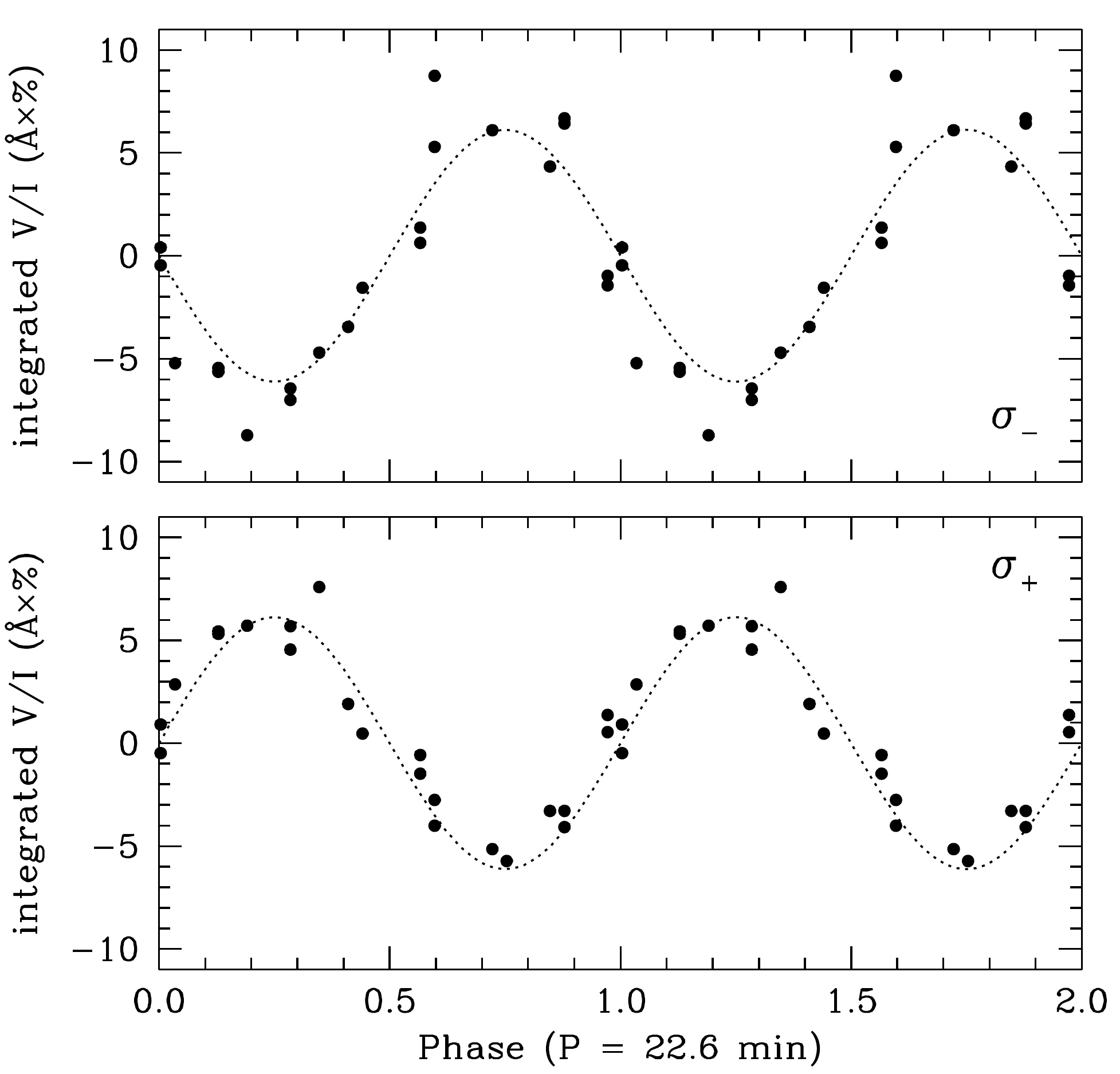}
\captionsetup{labelfont=bf, margin={0mm,12mm}, justification=justified}
\caption{Integrated polarization measurements of the two individual
$\sigma$ components phased on the rotation period of 22.6 minutes revealing
a complete reversal of the field vector. The top panel shows the measurements
for the blue-shifted $\sigma_{-}$ component and the bottom panel shows the
measurements of the red-shifted $\sigma_{+}$ component.}
\label{fig:RotPeriod}
\end{figure}

We measured the integrated polarization for both $\sigma$ components
and conducted a period search. Two significant periods at 22.6 minutes
and 9 minutes stand out. Since some of the exposure times were longer
than 9 minutes, it is unlikely that the 9 minutes period is real.
Figure~\ref{fig:RotPeriod} shows line polarization measurements obtained
by integrating $V/I$ over the wavelength range ($\approx \pm20$\AA)
covered by the individual $\sigma$ components phased on the 22.6
minute period. Both $\sigma$ components show sinusoidal behaviour
and a symmetry about the null polarization axis which imply that the
magnetic poles spend nearly equal time in the field-of-view.

\begin{figure}
\includegraphics[width=1.0\columnwidth]{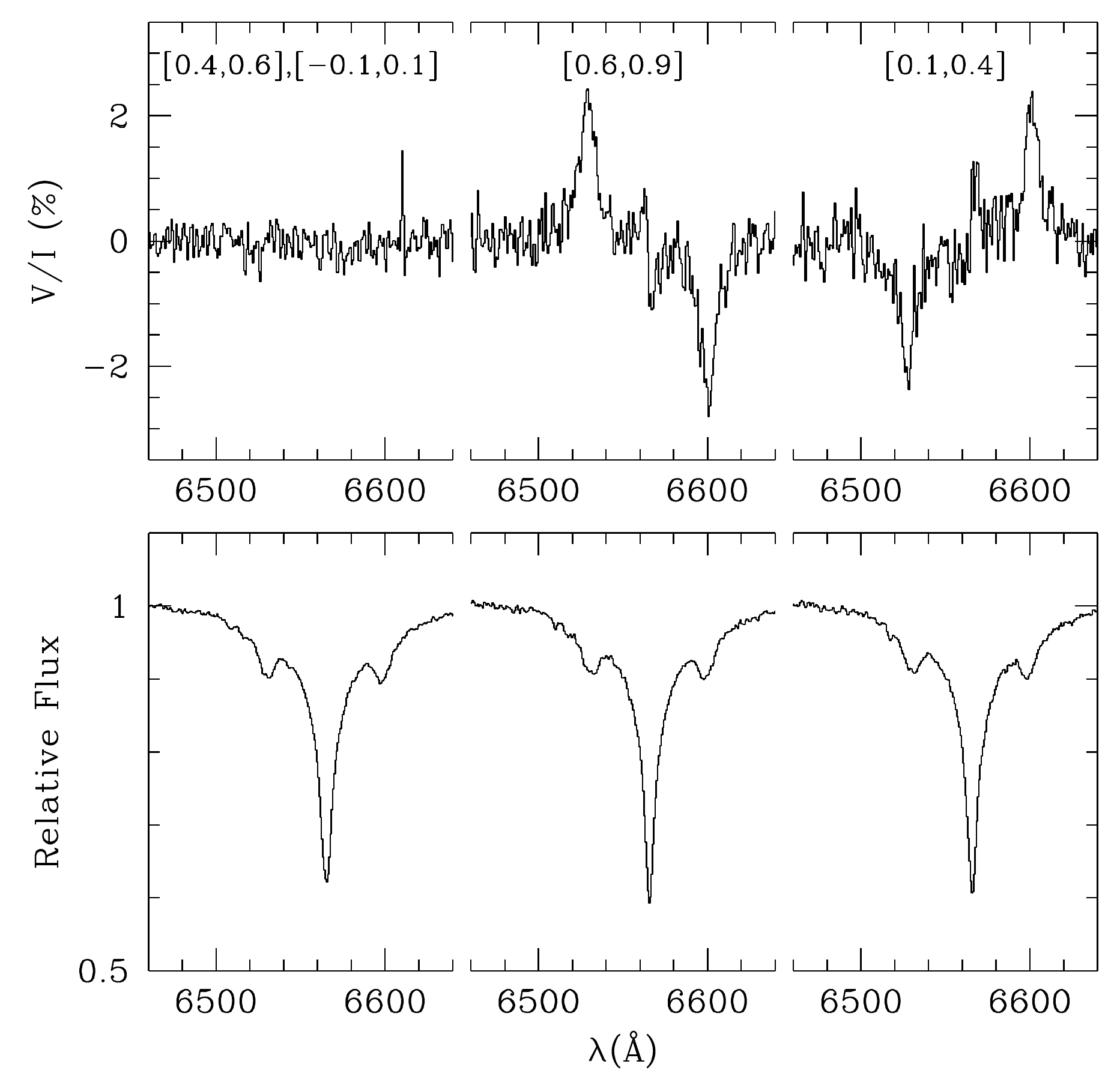}
\captionsetup{labelfont=bf, margin={12mm,0mm}, justification=justified}
\caption{Co-added FORS2 circular polarization spectra (top panel) and
  flux spectra (bottom panel) at three phase ranges showing the flip
  in the sign of the $\sigma$ components of H$\alpha$. The spectrum
  with zero polarization corresponds to a nearly orthogonal viewing
  angle to the magnetic axis.}
\label{fig:Forsx}
\end{figure}

Figure~\ref{fig:Forsx} shows the co-added FORS2 circular polarization
spectra over three separate ranges of a rotation cycle ($P=22.6$\,min)
highlighting the flipping of the sigma components.  The flip in the
sign of the H$\alpha$ $\sigma$ components at phases 0.1-0.4 and
0.6-0.9 and their anti-symmetric behaviour around the zero
polarization spectrum of phases 0.4-0.6, indicate that the magnetic
axis must be nearly perpendicular to the rotation axis of the white
dwarf.

\begin{figure}
\includegraphics[width=1.0\columnwidth]{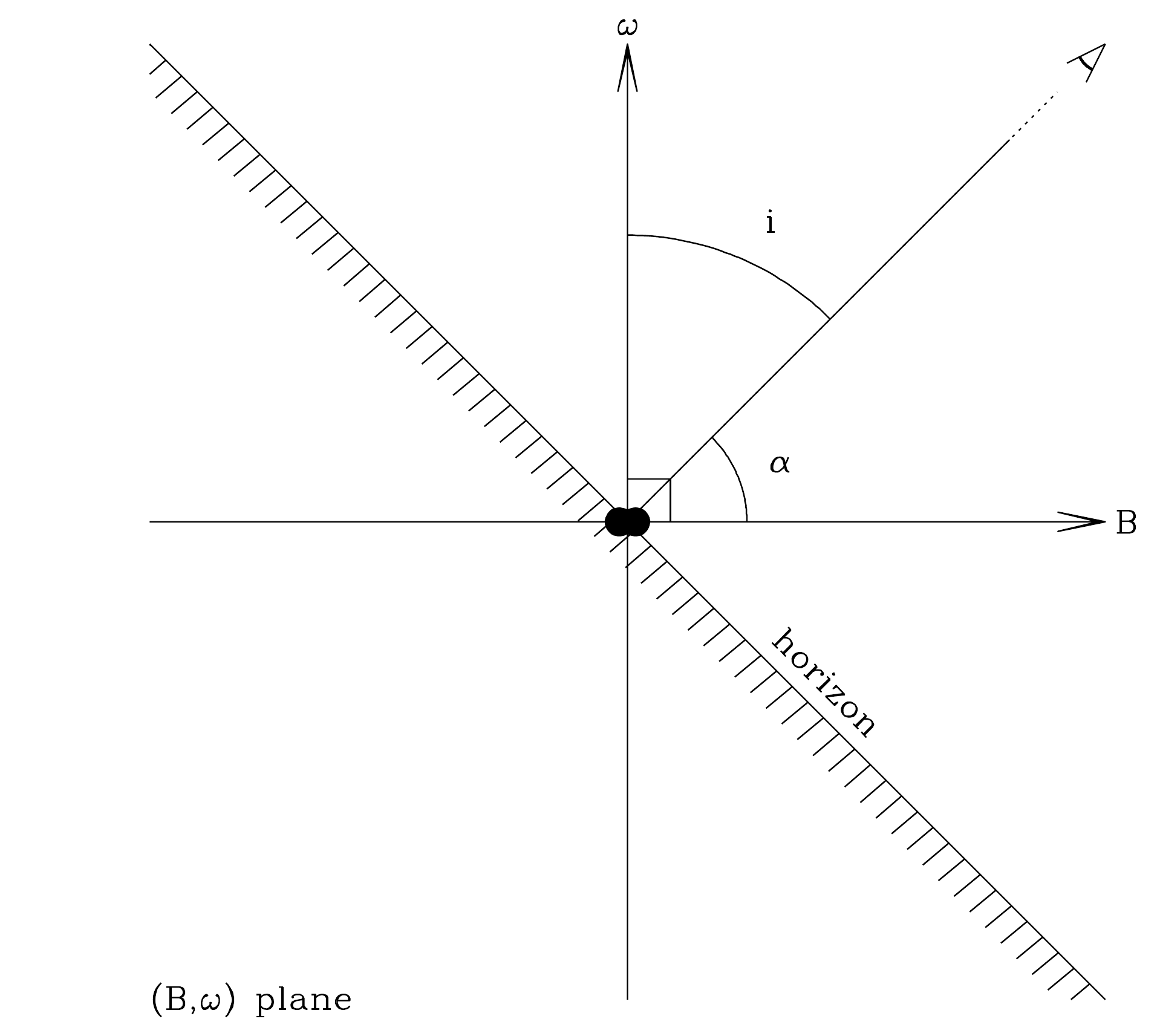}
\captionsetup{labelfont=bf, margin={0mm,12mm}, justification=justified}
\caption{Schematic view of the geometry of the double degenerate
  system NLTT\,12758. The rotation plane of the magnetic white dwarf
  is assumed to coincide with the orbital plane, and the spin axis is
  marked $\omega$. The spin axis is at an angle $i$ with respect to
  the observer and the magnetic field axis $B$ is at an angle $\alpha$
  with respect to the observer.}
\label{fig:Abs}
\end{figure}

Figure~\ref{fig:Abs} illustrates the geometry of the system with
$\alpha$ set at its minimum value ($90-i$).  Assuming $i=45^\circ$
(see Section~\ref{subsubsec:AtmParam}), the angle $\alpha$ will vary between
$90^\circ-45^\circ$ and $90^\circ+45^\circ$. When
$\alpha \approx 90^\circ$, the positive and negative polarization
contributions cancel each other and give rise to the unpolarized,
featureless spectrum observed in the phase range 0.4-0.6.  This can be
explained by the change, due to stellar rotation, between the magnetic
field direction and the line of sight to the observer averaged over
the visible hemisphere of the star \citep{wic1979}. The $\pi$
component in the circular polarisation spectra shows the presence of
narrow antisymmetric circular polarisation features. These are caused
by Faraday mixing due to magneto-optical effects which converts linear
polarisation into circular polarisation \citep{mar1981,mar1982} during
the radiation transfer.

\subsubsection{Photometric variations}
\label{subsubsec:PhotoVar}
\vspace{-6pt}
The photometric observations were analysed using three different
methods described in detail by \citet{pau2016}.  First, we employed
periodic orthogonal polynomials which are particularly useful for the
detection of non-sinusoidal oscillations. We fitted the observations
to identify the period and employed the analysis of the variance
(ANOVA) statistic to evaluate the fit quality \citep{sch1996}.

Next, we employed the string-length methods which simply minimize the
separation between consecutive phased data points at trial periods.
The best-fitting period corresponds to a minimum in the
''string-length'' which consists of the sum of data separations.  The
methods are useful for sparse data sets.

Finally, The Phase Dispersion Minimization (PDM) method is similar to
the string-length method \citep{ste1978}.  In this method, the data
are sorted into phase bins at trial periods and the variance within
each bin is calculated. The sum of the variances is minimized at the
best-fitting period.

\begin{figure}
\includegraphics[width=1.0\columnwidth]{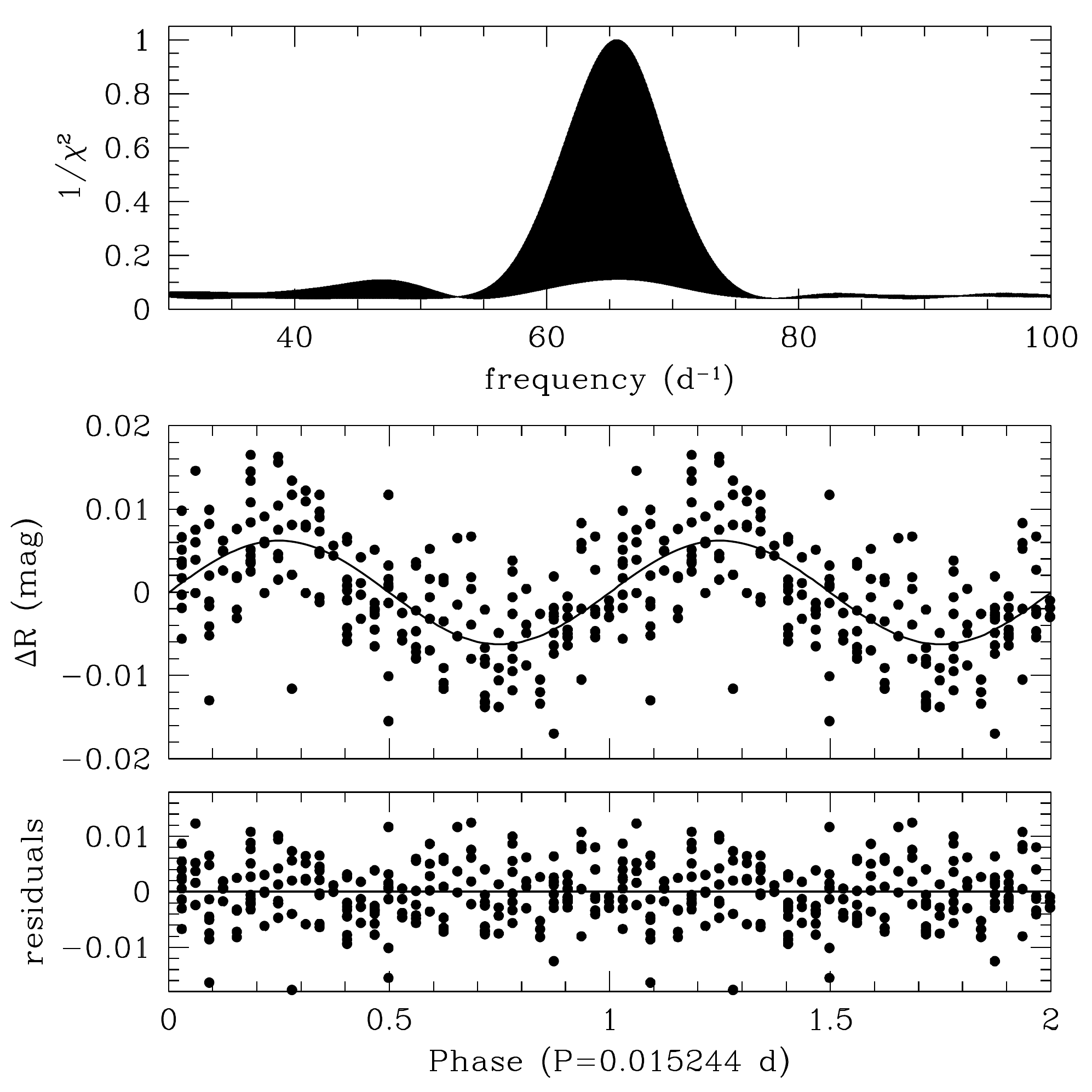}
\captionsetup{labelfont=bf, margin={2mm,20mm}, justification=justified}
\caption{(Top panel) period analysis of the measured $R$ photometric
  measurements.  (Middle panel) photometric $R$ magnitudes phased on
  the best rotation period and (bottom panel) residuals.}
\label{fig:Phot}
\end{figure}

We found that the photometric observations in the $R$ band show
variations. The calculated frequencies and their errors for the three different 
nights are 65.4\,$\pm$\,1.3, 65.3\,$\pm$\,0.6, and 65.6\,$\pm$\,1.2\, cycles
per day, respectively. The errors depend on the individual data set
lengths and the overall quality of the nights. Within the errors,
these values transform to a period of 22\,$\pm$\,0.5\,min. The semi-amplitude
of the variations is 6.2 mmag.  Figure~\ref{fig:Phot} shows the photometric 
magnitudes phased on the best rotation period of 22.0 minutes with the 
periodogram.

We conclude that the variations in spectropolarimetry and photometry
coincide and are phased on the rotation period of the magnetic white
dwarf. The photometric variation may be explained in terms of
magnetic dichroism which is caused by the different absorption
coefficients of left and right handed circularly polarised
radiation. A formulation for magnetic dichroism of hydrogen in
magnetic white dwarfs was first obtained by \citet{lam1974} and used
to explain the photometric variations of the high field magnetic white dwarf
EUVE\,J0317-855 \citep{Ferrario1997b}. However the magnetic field of the DAP
component of NLTT\,12758 is relatively low ($B < 20$ MG) for this
effect to be important. An alternate explanation for the photometric 
variations could be stellar spots \citep{Brinkworth2005}. Such a spot could be formed
by the inhibition of convection in the atmosphere by the magnetic field.
\citet{tre2015} show that convection is inhibited at the surface of objects
such as the magnetic component of NLTT~12758, however
their models are not able to explain flux variations like those observed 
in NLTT~12758 and other cool white dwarfs with low magnetic fields observed by
\citet{bri2013}.

\subsection{Stellar and atmospheric parameters}
\label{subsec:SandAParms}

\subsubsection{Modelling the field structure}
\label{subsubsec:FieldModel}

The appearance of the spectra of magnetic white dwarfs changes
dramatically as the field increases in strength.  If we indicate with
$(n,l,m_l)$ the zero field quantum numbers, the linear Zeeman regime
arises through the removal of the $m_l$ degeneracy, which for the
Balmer series occurs at a field strength of $\sim 1-4$\,MG.  As the
field increases, or the principal quantum number $n$ increases, the
quadratic effect becomes more important until the $l$ degeneracy is
also removed. This is called the quadratic Zeeman regime. In this
regime, the wavelength shift depends on the electron excitation level
and the $\pi$ and $\sigma$ Zeeman components are all displaced from
their zero field positions by different amounts. The quadratic shift is
of similar importance to the linear shift at $B\sim 4$\,MG for the
higher components of the Balmer series (e.g. $H\delta$). The spectra
of NLTT\,12758 indicate that the magnetic component of this system
belongs to the low field regime, as first reported by \citet{kaw2012}.

Before outlining our modelling methods, we need to stress that an
important and as yet unsolved problem regarding the modelling of
magnetic atmospheres, particularly in the high magnetic field regime,
is that concerning line broadening. However in the low field regime,
which is appropriate to the study of the spectrum of NLTT\,12758, it
is possible to assume, as a first approximation, that each Zeeman
component is broadened as in the zero field case. This approach has
been used successfully for the Zeeman modelling of hot white dwarfs
and has allowed the determination of the mass of the hot and
ultra-massive magnetic white dwarfs 1RXS\,J0823.6-2525
\citep[$B\sim 2.8-3.5$\,MG][]{Ferrario1998} and PG\,1658+441
\citep[$B\sim 3.5$\,MG][]{Ferrario1998,Schmidtetal92}. In the case of
PG\,1658+441, the spectroscopic mass was found to be in good agreement with 
that determined by the trigonometric parallax method
\citep{dah1999,ven2008}. No trigonometric parallax is as yet
available for 1RXS\,J0823.6-2525. On the other hand, in cool white dwarfs
such as NLTT\,12758, the contribution due to Stark broadening is negligible
and spectral line broadening is dominated by resonance. For H$\alpha$ to
H$\gamma$, we used parameters from the comprehensive self-broadening theory of 
\citet{bar2000}, and for the upper Balmer lines we combined the impact 
parameters from \citet{ali1965,ali1966} with the van der Waals parameters 
as described in \citet{kaw2012b}.

The modelling of the magnetised spectrum of NLTT\,12758 has been
conducted as follows. First, we have computed a zero-field grid of
pure hydrogen white dwarf model atmospheres \citep[see ][]{kaw2012}.
We used the ML2 parameterization of the mixing length theory with
$\alpha=0.6$, where $\alpha$ is the ratio of the mixing length to the
pressure scale height. Convection is predicted to be suppressed in cool
magnetic white dwarfs \citep{tre2015}, however we will investigate the effect 
of suppressed convection on the spectral lines of stars 
such as NLTT~12758 in future work.
This grid of models was then used as input for the
magnetic atmosphere program of \citet{wic1979}, modified to allow for Doppler,
resonance and Stark broadenings and magneto-optical effects which 
take into account the different refractive indices for radiation with different 
polarisation state \citep{mar1981}. The shifts and strengths in hydrogen lines, 
caused by the magnetic field, are included using the results of Zeeman
calculations by \citet{kem1974}.
Atmospheric models were then constructed at selected points on the
visible hemisphere of the white dwarf taking into consideration the
changes in field strength and direction. The resulting Stokes
intensities were then appropriately summed to yield a synthetic
spectrum.

The field geometry is strongly dependent on field strength and
structure and models built on observations obtained at different
phases, if the star rotates around its axis, are better constrained
than those restricted to one single intensity spectrum corresponding
to only one magnetic phase. The best constrained models are those
based on observations at different rotational phases and for which
\emph{both} intensity and polarisation spectra are available as it is
the case for NLTT\,12758.

The modelling of a magnetic white dwarf usually starts with the
assumption that the magnetic field configuration is that of a centred
dipole. Then, if necessary, more complex structures are
investigated. These usually consist of offset dipoles or combinations
of higher order multipoles. For the present set of observations of
NLTT\,12758 we found that a centred dipole model was inadequate to
model the rotationally modulated Zeeman components by simply changing
the viewing angle. This is because a centred dipole allows a field
spread of at most of a factor 2, which is not sufficient to model the
observations of NLTT\,12758.  It is possible to achieve a larger
magnetic field spread by offsetting the dipole from the centre of the
star. If the dipole is shifted by a fraction $a_z$ of the stellar radius
along the dipole axis, then the ratio of the field strengths $B_{p1}$
and $B_{p2}$ at the two opposite poles become
\begin{equation}
\frac{B_{p1}}{B_{p2}}=\left(\frac{1-a_z}{1+a_z}\right)^3
\end{equation}
We describe in detail how we have achieved the best-fit model for
NLTT\,12758 in the sections that follow.

\subsubsection{Spectroscopic analysis}
\label{subsubsec:AtmParam}

We fitted the X-shooter spectra with two sets of model spectra. The first
set of model spectra are for non-magnetic hydrogen-rich white dwarfs as 
described in \citet{kaw2012}. The Balmer line profiles used in the
synthetic spectra calculations are described in \citet{kaw2012b}.
The second set of model spectra include a 
magnetic field (as described above).

The procedure fits simultaneously the effective temperature and
surface gravity of both white dwarfs (4 parameters). We used the
mass-radius relations of \citet{ben1999} to scale the flux for both
stars and ensure that the relative flux contribution of each star is
preserved imposing a common distance for both stars. A similar decomposition
method was adopted in the analysis of the hot double degenerate EUVE~J1439+750
\citep{Vennes1999} and in the analysis of a sample of double degenerates by 
\citet{rol2014} and \citet{rol2015}. The results are model dependent due to
uncertainties in the treatment of line broadening in the presence of a magnetic
field as previously noted by \citet{kul2009}. However, the presence of a 
non-magnetic DA companion with a reliable radius measurement, as in the case 
of NLTT~12758, helps constrain the radius of the magnetic component.
A direct constraint on the stellar radii would be achieved
with a parallax measurement.

The Zeeman
splitting observed in the X-shooter spectra (H$\alpha$ and H$\beta$)
implies an averaged surface field of $B_S = 1.70\pm0.04$\,MG. We used
this value as a starting point to calculate sets of magnetic field
spectra with varying polar field strength and offset. 
We fitted the spectra with the following magnetic field
strengths and offsets: offset $= -0.1$ at $B_P = 2.8, 3.1, 3.4, 3.6$
MG; offset $= 0$ at $B_P = 2.6, 3.0, 3.2, 3.4$ MG; offset $= +0.1$ at
$B_P = 2.4, 2.9, 3.1, 3.3$ MG. We also fitted the X-shooter spectra at
viewing angles of $50^\circ$ and $80^\circ$ for each offset and field
strength value. Note that the total exposure time covers nearly two
complete rotation cycles and the viewing angle represents a cycle average.

\begin{figure}
\centering
\includegraphics[width=140mm]{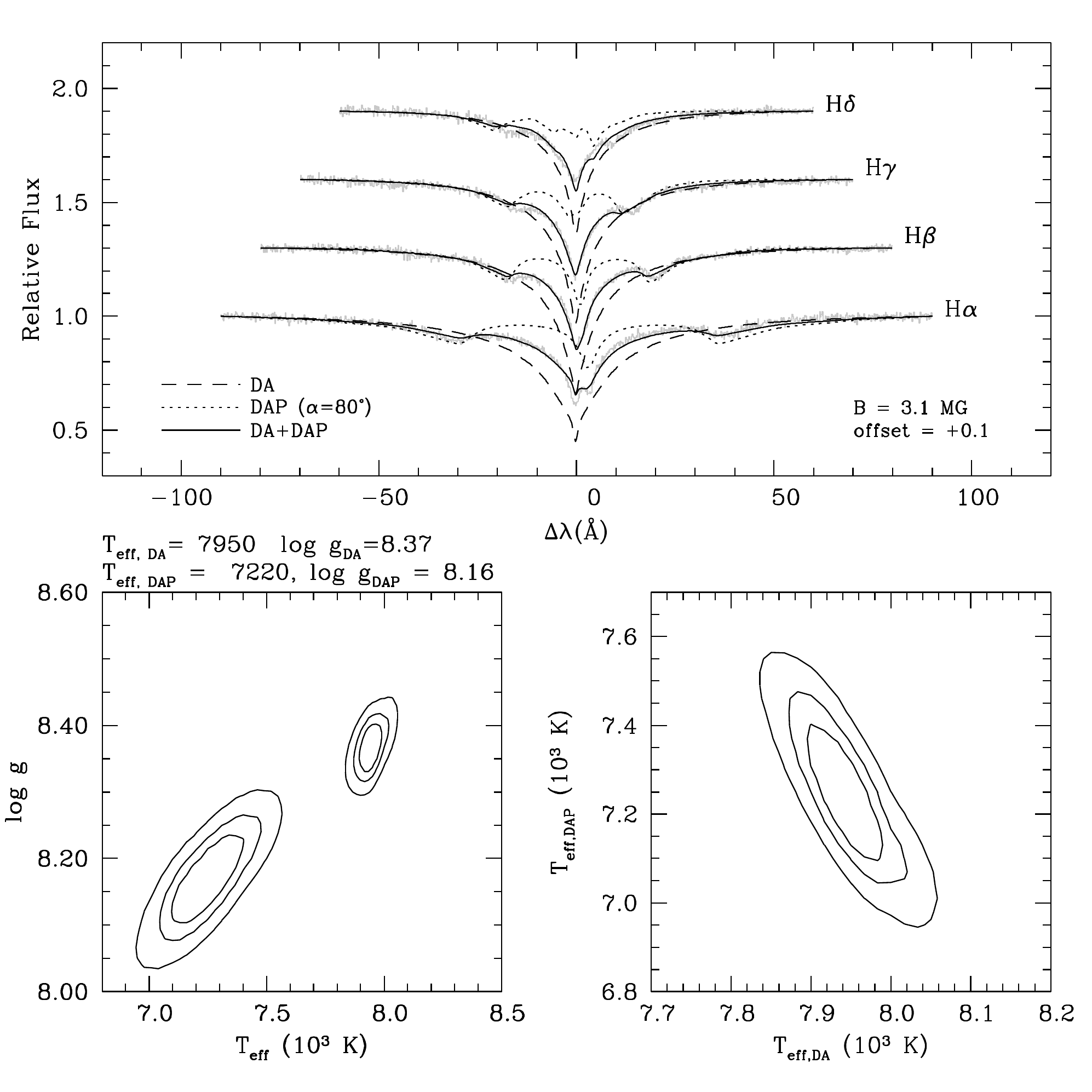}
\captionsetup{labelfont=bf, margin={8mm,5mm}, justification=justified}
\caption{(Top panel) Observed Balmer line profiles of NLTT\,12758 compared to 
the best-fitting models. The best-fit shows that the components of NLTT\,12758 
are a non-magnetic DA white dwarf (dashed lines) paired with a magnetic DA 
white dwarf (dotted lines). Confidence contours at 66, 90, and 99\% are shown 
in the $T_{\rm eff, DAP}$ vs $T_{\rm eff, DA}$ plane (bottom right) and 
$\log{g}$ vs $T_{\rm eff}$ for both stars (bottom left).}
\label{fig:BalmerFit}
\end{figure}

Figure~\ref{fig:BalmerFit} compares the X-shooter spectrum and the
best-fitting models for the two stars. The magnetic white dwarf has a
polar magnetic field $B_P =3.1$ MG offset by $a_z = +0.1$ from the
stellar centre. The magnetic white dwarf appears to be slightly
cooler with $T_{\rm eff, DAP} = 7220\pm180$ K and a surface gravity of
$\log{g}=8.16\pm0.08$. The non-magnetic white dwarf is a little hotter and
more massive with $T_{\rm eff, DA} = 7950\pm50$ K and
$\log{g} = 8.37\pm0.04$.  The best-fitting viewing angle to the dipole
axis is on average $\alpha=80^{\circ}$.
Table 4 lists the stellar parameters. We computed the mass and cooling age
of each component using the evolutionary models of \citet{ben1999}. The
spectroscopic mass ratio $M_{\rm DA}/M_{\rm DAP}=1.1-1.3$ is consistent 
with the orbital mass ratio, but also more accurate, and implies that the mass 
of the DA star may be slightly higher than the mass of the DAP star. We 
then estimated the absolute magnitude of each component and calculated the
distance to the system.

\citet{rol2014} and \citet{rol2015} measured the stellar parameters of NLTT\,12758 by
fitting H$\alpha$ together with the spectral energy distribution (SED)
including only $VJHK$. They obtained $T_{\rm eff, DAP} = 6041$ K and
$T_{\rm eff, DA} = 8851$ with a radius ratio of
$R_{\rm DA}/R_{\rm DAP} = 0.908$. Although our radius ratio is in agreement 
with theirs, our effective temperatures differ from their effective 
temperatures.

\begin{landscape}
\begin{figure*}
\centering
\vspace{5mm}
\includegraphics[width=220mm]{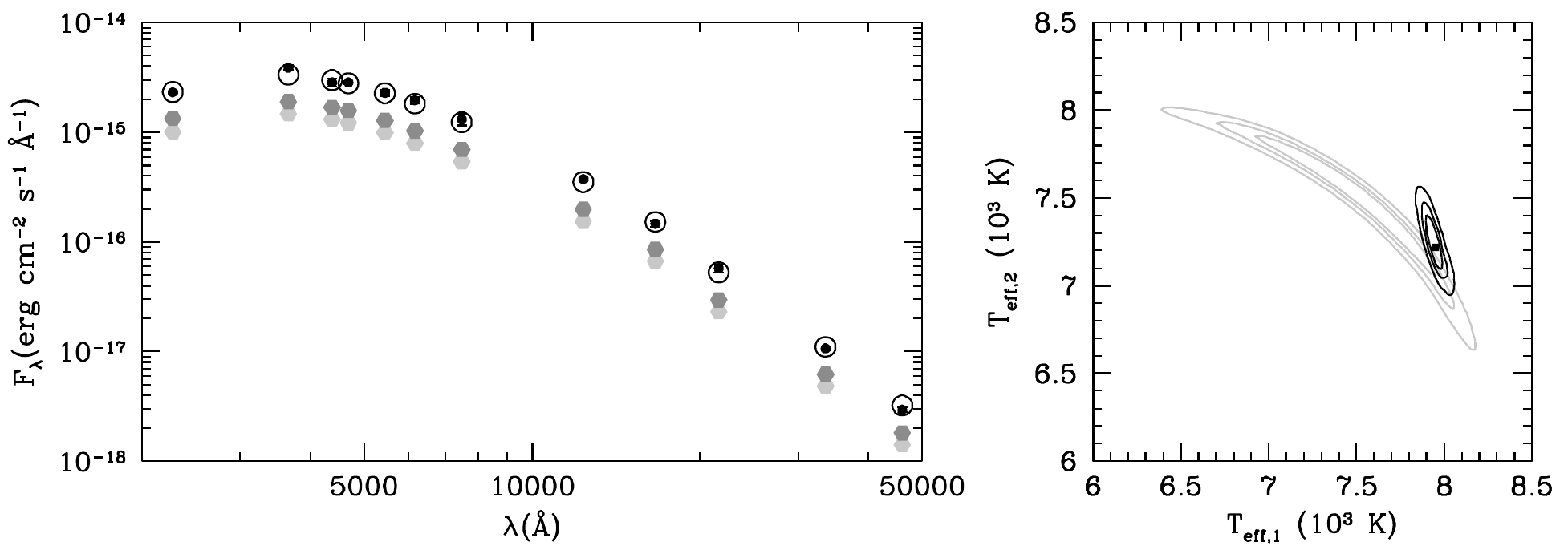}
\vspace{5mm}
\captionsetup{labelfont=bf, margin={5mm,15mm}, justification=justified}
\caption{The left panel compares the best-fitting photometry (open circle)
to the observed photometry (solid black circles). The contribution of 
individual stars are plotted in different grey shades as hexagonals. The right
panel plots the confidence contours (66, 90, and 99\%) of the spectroscopic 
fit (in black) and the contours of the SED fit (grey full lines). Note that 
$\log{g}=8.4$ for star 1 (DA) and $\log{g}=8.2$ for star 2 (DAP).}
\label{fig:SedFit}
\end{figure*}
\end{landscape}

Taking advantage of a broader wavelength coverage, we re-analysed the
SED. First, we fitted the photometric data set
($NUV$,$UBV$,$gri$,$JHK$ and $W1$,$W2$) by fixing the surface gravity
measurements to those obtained in the spectroscopic analysis. We
allowed for both temperatures to vary and assumed null interstellar
extinction. The resulting effective temperatures are nearly in
agreement with the spectroscopic analysis showing that interstellar 
extinction in the line of sight toward NLTT~12758 is negligible when
compared to the totalextinction in the same line of sight, $E(B-V) = 0.06$
\citep{sch1998}. Figure~\ref{fig:SedFit} shows the model photometry
fitted to the measured photometry and compares the confidence contours
for the SED fit, as well as the confidence contours for the Balmer line fit
(Fig.~\ref{fig:BalmerFit}). The overlapping contours show that the two
methods are consistent and imply that the two objects share similar
stellar parameters. In the following discussion we adopt the results
of the spectroscopic analysis.

Our results differ markedly from those of \citet{rol2014} and 
\citet{rol2015} who reported a temperature difference 
$\Delta T=T_{\rm eff, DA}$-$T_{\rm eff, DAP}\approx2800$~K while we estimated 
much closer temperatures for the components ($\Delta T\approx 700$~K). On the 
other hand we estimated a similar mass ratio. Our spectroscopic analysis 
includes the first four members of the Balmer line series (H$\alpha$ to 
H$\delta$), thereby lifting potential degeneracy in the $T_{\rm eff}/\log{g}$ 
solution, while \citet{rol2014} and \citet{rol2015} only include H$\alpha$. 
However, both solutions are model dependent and part of the discrepancy may
also be attributed to different line-broadening prescriptions used in 
calculating magnetic synthetic spectra. The large temperature difference 
reported by \citet{rol2014} and \citet{rol2015} should also be noticeable in 
the SED, particularly in the near ultraviolet. Our own analysis based an 
extensive data set implies a temperature difference no larger than 
$\approx 1100$~K ($1\sigma$) while a larger temperature difference would be 
incompatible with the $GALEX$ $NUV$ measurement.

Using the evolutionary mass-radius relations of \citet{ben1999}, we
find that the cooling ages of the two white dwarfs in NLTT\,12758 are
comparable.  However, \citet{val2014} have proposed that convection in
cool white dwarfs is suppressed by magnetic fields, and therefore
magnetic white dwarfs may appear younger than they are. The 3D
radiation magnetohydrodynamic simulations of \citet{tre2015} have
confirmed that magnetic fields do suppress convection, however they do
not affect the cooling of the white dwarf until temperatures have dropped 
below 6000\,K. Since the magnetic white
dwarf is hotter than this upper limit, it is likely that its age is
not affected and that the two objects formed around the same time.

We derived an orbital inclination of $i = 45^\circ$ for NLTT\,12758 by combining 
the component masses with the orbital parameters
and using:

\vspace{-8pt}
\begin{equation}
\frac{M_{\rm DA}^3 \sin^3{i}}{(M_{\rm DA}+M_{\rm DAP})^2} = \frac{P K_{\rm DAP}^3}{2\pi G}
\vspace{1pt}
\end{equation}

where $M_{\rm DA}$ and $M_{\rm DAP}$ are the masses of the
non-magnetic and magnetic white dwarfs respectively, $P$ is the
orbital period, $K_{\rm DAP}$ is the velocity semi-amplitude of the
magnetic white dwarf and $G$ is the gravitational constant.  Setting
the system inclination at $i=90^\circ$, the maximum orbital period is
$P\approx3.3$~d.

\begin{table}
\centering
\captionsetup{labelfont=bf, margin={0mm,0mm}, justification=justified}
\caption{Summary of NLTT\,12758 parameters
\label{tab:Param}\vspace{0pt}}

\begin{tabular}{lcc}
\toprule
Parameter & DA & DAP \\
\midrule
$T_{\rm eff}$ (K)      & $7950\pm50$   & $7220\pm180$ \\
$\log{g}$ (c.g.s)      & $8.37\pm0.04$ & $8.16\pm0.08$ \\
Mass (M$_\odot$)       & $0.83\pm0.03$ & $0.69\pm0.05$ \\
Cooling age (Gyrs)     & $2.2\pm0.2$   & $1.9\pm0.4$ \\
$M_V$ (mag)            & $13.65\pm0.06$& $13.69\pm0.18$ \\
Period (d)             & \multicolumn{2}{c}{$1.15401\pm0.00005$} \\
$K$ (km~s$^{-1}$)      & $81.9\pm17.3$ & $89.7\pm3.8$ \\
$\gamma$ (km~s$^{-1}$) & $94.2\pm17.3$ & $96.4\pm2.6$ \\
$d$ (pc)               & \multicolumn{2}{c}{$32.6\pm3.5$} \\
$v_r$ (km~s$^{-1}$)    & \multicolumn{2}{c}{$58.0\pm3.9$} \\
\bottomrule
\end{tabular}
\vspace{-14pt}
\end{table}

The calculated white dwarf gravitational redshifts ($\gamma_{g,DAP}=38.4\pm2.9$,
$\gamma_{g,DA}=53.6\pm1.7$~km~s$^{-1}$) may be subtracted from their
respective systemic velocities to obtain an estimate of the radial
velocity of the system. Using the more precise velocity of the DAP
star we obtain $v_r=58.0\pm3.9$\,km\,s$^{-1}$.  Combining the proper
motion measurements \citep{kaw2012}, the photometric distance estimate
($d$) and the radial velocity ($v_r$) of the system we determine the
Galactic velocity components
$(U, V, W)=(-40\pm4,-48\pm5,-3\pm6)$~km~s$^{-1}$ which suggest that
the system is relatively young and belongs to the thin disc
\citep{pau2006}.

\subsection{Evolution of NLTT\,12758}
\label{subsec:Evolution}

In order to understand the evolution of NLTT\,12758 we have used the
rapid binary star evolution algorithm, {\sc bse}, of
\citet{Hurley2002}. We have evolved a number of binaries from the Zero
Age Main Sequence (ZAMS) to the age of the Galactic disc \citep[9.5
Gyr, e.g.~][]{Oswalt1996,liu2000}.  This code is a derivation of the
single star evolution code of \citet{Hurley2000} where the authors use
analytical formulae to approximate the full evolution of stars.  The
{\sc bse} takes into consideration stellar mass-loss, mass transfer,
Roche lobe overflow, CE evolution, tidal interaction,
supernova kicks and angular momentum loss caused by gravitational
radiation and magnetic braking.  In order to model the CE
evolution, the {\sc bse} uses the $\alpha_{\rm CE}$ formalism, where
$\alpha_{\rm CE}$ is a parameter with values in the range $0.1-0.9$.
In our calculations we have adopted $\eta=1.0$ for the Reimers'
mass-loss parameter, as outlined in chapter 2 \citep[and][]{bri2015} and a stellar
metallicity, $Z = 0.02$.

We have then generated a synthetic population of binaries with ZAMS
conditions of the mass of the primary star, $M1$, between 3.5 and
4.5\,$M_\odot$, the mass of the secondary star, $M2$, between 2.5 and
3.5\,$M_\odot$ and the initial period $P_0$ in the range
$2000-3500$\,days, as these values were in the region of the expected
initial conditions for the final properties of the components of
NLTT\,12758.  We allowed 200 steps in each parameter in the {\sc bse}
evolution of the population through to the age of the Galactic disk.
In all cases we assumed an initially circular orbit for the progenitor
binary, that is, an eccentricity of zero.  The calculations were
repeated for values of $\alpha_{\rm CE}$ = {\{}0.10, 0.20, 0.25, 0.30,
0.40, 0.50, 0.60, 0.70, 0.80, 0.90{\}}.

A number of stellar types are recognised by {\sc bse} within its logic 
throughout the stages of evolution. These types are set out in Table~1 
of \citet{bri2015}.

The evolved populations were searched for systems that resulted in a
pair of Carbon/Oxygen white dwarfs (CO\,WDs), that is, type 11s in the
{\sc bse} system.  We have found that as $\alpha_{\rm CE}$ increases
the number of CO\,WD double degenerate systems increases.  However, not
all of these systems correspond with the type of evolution path that
would lead to the final parameters of NLTT\,12758, i.e. cooling ages,
period, masses.  

A suitable near match to NLTT\,12758 was achieved at
$\alpha_{\rm CE}=0.15$ with initial masses of 3.75\,M$_\odot$ and
2.80\,M$_\odot$ and with an initial period of 2656\,days.  As {\sc bse}
consists of many approximations, the resulting solution is
considered to be satisfactory and within the errors on the
parameters of NLTT\,12758 given in Table\,\ref{tab:Param}.

The evolution shown in Table\,\ref{tab:Evolve} starts with two stars, S1
and S2, and follows each of them through their normal evolution until
256\,Myr. Up until this time the only interactions between the two
stars are small mass losses due to winds and the consequent small
changes in orbital separation and period. At 256\,Myr the stars start
to interact by common envelope evolution. First, the more massive
star, S1, develops an extended envelope which overflows the Roche
lobe. This draws the stars closer together by friction eroding the
orbit. When Roche lobe overflow ceases and S1 reveals its core as a
CO WD, the two stars are about 588\,R$_\odot$ apart with a period of
around 864\,days. At this point, S2 is still a main sequence star. About
315\,Myr later, S2 initiates its own common envelope evolution
resulting in a second CO WD, an orbital separation of 5.3\,R$_\odot$
and a period of only 1.161\,days. One of the pair, S2, is now a
magnetic WD resulting from the dynamo effect within the common
envelope. S1 loses about 2.8\,M$_\odot$ during the first common
envelope phase while S2 loses about 2\,M$_\odot$. As the second common
envelope evolution brings the two stars very close together by
shrinking the orbit from about 500 to 5\,R$_\odot$, it is S2 that
develops the magnetic field and the rapid rotation.

From this time the pair interact by gravitational radiation and
magnetic braking with consequent orbital shrinkage until at 2791\,Myr
they reach the present day with a separation of 5.278\,R$_\odot$ and
an orbital period of 1.154\,days.  The cooling ages are 2535\,Myr and
2220\,Myr for the non-magnetic and magnetic white dwarfs respectively
\citep[for details of the method see ][and chapter 2 of this work]{bri2015}. Further evolution will see the orbit shrinking further until at some stage (over a time much longer than a Hubble time) Roche lobe overflow restarts and the two stars merge. The possible final fate of double degenerate white dwarf
systems, such as NLTT\,12758, is discussed in the section below.

\begin{landscape}
\begin{table*}
\centering
\footnotesize
\captionsetup{labelfont=bf, margin={15mm,15mm}, justification=justified}
\vspace{12pt}
\caption{Evolution of a binary star system of approximately the size of 
NLTT\,12758 starting from ZAMS through to the end of their interaction 
and the production of a double degenerate WD pair. $M1$ and $M2$ are 
the masses of the primary and secondary stars respectively (in solar 
masses), $S1$ and $S2$ are the stellar types varying throughout their 
evolution as shown in Table 1 in \citet[][and chapter 2]{bri2015}. $Sepn$ is the stellar 
separation in solar radii, $Period$ is the orbital period in days and the 
$Event-Type$ is the event happening to the system at the time given 
in column 2.\vspace{6pt}}
\label{tab:Evolve}
\begin{tabular}{r r r r r r r r l}
\toprule
 Step & $Time$  &   $M1$    &   $M2$    & $S1$  & $S2$ & $Period$ & $Sepn$ & $Event-Type$\\
        &  (MYr) &(M$_\odot$)&(M$_\odot$)&      &     &   (days)  &(R$_\odot$)\\
\midrule
{ }1&      0.000 &  3.750 &  2.800 & 1 & 1 &   2656.000  &  1510.578 & ZAMS  \\
{ }2&    210.988 &  3.750 &  2.800 & 2 & 1 &   2653.321  &  1509.562 & S1$\Rightarrow$Hertzsprung Gap \\
{ }3&    212.057 &  3.750 &  2.800 & 3 & 1 &   2653.713  &  1509.674 & S1$\Rightarrow$RGB  \\
{ }4&    212.955 &  3.747 &  2.800 & 4 & 1 &   2655.573  &  1510.204 & S1$\Rightarrow$He core burning \\
{ }5&    253.754 &  3.676 &  2.800 & 5 & 1 &   2714.292  &  1526.805 & S1$\Rightarrow$Early AGB \\
{ }6&    255.551 &  3.597 &  2.801 & 6 & 1 &   2668.247  &  1503.396 & S1$\Rightarrow$Late AGB \\
{ }7&    255.989 &  2.787 &  2.827 & 6 & 1 &   2819.839  &  1493.233 & Begin Roche lobe overflow \\
{ }8&    255.989 &  0.827 &  2.827 &11& 1 &    864.356   &   588.342  & CEE, S1$\Rightarrow$CO WD\\
{ }9&    255.989 &  0.827 &  2.827 &11& 1 &    864.356   &   588.342  & End Roche lobe overflow \\
  10&    443.089 &  0.827 &  2.827 &11& 1 &    864.356   &   588.342  & S2$\Rightarrow$Blue straggler\\
  11&    449.391 &  0.827 &  2.827 &11& 2 &    864.356   &   588.342  & S2$\Rightarrow$Hertzsprung Gap \\
  12&    452.151 &  0.827 &  2.826 &11& 3 &    864.520   &   588.398  & S2$\Rightarrow$RGB  \\
  13&    455.303 &  0.827 &  2.824 &11& 4 &    865.691   &   588.810  & S2$\Rightarrow$He core burning \\
  14&    567.390 &  0.827 &  2.774 &11& 5 &    889.679   &   596.911  & S2$\Rightarrow$Early AGB \\
  15&    570.808 &  0.828 &  2.725 &11& 6 &    757.173   &   533.592  & S2$\Rightarrow$Late AGB \\
  16&    571.109 &  0.828 &  2.662 &11& 6 &    689.277   &   498.253  & Begin Roche lobe overflow \\
  17&    571.109 &  0.828 &  0.652 &11&11&        1.161   &       5.297  & CEE, S2$\Rightarrow$CO WD \\
  18&    571.109 &  0.828 &  0.652 &11&11&        1.161   &       5.297  & End Roche lobe overflow \\
  19&   2791.209&  0.828 &  0.652 &11&11&        1.154   &       5.278  & Present Day \\
\bottomrule
\normalfont
\end{tabular}
\end{table*}
\end{landscape}

\section{Discussion}
\label{sec:Discuss}

NLTT\,12758 is a member of a growing class of double degenerate
systems consisting of two white dwarfs, one magnetic and one
not. Table~\ref{tab:DD} lists the currently known double degenerate
systems containing at least one magnetic white dwarf. The table lists
the names, orbital and rotational periods, the magnetic field
strength, effective temperatures and masses of the components. It
includes both close binaries and common proper motion (CPM) systems.
Most of the systems for which effective temperatures and masses
are determined appear to have formed, within uncertainties,
at the same time. In the case of CPM systems, where it is assumed
that the stars did not interact during their evolution, there are systems
with inconsistencies in their ages
if one assumes single star evolution for each star. Apart from the
well documented case of EUVE\,J0317-855 \citep{Ferrario1997b}, another more
recent example is given by PG\,1258+593 and its common proper motion
magnetic white dwarf companion SDSS\,J1300+5904.  \citet{Girven2010}
found that the masses of these white dwarfs are
0.54$\pm$0.06\,M$_\odot$ for the non-magnetic and
0.54$\pm$0.01\,M$_\odot$ for the magnetic component.  Despite their
very similar masses, SDSS\,J1300+5904 is a cool white dwarf
($T_{\rm eff}=6300\pm300$\,K) while PG\,1258+593 is substantially
hotter ($T_{\rm eff}=14790\pm77$\,K). \citet{Girven2010} find that the
temperature discrepancy gives a difference in cooling age (and thus in
formation age of the white dwarfs) of $1.67\pm 0.05$\,Gyr. If one
makes the plausible assumption that the progenitors of these CPM white
dwarfs formed in the same protostellar cloud at roughly the same time,
then the similar white dwarf masses and their large age discrepancy
give rise to a paradox. A possible solution is that this system was
initially a triple system where two stars interacted and merged to
form the magnetic white dwarf SDSS\,J1300+5904 about 1.67\,Gyr before
the third non-interacting object evolved into the non-magnetic white
dwarf PG\,1258+593.

However, the situation appears to be rather different for the double
degenerate system NLTT\,12758, as reported in section
\ref{subsec:Evolution}. Since NLTT\,12758 is a close binary system, it is
highly unlikely that the field of the magnetic component was caused by
the merging of two stars in an initially triple system. Instead, the
magnetic field must have originated during CE evolution
in a manner very similar to that occurring during the formation of a
magnetic cataclysmic variable, as proposed by \citet[][submitted to MNRAS]{bri2018a}.
In this scenario, the closer the cores of the two stars
are drawn during CE evolution, the greater the
differential rotation and thus the larger the dynamo generated field
will be. If CE evolution leads to the merging of the two
stellar cores the resulting object would be an isolated highly
magnetic white dwarf \cite[see][]{wickramasinghe2014}. If the two stars do not
coalesce they are expected to emerge from the CE as close
binaries that are already interacting, and thus appear as magnetic
cataclysmic variables, or are close to interaction. The low-accretion
rate polars, where a magnetic white dwarf accretes matter from its
companion through a stellar wind, have been suggested by
\citet{sch2009} to be the progenitors of the polars, which are the
highest field magnetic cataclysmic variables. In the polars a MWD
accretes matter from an un-evolved low-mass (M-dwarf)
companion via magnetically confined accretion flows. The orbital
periods are typically between 70 minutes to a few hours and Zeeman and
cyclotron spectroscopy from the UV to the IR bands have revealed the
presence of fields between 7 to 230\,MG
\citep[e.g. see][]{fer1992,fer1993,fer1996,sch2001b} in the case of the 
polars, and 1 to 20\,MG in the case of intermediate polars \citep{fwk1993}. 
The difference between these systems and NLTT\,12758 is that both progenitor 
stars of NLTT\,12758 were too massive to evolve into a magnetic cataclysmic
variable. However, the indications
seem to be that the magnetic white dwarf component of NLTT\,12758
acquired its field via a mechanism similar to that propounded to
explain the origin of magnetic cataclysmic variables.

The properties of NLTT~12758 mean that the two white dwarfs 
will coalesce in a time much longer than a Hubble time
\citep[$\sim140$\,Gyr;][]{rit1986}, however it is still interesting to
speculate what the final fate of a system like this might be.

The first simulations of two merging CO WDs were conducted by
\citet{sai1985} and showed that the fast mass accretion rate
($\gtrsim 10^{-5}$\,M$_\odot$yr$^{-1}$) from the less massive to the
more massive white dwarf ignites an off-centre carbon flash. The
carbon nuclear burning then propagates toward the stellar centre
turning the CO WD into an ONe WD quiescently. The outcome of such an
event would not be a carbon deflagration but an accretion induced
collapse (AIC) triggered by electron captures on $^{24}$Mg and
$^{20}$Ne. The result would be a rapidly spinning neutron star that
would appear as an isolated millisecond pulsar \citep[MSP,
e.g.][]{lor2008}. The low space velocities of isolated MSPs suggest
that there could not have been a substantial SNII kick imparted to the
emerging neutron star, thus supporting the AIC hypothesis
\citep{fer2007,hur2010}. The calculations of \citet{chen2013} lend
further support to this idea since they show that it is unlikely that
the isolated MSPs may be generated via the LMXB recycling scenario because
this would require the total ablation of their donor star. Thus, merging
events of systems similar to NLTT\,12758, but with initial
parameters that would allow faster evolutionary timescales, could
provide a simple explanation for the existence of isolated MSPs. 

On the other hand, the merging of the two stars in NLTT\,12758 may
give rise to a supernova event. Recent simulations conducted by
\citet{dan2014} and \citet{dan2015} showed that a merging system with
a total mass M$_{\rm tot} \ge 2.1$\,M$_\odot$ and comprised of two
white dwarfs of similar mass may result in a Type Ia supernova; The
total mass of NLTT\,12758, $M_{\rm tot} = 1.52$\,M$_\odot$, would be
below the predicted cutoff for this event to occur. However,
other studies conducted by \citet{pak2011} and \citet{sat2016} found
that systems with a mass ratio greater than $\sim 0.8$ could indeed
result into a SNIa explosion.  Clearly, a consensus in this area of
research still needs to be reached \citep[e.g.][]{fer2013}.

\section{Conclusions}
\label{sec:Conclude}
In this chapter we have reported our studies on the close,
super-Chandrasekhar double degenerate system NLTT\,12758 consisting of
two CO WDs of similar masses and ages and with one of the two
components highly magnetic. The magnetic white dwarf spins around its
axis with a period of 23 minutes and they orbit around each other with
a period of 1.15 days. Although the components of NLTT\,12758 will not merge
over a Hubble time, systems with very similar initial parameters will
come into contact and merge thus undergoing either an accretion
induced collapse to become a rapidly spinning neutron star (an
isolated MSP) or a Type Ia supernova explosion. Given the
theoretical uncertainties, the jury is still out on the fate of such
systems.

\section*{Acknowledgements}
AK and SV acknowledge support from the Grant Agency of the Czech
Republic (P209/12/0217 and 15-15943S) and the Ministry of Education of 
the Czech Republic (LG14013). This work was also supported by
the project RVO:67985815 in the Czech Republic.  SV acknowledges
support from the Mathematical Sciences Institute of the Australian
National University. EP acknowledges support by the Ministry of Education 
of the Czech Republic (grant LG15010).
GPB gratefully acknowledges receipt of an Australian Postgraduate
Award. We thank the referee, Pier-Emmanuel Tremblay, for a thorough report 
and helpful comments on line-broadening theory.

\begin{landscape}
\begin{center}
\footnotesize
\begin{longtable}{llcccccccc}

\captionsetup{labelfont=bf, justification=justified}
\caption{Known double degenerates containing a magnetic white dwarf}
\label{tab:DD}
\vspace{-4pt}\\
\toprule
Name & Alternate name & $P_{\rm orb}$  & $P_{\rm rot}$  & $B$  & \multicolumn{2}{c}{$T_{\rm eff}$ (K)} & \multicolumn{2}{c}{Mass ($M_\odot$)} & Reference \\
         &                        &                       &                       & (MG)&  Magnetic           &   Companion           &   Magnetic         & Companion             &            \\
\midrule
\endfirsthead

0040+000 & SDSS~J004248.19+001955.3 & ...    & ...    & 14         & \multicolumn{2}{c}{11000}   & ...           & ...           & 1 \\
0121-429$^a$ & LHS 1243             & ...    & ...    & 10.3       & 6105         & 5833         & 0.7$^d$       & 0.54$^d$      & 2,3,4 \\
0239+109$^a$ & G 4-34, LTT 10886    & ...    & ...    & 0.7        & 10060        & 7620         & ...           & ...           & 5,6 \\
0325-857 & EUVE~J0317-855           & $\sim2095$ yr & 725 s & 185-425 & 33000     & 16360        & 1.3           & 0.85          & 7,8,9,10 \\
0410-114 & NLTT\,12758, G160-51      & 1.15 d & 23 min & 3.1       & 7220         & 7950         & 0.69          & 0.83          & This work \\
0512+284$^a$ & LSPM~J0515+2839      & ...    & ...    & 2.15       & 5940         & 6167         & 0.81$^d$      & 0.61$^d$      & 3,4 \\
0745+303 & SDSS~J074853.07+302543.5 & CPM    & ...    & 11.4       & 21000        & 22702        & 0.81          & 0.88          & 11 \\
0843+488$^b$ & SDSS~J084716.21+484220.4 & ...& ...    &            & \multicolumn{2}{c}{19000}   & ...           & ...           & 1 \\
0924+135 & SDSS~J092646.88+132134.5 & CPM    & ...    & 210        & 9500         & 10482        & 0.62          & 0.79          & 12 \\
0945+246 & LB~11146                 & $\sim130$ d &...& $\sim 670$ & 16000        & 14500        & 0.90          & 0.91          & 13,14,15 \\
1026+117$^a$ & LHS~2273             & ...    & ...    & 17.8       & 5691         & 7350         & 0.75$^d$      & 0.64$^d$      & 3,4 \\
1258+593 & SDSS~J130033.48+590407.0 & CPM    & ...    & 6          & 6300         & 14790        & 0.54          & 0.54          & 16 \\
1330+015$^a$ & G~62-46              & ...    & ...    & 7.4        & 5712         & 7618         & 0.82$^d$      & 0.64$^d$      & 3,4 \\
1440+753 & EUVE~J1439+750           & ...    & ...    & 10         & 42000        & 30000        & 0.9           & 1.1           & 17 \\
1503-070$^a$ & GD~175               & ...    & ...    & 2.9        & 6062         & 7051         & 0.95$^d$      & 0.73$^d$      & 3,4 \\
1506+399 & CBS~229                  & CPM    & ...    & 18.9       & 18000        & 16761        & 0.81          & 0.82          & 11 \\
1506+523 & SDSS~J150746.80+520958.0 & CPM    & ...    & 65.2       & 18000        & 17622        & 0.99          & 0.70          & 12 \\
1514+282$^a$ & SDSS~J151625.07+280320.9 &... & ...    & 2.05       & 7168         & 7662         & 0.77$^d$      & 0.54$^d$      & 3,4 \\
1713+393$^a$ & NLTT~44447           & ...    & ...    & 2.1        & 6204         & 6556         & 0.94$^d$      & 0.54$^d$      & 3,4,18 \\
1814+248$^c$ & G~183-35             & ...    & ...    & 12.05/7.8  & 5998         & 5849         & 0.85$^d$      & 0.74$^d$      & 3,4,19 \\
1818+126$^a$ & G~141-2              & ...    & ...    & 3.75       & 5215         & 6451         & 0.64$^d$      & 0.54$^d$      & 3,4 \\
\bottomrule
\end{longtable}
\end{center}

\begin{minipage}{210mm}
\vspace{-10mm}
\footnotesize
{$^a$ DAH+DC},\,\,\,\,{$^b$ DAH+DB},\,\,\,\,{$^c$ DAH+DAH}\\
$^d$ Masses are calculated using the mass-radius relations of \citet{ben1999}, 
the published parameters of the magnetic star and ratio of the stellar radii.
\normalsize
\end{minipage}
\newpage
\begin{minipage}{210mm}
\normalsize
{\bf Table \ref{tab:DD}} continued,\\
\small
{\bf References:} (1) \citet{sch2003}; (2) \citet{sub2007}; (3) \citet{rol2014}; (4) \citet{rol2015}; (5) \citet{koe2009}; (6) \citet{gia2011};
(7) \citet{Ferrario1997b}; (8) \citet{Vennes2003}; (9) \citet{kul2010}; (10) \citet{law2013}; (11) \citet{Dobbie2013}; (12) \citet{Dobbie2012};
(13) \citet{nel2007}; (14) \citet{Glenn1994}; (15) \citet{Liebert1993}; (16) \citet{Girven2010}; (17) \citet{Vennes1999}; (18) \citet{kaw2006};
(19) \citet{put1995}
\end{minipage}
\normalsize
\end{landscape}

\blanknonumber
\chapter{Conclusions}
\label{Chapter 6}
\vspace{-9mm}

The realisation that there is a population of isolated WDs with extremely high magnetic fields led to attempts to
explain their origin. The aim of this study was to model the evolution of a synthetic population of binary stars to test the
hypothesis of \citet{tout2008} regarding the origin of high magnetic fields in WDs. Two main mechanisms
for the origin of these intense fields have been proposed. They are the fossil field model and the merging star model.
The fossil field model \citep[as first proposed by][]{Woltjer64, landstreet1967} is based upon the fact that early type
stars of classes Ap and Bp have maximum poloidal magnetic fields of similar strength to those observed in the magnetic
WD.  It was therefore supposed that the strong fields are maintained through the evolution of the star. However no model
of stellar evolution has been able to explain how a strong fossil magnetic field can survive through the various stages of
evolution from main sequence to WD.  Nor has it been possible to show that there is a common cause for the observed
maximum magnetic field strengths in the two classes of stars.

In the case of the WDs, a clue to the origin of their fields, in both isolated and binary systems, is given by the study of their
binary properties \citep{liebert2005, lie2015}.  This led \citet{tout2008} to propose that the origin of
magnetic fields in WDs is related to their duplicity and stellar interaction during CEE \citep{Pac1976}.
Additionally the dynamo model of \citet{wickramasinghe2014} provides physical reasons for similar maximum magnetic fluxes
in the magnetic main sequence stars and the isolated MWDs if the fields are generated from differential rotation
caused by merging.

The two stars that comprise a binary system are born from the protostellar cloud at about the same time. The more massive
of the two, initially both main sequence stars, evolves faster and it is the first to evolve along the RGB.
Here, its envelope expands greatly and engulfs the secondary star. The drag encountered by the secondary star, now within
the envelope of the primary, causes the orbit of the two stars about the common centre of gravity to decay. As a consequence
the two stars may merge at this stage. If they do not merge, more CE episodes can occur perhaps during the evolution of
the binary on the AGB and merging may occur at this point.  

The proposal of \citet{tout2008} posits that the high magnetic fields found in the sub population of isolated MWDs are
caused by differential rotation of the envelope caused by the orbital revolution of the secondary within the CE.  The first condition
for the formation of high fields is that during CEE the core of the pre--WD star is degenerate.  The second condition is no nuclear
burning should occur from CEE until the time the star reveals its core as a WD.  The requirement is necessary for the formation
of high magnetic fields because any nuclear burning would create convection and destroy any strong ordered field created.

\vspace{-3mm}
\section{The Study Method in Review}
{\vspace{-4mm}
In order to test the viability of this model, I have modified the {\sc{bse}} (Binary Star Evolution) code of \citet{Hurley2002}.
This code builds on the {\sc{sse}} (Single Star Evolution) code of \citet{Hurley2000} by adding the major phenomena of
binary evolution comprising Roche lobe overflow, CE evolution, tidal interaction, collisions, gravitational radiation
and magnetic braking.  For the evolution through the CE phase {\sc{bse}} uses the energy formulation with the
variable $\alpha$ as the efficiency of envelope removal ($0.0\le\alpha\le1.0$). In this study I have investigated how $\alpha$
affects the binary evolution outcome.

I created a synthetic population of binaries by three parameters at the ZAMS, mass of the primary,
mass of the secondary and the orbital period, each with two hundred members in a log scale.  The mass of the secondary
was constrained to be less than the mass of the primary.  In all cases I assumed an initially circular orbit for the progenitor binary.
The calculations were repeated for all $\alpha\in$\,\{0.10, 0.20, 0.25, 0.30, 0.40, 0.50, 0.60, 0.70, 0.80 and 0.90\}.  I took
the metallicity to be solar, $Z$\,=\,0.02, and $\eta$\,=\,1.0 for the Reimers' mass loss parameter as outlined in chapter 2.

This gave me a synthetic population of about seven million binaries covering all combinations of the above parameters. The evolution
of each binary system was examined for satisfaction of the CE degeneracy and nuclear burning criteria detailed above.
The qualifying systems gave the required CEE population of binaries numbering about one million.  The matching
binaries were then weighted according to the \citet{Salpeter1955} initial mass function and integrated from the ZAMS to the age of the
Galactic disk (9.5GYr, \citet{Kilic2017}). The properties of the pre and post CE binaries and/or remnant WDs could then be
extracted and analysed.

\vspace{-4mm}
\section{The Isolated Magnetic WDs}
\vspace{-4mm}
In chapter 2 of this work the mass distributions of the theoretical high field magnetic WDs (HFMWDs) were reported.
It was found that for models with $0.1\le\alpha\le0.3$ the mean predicted mass of isolated HFMWDs is 0.88\,\msun. Observations
indicate mean masses of $0.85\,\msun$ \citep{Kepler2013} for HFMWDs and $0.62\,\msun$ \citep{Kleinman2013} for all non-magnetic
Galactic field WDs. A K--S test \citep{Press1992} on the masses of the theoretical HFMWDs against the few reliably measured observed HFMWDs gave a probability of 0.71 that they were from the same distribution, while a K--S test of theoretical HFMWDs
against observed non--magnetic WDs only gave a probability of $3\times10^{-5}$ that the masses had the same distribution. 

One of the main results from this work is that there are two possible paths to HFMWDs. In the CE path merging occurs during
CEE. In the other path the merging occurs after both stars have evolved to the WD stage and merge as double degenerates (DD path).
The CE path greatly outnumbers the DD path for all $\alpha$.  The CE path yields mainly CO WDs with small numbers of He and ONe WDs
while the DD path yields only CO WDs which populate the high end of the mass distribution. These calculations, when taken together
with the observation that there are no examples of HFMWDs in detached binary systems, argue strongly in favour of the CE merging
hypothesis for the formation of HFMWDs. Thus the progenitors of HFMWDs are the RGB or AGB 
degenerate cores of stars that merge with their lower-mass companions and then continue their evolution, as single stars no longer
undergoing nuclear fusion, to the WD stage.

In chapter 3 I have assumed that the magnetic field strength induced in the core of the single coalesced star emerging from
CEE is proportional to the orbital angular velocity of the binary at the point that the secondary is disrupted by
tidal forces and merges with the primary as first proposed by \citet{Reg95, Tout95b, tout2008, wickramasinghe2014}. The maximum field strength
that can be achieved by a compact core during a merging process is limited by the break--up angular velocity and this can only be
reached if the merging stars are in a very compact binary, such as a merging DD system.

In this model two parameters must be empirically estimated.  These are $B_0$, which is linked to the efficiency with which the poloidal
field is regenerated by the decaying toroidal field \citep[see][]{wickramasinghe2014} and the CE efficiency parameter $\alpha$.
Having calculated the CDFs of the observed and theoretical field distributions, a K--S test was carried
out between the observed and theoretical field distributions for a wide range of $B_0$.  From the K--S probability, $P$, I determined
that the observed field distribution is best fitted by models characterised by $B_0=1.35\times10^{10}$\,G and $\alpha=0.2$ with $P$=0.61,

It is also speculated that close stellar encounters can send a giant gaseous planet from the outer regions of a WD's planetary system
into a highly eccentric orbit.  The plunging of this super--Jupiter into the WD can generate a magnetic field and thus provide an
answer as to why magnetism among cool WD, and particularly among cool DZ WDs, is higher than among hot WDs.

\vspace{-2mm}
\section{The Magnetic Cataclysmic Variables}
\vspace{-4mm}
Population synthesis studies of binary systems that survive the CE environment without merging can go on
to form MCVs.  By making use of synthetic population modelling similar to that for the
isolated HFMWDs it was shown that the same $B_0$ can also explain the magnetic field distribution of magnetic binaries.

The population synthesis study of binary systems carried out for the HFMWDs was modified to explain the origin of fields
in the accreting WD in MCVs and the results are shown in chapter 4.  The CE efficiency parameter
$\alpha$ was again varied to investigate its effects on the resulting synthetic population of MCVs. This showed that models
with $\alpha\,>\,0.4$ are not able to reproduce the large range of WD masses, field strength, secondary types and
masses that are observed in MCVs.

Quantitative support in favour of models with $\alpha \le 0.4$ was given by K--S tests conducted to compare the synthetic
WD mass and magnetic field distributions with the observed populations. However, once again, it must be stressed
that there are some shortcomings of this work and in particular those that arise from our comparison to observations.

Many of the parameters of the Galactic populations of MCVs and pre-polars (PREPs) that are needed for comparison studies
(e.g., WD mass, magnetic field, secondary star mass and type, and orbital period) are often hard to determine owing
to evolutionary effects and observational biases that are difficult to separate.  For example, magnetic WDs in PREPs
are the best objects with which to compare our theoretical results and in particular the mass distribution, because their mass
has not been corrupted by accretion processes.  However the members of this population are scarce.  Owing to
accretion and nova explosions nor can the much larger sample of MCVs be used for comparison purposes of the WD mass
distribution, because masses vary over time. Instead the sample provided by the non magnetic pre CVs of \citet{Zorotovic2011}
was used.  This is not an ideal sample but it is the best currently at our disposal until a statistically significant number of PREPS
has been discovered.  A K--S test on the observed Pre CV masses against the theoretical population at the start of Roche lobe overflow
(RLOF) gave a probability of a match of the two populations of $P$=0.95 at $\alpha$=0.10 while the K--S test of the observed
population of MCVs from \citet{fer2015a} against the theoretical population only gave an $P$ = 0.02 at $\alpha$ = 0.1
indicating that WD masses do vary over time.

The situation is somewhat alleviated when we consider the magnetic field distribution because fields are not expected to
change over time \citep{fer2015a}.  However, the observed magnetic field distribution of MCVs may not be reliable because
it may be afflicted by observational biases. For example, at field strengths below a few $10^7$\,G, most systems (the
intermediate polars) have an accretion disc from which continuum emission and broad emission lines swamp the Zeeman
and cyclotron spectral features that are essential to determine their field strengths arising from the WD surface \citep{fer1992}.
Very high field polars are also likely to be under represented in the observations because mass accretion from the companion
is hampered by the presence of strong fields \citep{fer1989, liu2000} making these systems very dim wind accretors. Despite
these restrictions, the characteristics of the MCVs are generally consistent with those of a population of binary systems that
are born already in contact (exchanging mass) or close to contact, as suggested by \citet{tout2008} and is generally in
agreement with the speculation of \citet{sch2009} that the binaries known as PREPs, where a HFMWD accretes matter from
the wind of a low mass companion, are the progenitors of the MCVs.

\vspace{-2mm}
\section{Evolution of a Double Degenerate System: NLTT\,12578}
\vspace{-3mm}
The binary system NLTT\,12758 is a close super Chandrasekhar double degenerate system consisting of two CO\,WDs of similar
masses ($0.83\,\msun$ and $0.69\,\msun$). One of the two components is highly magnetic and spins around its axis with a period
of 23\,min. The orbital period was determined to be 1.15\,d (See chapter 5).

In order to understand the evolution of NLTT\,12758 I created a number of binaries and evolved them from the ZAMS to the age
of the Galactic disc. The mass of the primary star was chosen to vary between 3.5\,$\msun$ and 4.5\,$\msun$ and the mass of
the secondary star between 2.5\,$\msun$ and 3.5\,$\msun$ and the initial period in the range 2000 to 3500\,d. These values
were chosen because they were in the region of the expected initial conditions that would lead to the final properties of the
components of NLTT\,12758.

The evolved populations were searched for systems that resulted in a pair of CO WDs at the correct stage of evolution.  It was
found that, as $\alpha$ increases, the number of CO WD double degenerate systems increases.  However, not all of these systems
correspond with the type of evolution path that would lead to the final parameters of NLTT\,12758, its cooling age, period and
mass.  A suitable near match to NLTT\,12758 was achieved at $\alpha$\,=\,0.15 with initial masses of 3.75\,{\msun} and
2.80\,{\msun} and with an initial period of 2656\,d.

The future evolution of NLTT\,12758 will see the orbit shrinking further until at some stage Roche lobe overflow will restart
and the two stars will coalesce in approximately 140\,Gyr. At this time they will undergo either an accretion induced collapse
to become a rapidly spinning neutron star (an isolated millisecond pulsar) or undergo a Type\,Ia supernova explosion.

\vspace{-1mm}
\section{A Final Word}
\vspace{-4mm}
The CE merging theory for the formation of high fields in magnetic WD as proposed by \citet{tout2008}
has successfully predicted the mass distribution of the observed HFMWDs and shown that it is significantly different from that
of Galactic field WDs.  My modelling has also successfully computed the magnetic field distribution of HFMWDs.  In
addition it has accurately predicted the post CE characteristics of those systems that become MCVs.

For the double degenerate WD system NLTT\,12758, I was able to successfully construct the history of the evolution
of the system and explained the genesis of the magnetic field in one of the components of the binary.

I can therefore conclude that the CEE theory of the genesis of high magnetic fields in isolated WDs, in cataclysmic variables
and in double degenerate systems is the most likely so far to explain all observations of these systems.

\blanknonumber


\end{document}